\documentclass[%
 reprint,
superscriptaddress,
nofootinbib,
 amsmath,amssymb,
 aps,
prb,
floatfix,
]{revtex4-2}

\usepackage{graphicx}% Include figure files
\usepackage{dcolumn}% Align table columns on decimal point
\usepackage{bm}% bold math
\usepackage{accents}
\usepackage{xcolor}
\usepackage{mathrsfs}
\usepackage{physics}
\usepackage{braket}
\usepackage{microtype}
\usepackage{mathtools}
\usepackage{bigints}
\usepackage{amsfonts}
\usepackage{amsmath}
\usepackage{caption}
\usepackage{subcaption}
\usepackage{float}
\usepackage{esint}
\usepackage{nicefrac}
\usepackage{enumerate}
\usepackage{hyperref}% add hypertext capabilities
\usepackage[noabbrev]{cleveref}

\usepackage{multirow}
\usepackage[linesnumbered,ruled,vlined]{algorithm2e}

\newcommand{\qe}{\texttt{QE}}

\newcommand{\pawfe}{\texttt{PAW-FE}}
\newcommand{\abinit}{\texttt{ABINIT}}
\newcommand{\bR}{\boldsymbol{\textbf{R}}}
\newcommand{\br}{\boldsymbol{\textbf{r}}}
\newcommand{\bc}{\boldsymbol{\textbf{c}}}
\newcommand{\bv}{\boldsymbol{\textbf{v}}}
\newcommand{\bu}{\boldsymbol{\textbf{u}}}
\newcommand{\bT}{\boldsymbol{\textbf{T}}}
\newcommand{\bJ}{\boldsymbol{\textbf{J}}}
\newcommand{\bx}{\boldsymbol{\textbf{x}}}
\newcommand{\bX}{\boldsymbol{\textbf{X}}}
\newcommand{\bY}{\boldsymbol{\textbf{Y}}}
\newcommand{\by}{\boldsymbol{\textbf{y}}}
\newcommand{\bk}{\boldsymbol{\textbf{k}}}
\newcommand{\bL}{\boldsymbol{\textbf{L}}}
\newcommand{\bH}{\boldsymbol{\textbf{H}}}
\newcommand{\bS}{\boldsymbol{\textbf{S}}}
\newcommand{\btS}{\boldsymbol{\textbf{\tiny{S}}}}
\newcommand{\btH}{\boldsymbol{\textbf{\tiny{H}}}}
\newcommand{\bA}{\boldsymbol{\textbf{A}}}
\newcommand{\bB}{\boldsymbol{\textbf{B}}}
\newcommand{\bQ}{\boldsymbol{\textbf{Q}}}
\newcommand{\bZ}{\boldsymbol{\textbf{Z}}}
\newcommand{\bP}{\boldsymbol{\textbf{P}}}
\newcommand{\bM}{\boldsymbol{\textbf{M}}}
\newcommand{\bD}{\boldsymbol{\textbf{D}}}

\newcommand{\bC}{\boldsymbol{\textbf{C}}}
\newcommand{\bV}{\boldsymbol{\textbf{V}}}
\newcommand{\bq}{\boldsymbol{\textbf{q}}}
\newcommand{\feNodalVectorUtilde}{\widetilde{\boldsymbol{\mathsf{u}}}}
\newcommand{\bLambda}{\boldsymbol{\Lambda}}
\newcommand{\bPhi}{\boldsymbol{\textbf{$\Phi$}}}
\newcommand{\psiae}{\psi}
\newcommand{\psips}{\widetilde{\psi}}
\newcommand{\Psips}{\widetilde{\boldsymbol{\Psi}}}
\newcommand{\phiae}{\phi}
\newcommand{\phips}{\widetilde{\phi}}
\newcommand{\nTilde}{\widetilde{n}}
\newcommand{\uTilde}{\widetilde{u}}
\newcommand{\gTilde}{\widetilde{g}}
\newcommand{\pTilde}{\widetilde{p}}
\newcommand{\rhoTilde}{\widetilde{\rho}}
\newcommand{\bTilde}{\widetilde{b}}
\newcommand{\fePsips}{\widetilde{u}}
\newcommand{\bfePsips}{\widetilde{\boldsymbol{u}}}
\newcommand{\bfePsipsFull}{\widetilde{\boldsymbol{\mathsf{U}}}}
\newcommand{\DFTFE}{\texttt{DFT-FE}}

\newcommand{\cn}{\color{black}}
\newcommand{\cb}{\color{black}}

\begin{document}

%\preprint{APS/123-QED}

\title{Fast and scalable finite-element based approach for density functional theory calculations using projector-augmented wave method}

%\title{Higher order spectral finite-element based methodologies for projector augmented wave formalism in density functional theory calculations}% Force line breaks with \\

\author{Kartick Ramakrishnan}
\affiliation{% 
Department of Computational and Data Sciences, Indian Institute of Science Bengaluru, India
}%
\author{Sambit Das}%

\affiliation{%
Department of Mechanical Engineering, University of Michigan, Ann Arbor, Michigan 48109, USA
}%
\author{Phani Motamarri}%
\email{phanim@iisc.ac.in}
\affiliation{%
Department of Computational and Data Sciences, Indian Institute of Science Bengaluru, India
}%


\begin{abstract}
In this work, we present a computationally efficient methodology that utilizes a local real-space formulation of the projector augmented wave (PAW) method discretized with a finite-element (FE) basis to enable accurate and large-scale electronic structure calculations. To the best of our knowledge, this is the first real-space approach for DFT calculations, combining the efficiency of PAW formalism involving smooth electronic fields with the ability of systematically improvable higher-order finite-element basis to achieve significant computational gains. In particular, we developed efficient strategies for solving the underlying FE discretized PAW generalized eigenproblem by employing the Chebyshev filtered subspace iteration approach to compute the desired eigenspace in each self-consistent field iteration. These strategies leverage the low-rank perturbation of the FE basis overlap matrix in conjunction with reduced order quadrature rules to invert the discretized PAW overlap matrix while also exploiting the sparsity of both the local and non-local parts of the discretized PAW Hamiltonian and overlap matrices. Further, we employ higher-order quadrature rules to accurately evaluate integrals in these matrices involving PAW-atomic data, allowing the use of coarser FE meshes for various electronic fields. Using the proposed approach, we benchmark the accuracy and performance of various representative examples involving periodic and non-periodic systems with plane-wave-based PAW implementations. Furthermore, we also demonstrate a considerable computational advantage ($\sim$~5$\times$ -- 10$\times$) over state-of-the-art plane-wave methods for medium to large-scale systems ($\sim$~6,000 -- 35,000 electrons). Finally, we show that our approach (\pawfe) significantly reduces the degrees of freedom to achieve the desired accuracy, thereby enabling large-scale DFT simulations ( $>$ 50,000 electrons) at an order of magnitude lower computational cost compared to norm-conserving pseudopotential calculations using finite-element discretization.
\end{abstract}

\maketitle

\section{\label{sec:Intro}Introduction}
Electronic structure calculations based on Kohn-Sham density functional theory (DFT) have significantly advanced our understanding of material properties. The Kohn-Sham approach to DFT~\cite{kohn64,kohn65} reformulates the many-body problem of interacting electrons into an equivalent computationally tractable problem of non-interacting electrons within an effective mean field governed by electron density.  This reformulation has made it possible to predict a wide range of material properties, including mechanical, transport, electronic, magnetic and optical properties, over the past six decades. Pseudopotential approximations in DFT~\cite{bachelet82, chelik2000, peter2011} have made these calculations computationally efficient, as they focus only on electrons that participate in chemical bonding. 
To this end, these approximations transform the all-electron DFT calculations into an equivalent problem involving auxiliary smooth pseudo wavefunctions by lumping the effect of core electrons and the singular Coloumb potential within a smooth effective external potential (the pseudopotential). Numerically, pseudopotential DFT calculations solve for only those electronic wavefunctions that participate in chemical bonding and, hence, are computationally efficient. Additionally, this approach deals with smoother effective potentials, enabling the use of coarser grids or lower plane-wave energy cutoffs, which sufficiently resolve the smooth pseudo-electronic fields. Over the years, pseudopotentials have evolved from norm-conserving~\cite{hamann79,vanderbilt85,hamann89,tm91} to ultrasoft potentials~\cite{vanderbilt90}, and now to the state-of-the-art projector augmented wave (PAW)~\cite{PAWBlochl1994} method, enabling accurate predictions of material properties across diverse material systems without the need for expensive all-electron DFT calculations. However, the stringent accuracy required for these pseudopotential DFT calculations to compute meaningful properties, combined with cubic-scaling computational complexity with the number of electrons, requires significant computational resources, which typically limit their applicability to material systems with at most few thousands of electrons. Overcoming the current limits of \emph{ab-initio} calculations has been a decades-long pursuit by researchers around the world, and efforts over the past two decades have focused on developing accurate and computationally efficient methods for solving the Kohn-Sham DFT problem. These include efficient discretization methods for solving DFT equations, reduced-scaling algorithms for lowering the computational complexity of the DFT problem, and efforts to improve the parallel scalability of widely used codes. Among the various discretization methods, the plane wave (PW) basis is the most popular method of choice for metallic and heterogeneous material systems due to the computational efficiency afforded by these basis sets owing to its spectral convergence. Despite the popularity of the PW basis for DFT calculations, the limitations of these Fourier methods stemming from the poor scalability on parallel computing architectures,  inefficiency in treating non-periodic systems, and inability to handle generic boundary conditions led to the development of systematically improvable, efficient, and scalable real-space discretization techniques based on finite-difference~\cite{parsec2006,rescu2016,sparc2017a,octopus2015,GPAW2010}, finite-elements~\cite{white1989,tsuchida1995,tsuchida1996,pask1999,goedecker99,pask2005,bylaska,suryanarayana2010,MOTAMARRI2013308,SCHAUER2013644,zhou2014,denis2016,Bikash2017,kanungo2019real,dftfe0.6,Das2019FastSystem}, wavelets~\cite{bigdft2008,bigdftgpu2009,bigdftlinearscaling2014} and other reduced-order basis techniques~\cite{dgdft2015,motamarri2016,xu2018,lin2021a,lin2021b}.
 
\cb The finite-element (FE) basis, a compactly supported piece-wise polynomial basis, offer several advantages over other commonly used basis sets for DFT calculations\cn.  FE basis sets are systematically convergent for any material system and can accommodate generic boundary conditions including fully periodic, semi-periodic, and non-periodic. Their locality allows one to exploit fine-grained parallelism on modern heterogeneous architectures while providing excellent parallel scalability on distributed systems\cite{JPDC,Das2023}. As demonstrated \cb in the recent works\cn~\cite{Das2019FastSystem,dftfe0.6}, finite-element-based methods have significantly outperformed plane-wave basis in the case of norm-conserving pseudopotential DFT calculations, as the system size increases for a given accuracy of ground-state energy and forces. Furthermore, the adaptive resolution offered by FE basis sets enables all-electron calculations~\cite{MOTAMARRI2013308,Bikash2017}, mixed all-electron and pseudopotential calculations~\cite{Bikash2017,ghosh2021}. The open-source code DFT-FE~\cite{dftfe1.0}, the workhorse behind the ACM Gordon Bell Prize 2023, inherits these features while incorporating scalable and efficient solvers for the solution of the Kohn-Sham equations. DFT-FE code has shown excellent scalability on massively parallel many-core CPU and hybrid CPU-GPU architectures (tested up to ~200,000 cores on many-core CPUs and ~40,000 GPUs on hybrid CPU-GPU architectures), while simulating material systems as large as 600,000 electrons~\cite{Das2019FastSystem,Das2023}. Furthermore, DFT-FE has been recently employed in a range of scientific studies involving large-scale DFT calculations on material systems with tens of thousands of electrons: computing the electronic structure of large DNA molecules~\cite{Zhuravel2020}, predicting energetics of extended defects in crystalline materials~\cite{MENON2024119515,KUMAR2023103613,SHOJAEI2024105726,Das2019FastSystem,Das2023}, investigating phase transformations in doped nanofilms~\cite{yao2022modulating}, evaluating spin Hamiltonian parameters of systems with defects~\cite{Krishnendu2019,ghosh2021} and solving the inverse DFT problem~\cite{Kanungo2019} 

We note that the aforementioned finite-element-based DFT calculations are all limited to the use of norm-conserving pseudopotentials. These pseudopotentials can become very hard for material systems with elements involving $d$ or $f$ electrons, and these calculations can be computationally demanding. Moreover, these often require semi-core states to be treated as valence states for better transferability, resulting in harder pseudopotentials with more valence electrons, particularly for alkali, alkaline earth metals, or early transition metals. Previous works~\cite{dftfe0.6,dftfe1.0} have shown that these hard norm-conserving pseudopotentials require a large number of degrees of freedom (30,000 to 50,000 finite-element basis functions per atom) to achieve chemical accuracy, leading to high computational costs for solving the Kohn-Sham DFT eigenvalue problem. Although ultra-soft pseudopotentials relax the norm conservation condition~\cite{vanderbilt85}, resulting in softer potentials, their construction is complex and requires extensive testing to ensure accuracy and transferability. To address these issues, Blöchl~\cite{PAWBlochl1994} developed the projector augmented wave (PAW) method, formally an all-electron approach that generalizes the LAPW method and the pseudopotential approach. The PAW method recovers all-electron wavefunctions by introducing a linear transformation operator ($\mathcal{T}$) acting on smoothly varying pseudo wavefunctions. Within the frozen core approximation, it treats core states as frozen and retrieves the oscillatory behavior of all-electron valence wavefunctions near atomic centers through the action of $\mathcal{T}$ on the pseudo smooth valence wavefunctions. This allows one to reformulate the all-electron Kohn-Sham DFT eigenproblem into an equivalent PAW eigenvalue problem involving only pseudo smooth valence wavefunctions, striking a balance between accuracy and computational efficiency. Consequently, the PAW approach within the frozen core approximation has become widely adopted in popular plane-wave based DFT codes~\cite{Kresse1999,abinitPAW2008,Abinit2010,qe,Blochl2003,tackettpwpaw}.

The real-space discretization techniques have also been employed in the context of PAW implementations and are primarily based on finite-difference approximations~\cite{GPAW2010,mortensen2005}, wavelet based methods~\cite{rangel2016} and psinc basis~\cite{hine2016}. However, most of these PAW-based DFT calculations deal with material systems involving a few thousand electrons except the linear scaling DFT approach~\cite{hine2016} using the PAW formalism that was applied to material systems with a band-gap. The computational efficiency and scalability of these real-space implementations compared with state-of-the-art plane-wave-based PAW implementations remain unknown. This work introduces the local real-space formulation of the PAW method amenable to finite element (FE) discretization and proposes efficient strategies to solve the underlying FE-discretized PAW generalized eigenvalue problem using a self-consistent field iteration approach. Our methodology significantly reduces the FE degrees of freedom required to achieve chemical accuracy using the PAW formalism, outperforming the optimized norm-conserving (ONCV) pseudopotential based DFT calculations employed in previous FE based approaches for a given system size. Additionally, we demonstrate that our method can handle fully periodic, non-periodic, and semi-periodic boundary conditions for generic material systems as large as 50,000 electrons. Moreover, our method offers substantial computational efficiency over state-of-the-art plane-wave-based approaches for systems with more than 5,000 electrons.  Notably, to the best of our knowledge this represents the first development of a fast and scalable approach for the PAW method within the framework of finite-element discretization.

We adopt the original Bl\"{o}chl's~\cite{PAWBlochl1994} PAW method in our current work and do not employ the Kresse's modifications~\cite{Kresse1999} that include the compensation charge contribution for computing the exchange-correlation term, which can increase computational costs and introduce numerical instabilities~\cite{Abinit2010,abinitPAW2008}. The key aspects of the proposed finite-element (FE) approach for the PAW method include: (i) local reformulation of the extended PAW electrostatics involving pseudo smooth electron density and the compensation charge making it amenable for FE discretization; (ii) deducing the discretized governing equations by evaluating the variational derivative of the locally reformulated all-electron DFT energy functional in the PAW method; (iii) employing higher-order quadrature rules to efficiently and accurately evaluate integrals involving atom-centered functions, allowing the use of coarser FE grids for electronic fields and yet capture the relevant PAW atomic data; (iv) implementation of a self-consistent field (SCF) iteration approach based on Anderson mixing of total charge density employing an efficient real-space strategy to minimize the weighted total charge density residual norm; (v) an efficient scheme to evaluate the inverse of FE discretized PAW overlap matrix $\bS$ using Woodbury's formula in conjunction with Gauss-Lobatto-Legendre quadrature rules and a block-wise matrix inversion of the overlap matrix corresponding to the atomic projectors in real-space; and (vi) leveraging this efficient computation of $\bS^{-1}$ to design a Chebyshev polynomial based filter~\cite{SaadChFSIBook} in each SCF iteration to obtain a subspace rich in desired eigenvectors, and solving the PAW generalized eigenvalue problem by projecting onto this filtered space. The proposed formulation has been implemented in a distributed setting on CPUs using the message passing interface (MPI) for communication across multiple nodes to enable large-scale PAW based DFT calculations. We benchmark the accuracy and performance of our method against plane-wave-based PAW implementations on various representative non-periodic, semi-periodic, and fully periodic systems. Compared to reference data from plane-wave calculations, our results show excellent agreement in ground-state energies, energy variation with bond length and lattice parameters, formation energies, surface energies, and band structures. Relative to widely used plane-wave based PAW implementations, our finite-element based approach demonstrates significant computational advantage both in terms of CPU node-hrs and minimum wall time with increasing systems sizes reaching up to 5$\times$ and 10$\times$, respectively, for non-periodic systems ($\sim$18,000 electrons) and  3$\times$ and 7.5$\times$, respectively, for periodic systems ($\sim$34,000 electrons). Our implementation also exhibits excellent parallel scaling efficiency ($\sim$70\%) at 1200 degrees of freedom per MPI rank for material systems, achieving nearly a 10-fold speedup over the state-of-the-art DFT-FE\cite{dftfe0.6,dftfe1.0} code using norm-conserving ONCV pseudopotentials for system sizes close to 50,000 electrons, with a significant reduction in computational resource requirements.
 
 The remainder of this manuscript is organized as follows. Section~\ref{sec:realSpace} briefly introduces the Bl\"{o}chl PAW method followed by a discussion on the local real-space formulation of the PAW method and the governing equations to be solved. Subsequently, Sec.~\ref{sec:fem} introduces the finite-element discretization for the PAW governing differential equations, and Sec.~\ref{sec:scf} details the proposed computational methodology in achieving an efficient and scalable solution procedure for solving the FE discretized PAW generalized eigenproblem using a self-consistent field iteration approach. Section~\ref{sec:results} presents comprehensively accuracy and performance benchmarks of our implementation comparing with state of the art PAW based plane-wave(PW) codes on representative benchmark systems. Additionally, this section presents a comparative study of the current work and the norm-conserving pseudopotential (ONCV) calculations in the DFT-FE code in terms of parallel scalability and time to solution. Finally, we conclude with a short discussion summarizing the key findings and outlining future prospects arising from this work in Sec.~\ref{sec:concl}

\section{Real-space formulation of projector augmented wave method}\label{sec:realSpace}
%In this work, we refer to atom-centered fields with superscript '$a$', evaluated at $\bx-\bR^a$ where $\bx$ is any material point in the domain and $\bR^a$ is the atom position vector.
%In the infinite-periodic system, the index $a$ refers to atoms at all lattice sites with support in the unit cell of interest, while for a finite system, it refers to all the atoms in the system.

%The canonical electronic wavefunctions $\{ \psiae_i\}$ that are the minimizers of the all-electron Kohn-Sham energy functional exhibit high oscillations near the core while being smooth in the bonding region. Achieving accurate numerical solutions necessitates using a finer grid spacing (or equivalently, higher cut-off energies in Fourier space) to resolve these all-electron wavefunctions. Consequently, this leads to a huge increase in the number of basis functions per atom to reach the desired level of chemical accuracies.
%Bl\"ochl in his Projector Augmented Wave (PAW) formalism \cite{PAWBlochl1994} addresses this challenge 

 %In summary, the computation of the all-electron ground-state energy using an energy contribution involving smoother fields reduces the computational effort.
The ground-state properties of a material system comprising of $N_a$ nuclei and $N_e$ electrons are governed by the following all-electron Kohn-Sham DFT energy functional \cite{parr1979local,finnis2003interatomic}
%located at positions $\{ \bR^1, \bR^2,...\}$
%satisfying the orthogonality constraint $\int{\psiae^*_n(\bx)\psiae_m(\bx)d\bx} = \delta_{\text{nm}}$
\begin{multline}\label{eqn:allElecKS}
E\left[\{ \psiae_i \}, \{ \bR^a\} \right] = T_s\left[\{ \psiae_i \} \right] + E_\text{xc}[n(\bx)] + J\left[\rho(\bx),\{ \bR^a\}\right ]  
\end{multline}
where $\{ \psiae_i \} $ and $\{ \bR^{a}\}$ denote the set of single-electron orthonormal wavefunctions and the set of nuclear position vectors respectively. We restrict our discussion to the spin-unpolarized case, whereas the extension to spin-polarized systems is straightforward and not elaborated here. The term $T_s\left[\{ \psiae_i \} \right]$ in Eqn.~\eqref{eqn:allElecKS} denotes the kinetic energy of the non-interacting electrons where
\begin{equation}
 T_s\left[\{ \psiae_i \} \right] = 2\sum_i^N{f_i\int{\psiae_{i}^*(\bx)\left(-\frac{1}{2}\nabla^2\right)\psiae_{i}(\bx)d\bx}}, \label{eqn:Ts}
\end{equation}
with $f_i$ denoting the orbital occupancy function and the integer $N \geq N_e/2$ denotes the number of single-electron wavefunctions. Further, $E_\text{xc}[n(\bx)]$ in Eqn.~\eqref{eqn:allElecKS} is the exchange-correlation energy that accounts for the quantum-mechanical many-body effects. Adopting the generalized gradient approximation (GGA) \cite{Langreth19831809, martin2020electronic,GGA1997} for the exchange-correlation functional description in this work, we have
\begin{equation}
    E_{\text{xc}}[n(\bx)] = \int{\epsilon_{\text{xc}}\left[n(\bx),\nabla n(\bx)\right]d\bx}. \label{eqn:exc}
\end{equation}
where $n(\bx)= 2\sum_i^N{f_i|\psiae_i(\bx)|^2}$ represents the all-electron density. The total electrostatic energy in Eqn.~\eqref{eqn:allElecKS} is denoted as $J\left[\rho(\bx),\{ \bR^a\}\right] = E_\text{el}\left[\rho(\bx)\right] - E_\text{self}\left[ \{\bR^a\}\right]$. Here, $E_\text{el}$ is the electrostatics energy associated with the total charge $\rho(\bx)=n(\bx)+ \sum_a{b^a(\bx)}$, where $b^a(\bx)=-\mathcal{Z}^a\delta(\bx - \bR^a)$ denotes the nuclear charge density with $\mathcal{Z}^a$ being the atomic number of  $a^\text{th}$ atom.  This electrostatics energy $E_\text{el}$ involves a convolution in real-space and is given by:
\begin{equation}\label{eqn:Eel}
    E_\text{el}\left[\rho\right ] = \frac{1}{2}\int{\int{\frac{\rho(\bx)\rho(\by)}{|\bx-\by|}d\bx}d\by}
\end{equation}
where the domain of integration for $\by$ in the above Eqn.~\eqref{eqn:Eel} is $\mathbb{R}^3$, while the domain of integration for $\bx$ in the case of an infinite-periodic system is over the unit-cell of interest, whereas in the case of a finite system, it is over $\mathbb{R}^3$. Further, for a periodic system, the index $a$ refers to atoms at all lattice sites with the compact support of $b^a(\bx)$ lying in the unit cell of interest. In contrast, for a finite system, it refers to all the atoms in the system. Finally, denoting the self-interaction energy corresponding to the nuclear charge density $b^a(\bx)$ by $E^a_\text{self}[\bR^a]$, we have the total self-interaction energy as $E_\text{self}\left[ \{\bR^a\}\right] = \sum_a{E^a_\text{self}[\bR^a]}$.

Within the frozen-core approximation in PAW method, the valence all-electron wavefunctions($\{\psiae_i\}$) that are orthogonal to the core wavefunctions are recovered from pseudo-smooth(PS) wavefunctions ($\{\psips_i\}$) using a linear transformation operator($\mathcal{T}$) ~\cite{PAWBlochl1994} given by:
\begin{equation} \label{eqn:transf}
    \ket{\psiae_i} = \mathcal{T} \ket{\psips_i} = (\mathcal{I} + \sum_a \mathcal{T}^{a}) \ket{\psips_i}
\end{equation}
The index $i = \{1,2...,N\}$ in Eqn.~\eqref{eqn:transf} runs over the valence states with  $N \geq N_v/2$ where $N_v$ denotes the number of valence electrons. Furthermore, the linear transformation operator ($\mathcal{T}$) in Eqn.~\eqref{eqn:transf} is defined such that it differs from the Identity operator ($\mathcal{I}$) by a set of atom-specific transformations $\mathcal{T}^{a}$ that has an effect only inside a certain atom-centered ball $\Omega_{\text{a}}$ with radius $|\bx - \bR_a| < r_{c}^{a}$ referred to as augmentation sphere. Consequently, the all-electron valence wavefunctions can be expressed as
\begin{equation}\label{eqn:wfcdecomp}
\psiae_i(\bx) = \psips_i(\bx) + 
\sum_a{\sum_\alpha{\left(\phiae^a_\alpha(\bx)-\phips^a_\alpha(\bx)\right)}\braket{\pTilde^a_\alpha|\psips_i}} 
\end{equation}
where $\{\phiae^a_\alpha\}$ denote the all-electron partial waves, $\{ \phips^a_\alpha \}$ denote the pseudo-smooth partial waves that are used to expand $\{\psiae_i(\bx)\}$ and $\{\psips_i(\bx)\}$ inside the augmentation sphere respectively. Additionally, $\{ \pTilde^a_\alpha\}$ in Eqn.~\eqref{eqn:wfcdecomp} are the projector functions that are constructed to be orthogonal to the pseudo smooth partial waves i.e., $\braket{\pTilde^a_\alpha|\phips^a_\beta} = \delta_{\alpha \beta}$. In the above Eqn.~\eqref{eqn:wfcdecomp} and subsequently in the manuscript, the atom-centered fields are denoted with superscript `$a$', implicitly denoting their functional dependence on $(\bx - \bR^{a})$. Furthermore, the atom-centered functions $\{\pTilde^a_\alpha\},\, \{\phiae^a_\alpha\},\, \{\phips^a_\alpha\}$ in Eqn.~\eqref{eqn:wfcdecomp} are of the separable form $\vartheta_\alpha(r)S_{lm}  (\widehat{\boldsymbol{\theta}})$, with $\vartheta_\alpha(r)$  being a radial function and $S_{lm}(\widehat{\boldsymbol{\theta}})$ is the real-spherical harmonic function that denotes the angular part of these atom-centered functions.
We note that $r=|\bx-\bR^a|$ and  $\widehat{\boldsymbol{\theta}} \in \mathbb{R}^2$ represents the azimuthal and polar components of $(\bx-\bR^a)$. 
Moreover, we note that the compact support of the functions $\pTilde^a_\alpha(\bx)$ and $(\phiae^a_\alpha(\bx)-\phips^a_\alpha(\bx))$ lies in the augmentation sphere $\Omega_{\text{a}}$, and consequently the decomposition in Eqn.~\eqref{eqn:wfcdecomp} transforms the orthonormality requirement of all-electron Kohn-Sham wavefunctions to $\mathcal{S}$-orthonormality ($\mathcal{S}=\mathcal{T}^{\dagger}\mathcal{T}$) condition on the pseudo-smooth(PS) wavefunctions $\{ \psips_i\}$ i.e. $\bra{\psips_i}\mathcal{S}\ket{\psips_j}=\delta_{ij}$. 
 Eqn.~\eqref{eqn:wfcdecomp}, further allows the decomposition of the all-electron density $n(\bx)$ into a pseudo-smooth (PS) electron density $\nTilde(\bx)$ and an atom-centered correction which replaces the PS electron density inside the augmentation sphere with the true all-electron density and is given by
\begin{align}
  n(\bx) &= \nTilde(\bx) + \sum_a{\left(n^a(\bx) - \nTilde^a(\bx)\right)} \label{eq: Density decomposition}\\
\text{with,}\;\; \nTilde(\bx) &= 2\sum_i^N{f_i|\psips_i(\bx)|^2} + \nTilde_c(\bx) \label{eq: PS electron density} \\
\text{while,}\;\;n^a(\bx) &= n^a_c(\bx)+ \sum_{\alpha\beta}{\phiae^a_\alpha(\bx)\phiae^a_\beta(\bx)D^a_{\alpha\beta}} \label{eq: atom all electron density} \\
\text{and}\;\; \nTilde^a(\bx) &=\nTilde^a_c(\bx)+ \sum_{\alpha\beta}{\phips^a_\alpha(\bx)\phips^a_\beta(\bx)D^a_{\alpha\beta}} \label{eq: atom PS density}
\end{align}
In the above equations, $n^a(\bx)$ represents the atom-centered all-electron density and $\nTilde^a(\bx)$, the atom-centered PS electron density that matches with $n^a(\bx)$ outside the augmentation sphere. Furthermore, $n^a_c(\bx)$ and $\nTilde^a_c(\bx)$ represent the all-electron core density and PS core density while $n_c(\bx)=\sum_a{n^a_c(\bx)}$ and $\nTilde_c(\bx)=\sum_a{\nTilde^a_c(\bx)}$ denote the total all-electron core density and total PS-core density respectively. Additionally, the matrix-components of the atom-dependent Hermitian matrix $\bD^a$ (often referred to as channel occupancy matrix~\cite{Kresse1999} or spherical density matrix~\cite{Abinit2010}) in Eqn.~\eqref{eq: atom all electron density} and Eqn.~\eqref{eq: atom PS density} are given by
\begin{equation}
  D^a_{\alpha\beta} = 2\sum_i{f_i \braket{\psips_i|\pTilde^a_\alpha}\braket{\pTilde^a_\beta|\psips_i}}
  \label{eq: Dij entries}
\end{equation}
Using the decompositions in Eqns.~\eqref{eqn:wfcdecomp} and~\eqref{eq: Density decomposition}, the all-electron kinetic energy $T_s[\{\psiae_i\}]$ and the exchange-correlation energy $E_{\text{xc}}[n(\bx)]$ can be split into a smooth contribution  and an atom-centered correction term as follows:
\begin{align}
T_s[\{\psiae_i \}] &= T_s[\{ \psips_i\}] + \sum_a{T^a_{\text{core}}} + \sum_a{\sum_{\alpha\beta}\Delta T^a_{\alpha\beta}D^a_{\alpha\beta}} \nonumber \\ 
E_{\text{xc}}[n(\bx)] &= E_{\text{xc}}[\nTilde(\bx)] +
 \sum_a{\Delta E^a_{\text{xc}}}
\end{align}
where $T^a_{\text{core}}$ is the kinetic energy contribution from the core states while the terms $\Delta T^a_{\alpha\beta}, \Delta E^a_{\text{xc}} $ are the atomic corrections computed as
\begin{align}\label{eqn:atomCenteredCorrectionKEandXC}
   & \Delta T^a_{\alpha\beta} =  -\frac{1}{2}\int_{\Omega_a}{\left( \phiae^a_\alpha(\bx)\nabla^2\phiae^a_\beta(\bx) - \phips^a_\alpha(\bx)\nabla^2\phips^a_\beta(\bx)\right) d\bx} \nonumber \\
  &  \Delta E^a_{\text{xc}} = \int_{\Omega_a}{\left(\epsilon_{\text{xc}}[n^a,\nabla n^a]-\epsilon_{\text{xc}}[\nTilde^a,\nabla \nTilde^a]\right)d\bx}
\end{align}

Finally, the all-electron electrostatics energy is decomposed into a smooth contribution and an atom-dependent correction term by introducing a compensation charge $\bTilde^a(\bx)$ compactly supported in $\Omega_a$ such that the multipole moments of ($n^a(\bx) - \nTilde^a(\bx) + b^{a}(\bx) -\bTilde^a(\bx)$) vanish ~\cite{PAWBlochl1994} inside $\Omega_a$. To this end, the compensation charge $\bTilde^a(\bx)$ in terms of atom-centered spherical shape functions $\gTilde^a_{lm}(\bx)$ is given by
\begin{align}
    \bTilde^a(\bx) = \kappa^a\gTilde^a_{00}(\bx) + \sum_{lm,\alpha\beta}{\Delta^a_{lm\alpha\beta}D^a_{\alpha\beta}\gTilde^a_{lm}(\bx)}
    \label{eq: compensation charge}
\end{align}
The atom-centered shape functions $\gTilde^a_{lm}(\bx)$ in the above Eqn.~\eqref{eq: compensation charge} are compactly supported in $\Omega_a$ and are usually constructed in the form of Gaussians, sinc-squared or Bessel functions~\cite{PAWBlochl1994,abinitPAW2008,Abinit2010,GPAW2010,Kresse1999} to satisfy the condition $\int_{\Omega_{{a}}}{ r^lS_{lm}(\widehat{\boldsymbol{\theta}})\gTilde^a_{l'm'}(\bx)d\bx} = \delta_{ll'}\delta_{mm'}$. Furthermore $\kappa^a$ and $\Delta ^a_{lm\alpha\beta}$ in Eqn.~\eqref{eq: compensation charge} are given by
\begin{align}
\kappa^a &= \int_{\Omega_{a}}{(n^a_c(r)-\nTilde^a_c(r))r^2dr}-\nicefrac{\mathcal{Z}^a}{\sqrt{4\pi}} \nonumber \\ 
\Delta ^a_{lm\alpha\beta} &= \int_{\Omega_{{a}}}{r^l\left(\phiae^a_\alpha(\bx)\phiae^a_\beta(\bx) - \phips^a_\alpha(\bx)\phips^a_\beta(\bx) \right)S_{lm}(\hat{\boldsymbol{\theta}})d\bx}
\label{eqn: Kappa_a and Deltalm}
\end{align}
Now, the all-electron electrostatic energy can be computed as
\begin{equation}
 J\left[\rho(\bx),\{ \bR^a\}\right] =  E_\text{el}\left[\rhoTilde(\bx)\right ] + \sum_a{\Delta E^a_\text{el}}\label{eqn:electrodecomp}
\end{equation}
where the smooth total charge density is given by $\rhoTilde(\bx) = \nTilde(\bx) + \bTilde(\bx)$, with $\bTilde(\bx)=\sum_a{\bTilde^{a}(\bx)}$. The electrostatic energy involving $\rhoTilde(\bx)$ is given by
\begin{equation} \label{eqn:electrosmooth}
E_\text{el}\left[\rhoTilde(\bx)\right ] = \frac{1}{2}\int{\int{\frac{\rhoTilde(\bx)\rhoTilde(\by)}{|\bx-\by|}d\bx}d\by}
\end{equation}
and the atom-dependent correction term is given by
\begin{align}
&\Delta E^a_\text{el} = 
E^a_{\text{el}}\left[\rho^a(\bx)\right] - E^a_{\text{el}}\left[\rhoTilde^a(\bx)\right]  - E^a_{\text{self}}\left[\bR^a\right] \;\;\text{where} \nonumber \\
&E^a_{\text{el}}\left[\rho^a(\bx)\right] = \frac{1}{2}\int_{\Omega_a}{\int_{\Omega_a}{\frac{\rho^a(\bx)\rho^a(\by)}{|\bx-\by|}d\bx}d\by} \nonumber \\ &E^a_{\text{el}}\left[\rhoTilde^a(\br)\right]   = \frac{1}{2}\int_{\Omega_a}{\int_{\Omega_a}{\frac{\rhoTilde^a(\bx)\rhoTilde^a(\by)}{|\bx-\by|}d\bx}d\by}  
\label{eq: AtomCentered Electrostatics energy}
\end{align}
with $\rho^a(\bx) = n^a(\bx)+b^a(\bx)$ and $\rhoTilde^a(\bx) = \nTilde^a(\bx)+\bTilde^a(\bx)$. The choice of the compensation charge in Eqn.~\eqref{eq: compensation charge} allows for the evaluation of $\Delta E^a_\text{el}$ only within the augmentation sphere $\Omega_a$.
%one to reduce the domain of integration of the convolution involved in evaluating $E^a_{\text{el}}\left[\rho^a(\bx)\right]$ and $E^a_{\text{el}}\left[\rhoTilde^a(\bx)\right]$ to be within the augmentation sphere $\Omega_a$. 
By combining the various atom-dependent correction terms from different energy contributions into $\Delta E^a$, the all-electron energy in Eqn.~\eqref{eqn:allElecKS} in the PAW method is reformulated as follows:
\begin{equation}
 E\left[\{ \psiae_i \}, \{ \bR^a\} \right] = \widetilde{E}[\{ \psips_i\}, \{ \bR^a\}] +\sum_a{\Delta E^a[\left\{D^a_{\alpha\beta}\right\}]}
 \label{eq: PAW energy}
\end{equation}
 where we define $\widetilde{E}[\{ \psips_i\}, \{ \bR^a\}]$ and $\Delta E^a[\{D^a_{\alpha\beta}\}]$ as,
\begin{align*}
 \widetilde{E}\left[\{ \psips_i\}, \{ \bR^a\}\right] &=  T_s[\{ \psips_i \} ] + E_\text{xc}\left[\nTilde(\bx)\right] + E_{\text{el}}\left[\rhoTilde(\bx)\right ]  \\
\Delta E^a[\{D^a_{\alpha\beta}\}] &= \Delta E^a_{\text{xc}} + \Delta E^a_{\text{el}} + \sum_{\alpha\beta}\Delta T^a_{\alpha\beta}D^a_{\alpha\beta} + T^a_{\text{core}}. 
\end{align*}
The details of evaluating $\Delta E^a[\{D^a_{\alpha\beta}\}]$ are mentioned in Appendix-\ref{sec: appendix delta Ea}.
At this juncture, we note from Eqn.~\eqref{eq: PAW energy} that the all-electron Kohn-Sham energy functional of $\{\psiae_i\}$ has been decomposed into an energy functional of smooth wavefunctions $\{\psips_i\}$ and a term that includes atomic corrections. The validity of this decomposition is reliant on two key assumptions. First, it is assumed that the augmentation spheres are non-overlapping and to achieve this, specialised PAW data sets are generated based on the application of interest~\cite{ZWANZIGER201614}. Second assumption is that the set of partial waves ($\{\phips_\alpha \}$ and $\{\phiae_\alpha \}$) forms a complete basis within the augmentation sphere. However, in practice, only a finite number of partial waves are used. The error arising from this incompleteness is mitigated by introducing an atom-specific zero potential $\bar{v}^{a}(\bx)$  with a compact support only inside augmentation sphere \cite{PAWBlochl1994,Kresse1999,abinitPAW2008}. 

The electrostatic interaction energy involving the smooth total density $E_\text{el}\left[\rhoTilde(\bx)\right]$ in Eqn.~\eqref{eqn:electrosmooth} is extended in real-space and can be reformulated as a local variational problem in the spirit of \cite{SambitPRB2015,GAVINI2007669,MOTAMARRI2013308}. Subsequently, the variational problem involved in computing this electrostatic energy contribution is given by:
\begin{equation}
    E_\text{el}[\rhoTilde] = -\max_{\varphi \in \varsigma}{\left\{ \frac{1}{8\pi}\int{|\nabla \varphi(\bx)|^2 d\bx} - \int{\rhoTilde(\bx)\varphi(\bx)d\bx} \right\}} 
    \label{eqn:localelectro}
\end{equation}
where $\varphi(\bx)$ denotes the trial function for the electrostatic potential due to pseudo smooth electron density ($\nTilde(\bx)$) and the sum of atom-centered compensation charges ($\bTilde(\bx)$). Finally, the problem of determining the all-electron ground state energy and the all-electron density for a given position of nuclei is expressed as the following variational problem involving the smooth electronic fields:

\begin{widetext}
\begin{equation} \label{eqn:pawvar}
    E_{\text{GS}} = \min_{\{\psips_i\} \in\mathcal{X}}\max_{\varphi \in \varsigma}{\left\{ T_s[\{ \psips_i\}] - \frac{1}{8\pi}\int{|\nabla \varphi(\bx)|^2 d\bx} + \int{\rhoTilde(\bx)\varphi(\bx)d\bx} + E_{\text{xc}}[\nTilde(\bx)] + \sum_a{\Delta E^a}[\{D^a_{\alpha\beta}\}] \right\}}
\end{equation}
\end{widetext}
where the function space $\mathcal{X} = \left\{ \{\psips_i\} \bigg | \left<\psips_p,\psips_q \right>_{\mathbb{X}} = \delta_{pq}\right \}$ with $\left<\psips_p,\psips_q\right>$ denoting the $\mathcal{S}$-inner product $\braket{\psips_p|\mathcal{S}|\psips_q}$ defined on $\mathbb{X}$. The space $\mathbb{X}$ denotes a suitable function space that guarantees the existence of the minimisers. The domains used for numerical computations that are non-periodic corresponds to a large enough domain such that the wavefunctions $\{\psips_i\}$ decays to 0 and in periodic calculations, these correspond to supercells with periodic boundary conditions. Denoting such an appropriate domain by $\Omega$, the appropriate function spaces for $\mathbb{X}$ and $\varsigma$ are $\mathbb{X}=\varsigma=H^1_0(\Omega)$ in the case of non-periodic problems and $\mathbb{X} = \varsigma = H^1_{\text{per}}(\Omega)$ in the case of periodic problems.

\subsection{Governing Equations}
The stationarity condition with respect to the smooth fields $\{\psips_i\}$ corresponding to the PAW variational problem in Eqn.~\eqref{eqn:pawvar} yields nonlinear generalized Hermitian eigenvalue problem (GHEP) of the following form:
\begin{equation} \label{eqn:pawghep}
    \mathcal{H}\psips_i = \varepsilon_i\mathcal{S}\psips_i
\end{equation}
where $\mathcal{H}$ is a Hermitian operator with
\begin{align}\label{eqn:pawHamiltonian}
    &\mathcal{H} = -\frac{1}{2}\nabla^2 + V_{\text{eff}}(\rhoTilde,\{ \bR^a\}) + \mathcal{H}_{\text{nloc}} \nonumber \\
&\mathcal{H}_{\text{nloc}}\psips_i :=  \sum_a{\sum_{\alpha\beta}{{\pTilde^a_\alpha(\bx)}\Delta h^a_{\alpha\beta}\int{\pTilde^a_\beta(\by) \psips_i(\by)d\by}}}
\end{align}
and $\mathcal{S}$ is a positive-definite Hermitian operator with
\begin{align}\label{eqn:pawOverlap}
    &\mathcal{S} = \mathcal{I} + \mathcal{S}_{\text{nloc}} \nonumber \\
 & \mathcal{S}_{\text{nloc}}\psips_i :=  \sum_a{\sum_{\alpha\beta}{{\pTilde^a_\alpha(\bx)}\Delta s^a_{\alpha\beta}\int{\pTilde^a_\beta(\by) \psips_i(\by)d\by}}}    
\end{align}
The set $(\{\varepsilon_i\},\{ \psips_i\})$ are the eigenvalue and eigenfunction pairs of the generalized eigenvalue problem~\eqref{eqn:pawghep} and the $N$ smallest eigenvalue and eigenvector pairs are used to compute the PS density $\nTilde(\bx)$ and the matrix $\bD^a$ employing the Eqn.~\eqref{eq: PS electron density} and Eqn.~\eqref{eq: Dij entries} respectively. Further, the effective potential ($V_{\text{eff}}$) comprising of the total electrostatic potential, exchange-correlation potential and zero potential is computed as
\begin{align}
    &V_{\text{eff}}  = V_{\text{el}} + V_{\text{xc}} + \bar{V} =  \frac{\delta E_{\text{el}}}{\delta \nTilde} + \frac{\delta E_{\text{xc}}}{\delta \nTilde} + \sum_a{\bar{v}^a}
    \end{align}
 where the electrostatic potential $\frac{\delta E_{\text{el}}}{\delta \nTilde}=\varphi(\bx,\{\bR^a\})$ is computed as the solution of the Poisson equation associated with the smooth total density $\rhoTilde(\bx)$ as given by Eqn.~\eqref{eqn:localelectro}.
Furthermore, the atom-dependent contributions $\Delta h^a_{\alpha\beta}$  and $\Delta s^a_{\alpha\beta}$ in $\mathcal{H}_{\text{nloc}}$ and $\mathcal{S}_{\text{nloc}}$ are expressed as follows:
    \begin{align}    
    & \Delta h^a_{\alpha\beta} = \frac{\delta \Delta E^a}{\delta D^a_{\alpha\beta}} + \sum_{lm}{\Delta^a_{lm\alpha\beta}\int{\gTilde^a_{lm}(\bx)\varphi(\bx)d\bx}} \nonumber\\
   & - \int_{\Omega_{\text{a}}}{\bar{v}^a(\bx)\phips^a_\alpha(\bx)\phips^a_\beta(\bx)d\bx} 
    \label{eq: Hamiltonian coupling}
    \end{align}
    \begin{align}
    &\Delta s^a_{\alpha\beta}  = \int_{\Omega_a}{\left(\phiae^a_\alpha(\bx)\phiae^a_\beta(\bx)-\phips^a_\alpha(\bx)\phips^a_\beta(\bx)\right)d\bx} = \sqrt{4\pi}\Delta^a_{00\alpha\beta}
    \label{eq: overlap coupling}
\end{align}
Finally, the governing differential equations to be solved in the PAW method are:
\begin{align}
 &  \left(-\frac{1}{2}\nabla^2 + V_{\text{eff}}(\rhoTilde,\{ \bR^a\} )+\mathcal{H}_{\text{nloc}}\right)\psips_i =\varepsilon_i\left(\mathcal{I}+\mathcal{S}_{\text{nloc}} \right)\psips_i
  \nonumber \\
& 2\sum_i{f(\varepsilon_i,\mu)} = N_v,\;  f(\varepsilon_i,\mu) = \frac{1}{1+ \text{exp}\left(\frac{\epsilon_i-\mu}{\sigma} \right)}  \nonumber\\ 
 &  \nTilde(\bx) = 2\sum_i{f(\varepsilon_i,\mu)|\psips_i(\bx)|^2},\; \bTilde(\bx) = \sum_a{\bTilde^a(\bx)}  \nonumber\\
 & \bTilde^a(\bx) = \kappa^a\gTilde^a_0(\bx) + \sum_{lm,\alpha\beta}{\Delta^a_{lm\alpha\beta}D^a_{\alpha\beta}\gTilde^a_{lm}(\bx)} \nonumber \\
 & \hat{U}^a_{\alpha i} = \int{\pTilde_\alpha^a(\by)\psips_i(\by)d\by},\;D^a_{\alpha\beta} = 2\sum_i{f(\epsilon_i,\mu)\hat{U}^{a^{*}}_{\alpha i}\hat{U}^{a}_{i\beta}} \nonumber  \\
 & \rhoTilde(\bx) = \nTilde(\bx) + \bTilde(\bx),\; -\frac{1}{4\pi}\nabla^2\varphi(\bx,\{\bR^a\}) = \rhoTilde(\bx)  
   \label{eq: Governing PAW PDE}
\end{align}
where  $\mu$ is the Fermi-energy, $\hat{U}^{a^{*}}_{\alpha i}$ is the complex conjugate of $\hat{U}^{a}_{\alpha i}$ and $f_i:=f(\epsilon_i,\mu)$ is the orbital occupancy function lying in the interval $[0,1]$ which is computed using the Fermi-Dirac distribution with $\sigma = k_{B}T$ where $T$ is the smearing temperature and $k_B$ is the Boltzmann constant.

When dealing with periodic crystals, 2D periodic slabs, or surfaces, we invoke the Bloch theorem \cite{martin2020electronic,ashcroft2022solid} along the periodic axes. Instead of solving the DFT problem on a large periodic supercell, we solve the reduced problem on smaller unit cells with periodic boundary conditions. Using Bloch theorem, the eigenfunctions of the PAW generalized eigenvalue problem can be expressed as $\psips_{i,\bk}(\bx) = e^{\mathrm{i}\bk.\bx}\uTilde_{i,\bk}(\bx)$
% \begin{equation}
%     \psips_{i,\bk}(\bx) = e^{\mathrm{i}\bk.\bx}\uTilde_{i,\bk}(\bx)
%     \label{eq: Bloch wfc}
% \end{equation}
where $\mathrm{i} = \sqrt{-1}$ and $\uTilde_{i,\bk}(\bx)$ is a unit-cell periodic complex valued function satisfying $\uTilde_{i,\bk}(\bx+\bL_r) = \uTilde_{i,\bk}(\bx)$ for all lattice vectors $\bL_r$ with $\bk$ denoting a point in the first Brillouin zone(BZ) of the reciprocal lattice. The set of governing equations for the PAW formalism with Bloch wavefunctions $\uTilde_{i,\bk}(\bx)$ can be recast as:
\begin{widetext}
\begin{gather} 
\left(-\frac{1}{2}\nabla^2 - \mathrm{i}\bk\cdot \nabla + \frac{1}{2}|\bk|^2 + V_{\text{eff}}(\rhoTilde,\{ \bR^a\} )+\mathcal{H}^{\bk}_{\text{nloc}} \right)\uTilde_{i,\bk}    = \varepsilon_{i,\bk}\left(\mathcal{I}+\mathcal{S}^{\bk}_{\text{nloc}}\right)\uTilde_{i,\bk}\;\; \text{on}\;\; \Omega_p,\;\; \forall \bk \in BZ \nonumber \\
\mathcal{H}^{\bk}_{\text{nloc}} \uTilde_{i,\bk} :=  \sum_a{\sum_{\alpha\beta}{\sum_r{e^{-\mathrm{i}\bk\cdot (\bx-\bL_r) }\pTilde^a_\alpha(\bx-(\bR^a+\bL_r))}\Delta h^a_{\alpha\beta}U^{a\bk}_{\beta i}}}, \nonumber \\
 \mathcal{S}^{\bk}_{\text{nloc}} \uTilde_{i,\bk} :=  \sum_a{\sum_{\alpha\beta}{\sum_r{e^{-\mathrm{i}\bk\cdot (\bx-\bL_r) }\pTilde^a_\alpha(\bx-(\bR^a+\bL_r))}\Delta s^a_{\alpha\beta}U^{a\bk}_{\beta i}}} \nonumber \\
2\sum_i{\fint_{BZ}{2f(\varepsilon_{i,\bk},\mu) d\bk}} = N_v, \; f(\varepsilon_{i,\bk},\mu)  = \frac{1}{1+ \text{exp}\left(\frac{\epsilon_{i,\bk}-\mu}{\sigma} \right)},\;\nTilde(\bx) = \sum_i{\fint_{BZ}{2f(\epsilon_{i,\bk},\mu)|\uTilde_{i,\bk}(\bx)|^2 d\bk}},\; \bTilde(\bx) = \sum_a{\bTilde^a(\bx)}\nonumber \\
\widehat{U}^{a,\bk}_{\alpha i} := \int_{\Omega_p}{\sum_{r'}{e^{\mathrm{i}\bk\cdot (\by-\bL_{r'})}p_\alpha^a(\by - (\bR^a+L_{r'}))} \uTilde_{i,\bk}(\by)d\by},\; D^a_{\alpha\beta} = 2\sum_i{\fint_{BZ}{f(\epsilon_i^{\bk},\mu)\widehat{U}^{a,\bk*}_{\alpha i}\widehat{U}^{a,\bk}_{i \beta}d\bk}}  \nonumber \\
\bTilde^a(\bx) = \kappa^a\gTilde^a_0(\bx) + \sum_{lm,\alpha\beta}{\Delta^a_{lm\alpha\beta}D^a_{\alpha\beta}\gTilde^a_{lm}(\bx)},\; \rhoTilde(\bx) = \nTilde(\bx)+\bTilde(\bx),\;-\frac{1}{4\pi}\nabla^2\varphi(\bx,\{\bR^a\}) = \rhoTilde(\bx)
\label{eqn:PAW GDE}
\end{gather}   
\end{widetext}
where $\fint_{BZ}$ denotes the volume average of the integral over the first Brillouin zone (BZ) corresponding to the periodic unit cell $\Omega_p$. We further note that the equations corresponding to the non-periodic case in Eqn.~\eqref{eq: Governing PAW PDE} can be easily recovered from Eqn.~\eqref{eqn:PAW GDE} by considering ($\bk$ = 0, $\bL_r$ = 0) and replacing the periodic unit-cell domain $\Omega_p$ with a non-periodic simulation domain $\Omega$. We note that semi-periodic boundary conditions can be handled in a similar manner, depending on the number of periodic directions. Furthermore, we note that the set of equations in Eqn.~\eqref{eq: Governing PAW PDE} or Eqn.~\eqref{eqn:PAW GDE} involves a nonlinear generalized Hermitian eigenvalue problem (GHEP) that must be solved self-consistently in conjunction with the Poisson equation. In particular, we seek to solve the fixed point iteration  $[\nTilde(\bx),\{ \bD^a\}] = F([\nTilde(\bx),\{ \bD^a\}])$ using a self-consistent field (SCF) iteration procedure where the evaluation of the map $F([\nTilde(\bx),\{ \bD^a\}])$ involves the solution of the GHEP and the Poisson equation in Eqn.~\eqref{eq: Governing PAW PDE} or Eqn.~\eqref{eqn:PAW GDE} for every SCF iteration.

%\cred is a nonlinear generalized Hermitian eigenvalue problem(GHEP) as the Hamiltonian $\mathcal{H}$ depends on the eigenvalues ($\epsilon_{i,\bk}$) and eigenvectors ($\uTilde_{i,\bk}$). In this work, we solve the nonlinear GHEP as a fixed point iteration map $\rhoTilde(\bx) = F(\rhoTilde(\bx))$ of total charge density, where $F(\rhoTilde(\bx))$ involved solving the GHEP for an input $\rhoTilde(\bx)$ and computing the output $\rhoTilde(\bx)$. This procedure is referred to as self-consistent field iteration(SCF) and the details of which along with the FE discretization is discussed in the following section. \cn 
%\cred  Also make a similar change in the above governing equations just like you had in previous non-periodic case. We need to add the point that the above equations have to be solved using SCF strategy as a final remark before going to next section+\cn

\section{PAW-FE: finite element discretization for PAW method}\label{sec:fem}
The finite-element method involves decomposing the spatial domain of interest into non-overlapping subdomains called finite-elements (or cells) using a finite-element (FE) mesh. The underlying FE basis functions are piecewise polynomials that exhibit systematically convergent behaviour~\cite{brenner2008mathematical,hughes2012finite,bathe}. Furthermore, these basis functions are strictly local, having a compact support only in the finite-element cells sharing a finite-element node, making them well-suited for massive parallelization. In contrast to the conventional finite-element basis typically constructed from a tensor product of Lagrange polynomials interpolated through equidistant nodal points in a given finite-element, in this work, we specifically employ spectral finite-element basis functions that are $C^0$ continuous Lagrange polynomials generated using Gauss-Lobatto-Legendre(GLL) nodal points that are the roots of derivatives of Legendre polynomials\cb. Using GLL points for interpolation leads to a better conditioned basis as the finite-element Lagrange interpolating polynomial degree increases compared to uniformly distributed nodes\cn. We refer to previous studies~\cite{MOTAMARRI2013308,dftfe0.6,dftfe1.0,Bikash2017} regarding the advantages provided by higher-order adaptive spectral finite-element-based methods to solve the Kohn-Sham DFT problem in the context of norm-conserving pseudopotentials and all-electron DFT calculations, \cb wherein the systematic convergence afforded by these FE basis is achieved by decreasing the FE cell size and increasing the order of FE interpolating polynomial\cn. A crucial aspect of the computational methodology described below involves the use of these higher-order spectral finite-elements in conjunction with the reduced order Gauss-Lobatto Legendre (GLL) quadrature rules combined with the non-overlapping nature of augmentation spheres that enables the efficient inversion of the PAW overlap matrix. We discuss here the finite-element discretization of the governing equations deducing the algebraic generalized eigenvalue problem to be solved in the PAW method.

As seen from Eqn.~\eqref{eqn:pawvar}, the real-space formulation of the PAW method presented in this work results in a min-max problem involving the pseudo-smooth wavefunctions and the electrostatic potential. Employing a single FE discretization for both the wavefunctions and the electrostatic potential  can result in non-variational behaviour of the electronic ground-state energy. Therefore, we seek a solution for the electrostatics problem by solving the Poisson equation more accurately than the eigenvalue problem solution by employing different orders of FE interpolating polynomial basis for the electrostatic potential and the wavefunctions. To this end, we represent the electronic fields in Eqn.~\eqref{eqn:PAW GDE} in the FE basis and are given by 
\begin{align} \label{eqn:FEbasisexp}
    \fePsips^h_{i,\bk}(\bx) = \sum_{J=1}^{M}{N^{h,p}_J(\bx)\fePsips^J_{i,\bk}},\; \varphi^{h,p_{\text{el}}}(\bx) = \sum_{J=1}^{M_{\text{el}}}{N^{h,p_{\text{el}}}_J(\bx)\varphi^J}
\end{align}
where $\fePsips^J_{i,\bk}$, $\varphi^J$ denote the FE discretized fields, while $N^{h,p}_J:1 \leq J \leq M$ and $N^{h,p_{\text{el}}}_J:1 \leq J \leq M_{\text{el}}$ denote the strictly local Lagrange polynomial basis functions of degree $p$ and $p_{\text{el}}$, respectively spanning the finite-element subspace, generated using the nodes of the FE triangulation $\tau^h$ with the characteristic mesh size denoted by $h$ and $p_{\text{el}} > p$ as discussed. To this end, the finite-element discretization of the Kohn-Sham generalized eigenvalue problem in Eqn.~\eqref{eqn:PAW GDE} results in an algebraic generalized Hermitian eigenvalue problem $\bH^{\bk}\feNodalVectorUtilde_{i,\bk} = \varepsilon^h_{i,\bk}\bS^{\bk}\feNodalVectorUtilde_{i,\bk}$, where $\bH^{\bk}$ and $\bS^{\bk}$ denote the discretized Hamiltonian matrix and the PAW overlap matrices of size $M \times M$ with $\varepsilon^h_{i,\bk}$ denoting the $i$\textsuperscript{th} eigenvalue corresponding to the discrete eigenvector 
$\feNodalVectorUtilde_{i,\bk}$ ( see Appendix-\ref{sec: Appendix GEP} for derivation of the FE discretized algebraic EVP). As discussed in  Appendix-\ref{sec: Appendix GEP}, the finite-element discretized Hamiltonian and PAW overlap matrices has both local and non-local contributions and hence can be decomposed as $\bH^{\bk}=\bH^{\bk,\text{loc}}+ \bH^{\bk,\text{nloc}}$ and $\bS^{\bk}=\bM+ \bS^{\bk,\text{nloc}}$ where the matrix entries of the local contributions $\bH^{\bk,\text{loc}}$ and $\bM$ are given by
\begin{align}
&\text{H}_{IJ}^{\bk,\text{loc}} = \bigintsss_{\Omega_p}{\Biggl\{ \frac{1}{2}\nabla N^{h,p}_I(\bx)\cdot \nabla N^{h,p}_J(\bx)} \;\;\ \nonumber \\ 
&{+\;V^h_{\text{eff}}(\bx) N_I^{h,p}(\bx)N_J^{h,p}(\bx)} \nonumber \\
&{+\; \frac{1}{2}|\bk|^2N_I^{h,p}(\bx)N_J^{h,p}(\bx) - \mathrm{i}\bk \cdot N^{h,p}_I(\bx)\nabla N^{h,p}_J(\bx)}\nonumber \\
& {+\bV^h_{\text{GGA}}(\bx) \cdot \left(N_I^{h,p}(\bx)\nabla N^{h,p}_J(\bx)+N^{h,p}_J(\bx)\nabla N^{h,p}_I(\bx) \right)\Biggr\} d\bx} \label{eqn:hij} \\
 &\text{M}_{IJ} = \int_{\Omega_p}{N_I^{h,p}(\bx)N_J^{h,p}(\bx)d\bx}  \label{eqn:mij}
\end{align}
\vspace{-0.1in}
with
\begin{align*}
 &V^h_{\text{eff}} =  \left(\varphi^{h,p_{\text{el}}}(\bx)+ \bar{V}(\bx) +\frac{\partial \epsilon_{\text{xc}}(\nTilde, \nabla \nTilde)}{\partial \nTilde}\bigg\rvert_{\nTilde = \nTilde^h}  \right) \nonumber \\
 &\bV^h_{\text{GGA}}(\bx) = \frac{\partial \epsilon_{\text{xc}}(\nTilde, \nabla \nTilde)}{\partial \nabla \nTilde}\bigg\rvert_{\nTilde = \nTilde^h} \nonumber
\end{align*}
and the discrete electrostatic potential $\varphi^{h,p_{\text{el}}}(\bx)$ in $V^h_{\text{eff}}$ above is computed by solving the finite-element discretized Poisson equation $\bL \bPhi = \bc$ where 
$\bL$ denotes the FE discretized Laplace operator using the basis $N^{h,p_{\text{el}}}_J:1 \leq J \leq M_{\text{el}}$ while $\bPhi$ denotes the electrostatic potential vector whose entries are the expansion coefficients $\varphi^{J}$ in Eqn.~\eqref{eqn:FEbasisexp} and $\bc$ denotes the forcing vector corresponding to $\rhoTilde$, the right-hand side of the Poisson equation in Eqn.~\eqref{eqn:PAW GDE}. To this end, the entries of $\bL$ and $\bc$ are given by
\begin{align}
L_{IJ} &= \frac{1}{4\pi}\int_{\Omega_p}{\nabla N^{h,p_{\text{el}}}_I(\bx)\cdot\nabla N^{h,p_\text{el}}_J(\bx)d\bx},\label{eq: FELaplacian} \\
c_{I} &= \int_{\Omega_p}{\rhoTilde^h(\bx)N^{h,p_{\text{el}}}_I(\bx)d\bx} \label{eq:FERhs} 
\end{align}
Finally, the matrix entries corresponding to the non-local contributions in $\bH^{\bk}$ and $\bS^{\bk}$ are given by
\begin{align}
&\bH^{\bk,\text{nloc}} = \sum_a{\bC^{a,\bk}\boldsymbol{\Delta}^a_{\bH}\bC^{a,\bk^{\dagger}}},\; \bS^{\bk,\text{nloc}} = \sum_a{\bC^{a,\bk}\boldsymbol{\Delta}^a_{\bS} \bC^{a,\bk^{\dagger}}}\nonumber \\
&\text{with the matrix entries of }\; \bC^{a,\bk}\;\text{defined as}\nonumber \\
&\text{C}^{a,\bk}_{J \alpha} = \int_{\Omega_p}\sum_r{e^{-\mathrm{i}\bk\cdot(\bx-\bL_r)}\pTilde^a_{\alpha}(\bx-(\bR^a+\bL_r))N^{h,p}_J(\bx)d\bx}
\label{eq: FE nonlocal contribution}
\end{align}
where the atom-dependent coupling matrices $\boldsymbol{\Delta}^a_{\bH}$ and $\boldsymbol{\Delta}^a_{\bS}$ are of size $n^{a}_{\text{pj}}\cross n^{a}_{\text{pj}}$ with $n^{a}_{\text{pj}}$ denoting the number of projectors for atom index $a$ and the corresponding matrix entries denoted as $\Delta h^{a,h}_{\alpha\beta}$ and $\Delta s^a_{\alpha\beta}$ respectively are given by
\begin{align}
    & \Delta h^{a,h}_{\alpha\beta} = \frac{\delta \Delta E^{a,h}}{\delta {D^a_{\alpha\beta}}} - \int_{\Omega_{\text{a}}}{\bar{v}^a(\bx)\phips^a_\alpha(\bx)\phips^a_\beta(\bx)d\bx} \nonumber \\
    & + \sum_{lm}{\Delta^a_{lm\alpha\beta}\int_{\Omega_{\text{p}}}{\sum_r{\gTilde^a_{lm}(\bx-(\bR^a+\bL_r))\varphi^{h_{\text{el}}}(\bx)d\bx}}}\nonumber \\
    &\Delta s^a_{\alpha\beta}  = \sqrt{4\pi}\Delta^a_{00\alpha\beta} \label{eqn:deltaHanddeltaS}
\end{align}  
Since the FE basis functions are localised in real-space, the matrices  $\bH^{\bk,\text{loc}}$ and $\bS^{\bk,\text{loc}}$ are sparse. Since the projectors $\pTilde^a(\bx-\bR^a)$ have a compact support within the augmentation sphere $\Omega_{\text{a}}$, we note that the $M \times n^{a}_{\text{pj}}$ matrix $\bC^{a,\bk}$ is also a sparse matrix for every atom $a$. Consequently, we can now rewrite the discretized Hamiltonian matrix and the overlap matrix within the PAW formalism as 
 \begin{equation}\label{eqn: simplfied FE PAW}
        \bH^{\bk} = \bH^{\bk}_{\text{loc}} + \bP^{\bk}\boldsymbol{\Delta}_{\btH}{\bP^{\bk}}^{\dagger},\; \bS^{\bk} = \bM + \bP^{\bk}\boldsymbol{\Delta}_{\btS}{\bP^{\bk}}^{\dagger}
\end{equation}
where the $M \times n_{\text{proj}}$ matrix $\bP^{\bk} = \left[\bC^{1,\bk}... \bC^{a,\bk}...\bC^{N_a,\bk} \right]$, with $n_{\text{proj}} = \sum_a n^{a}_{\text{pj}}$ denoting the total number of atomic projectors in the system. In the above Eqn.~\eqref{eqn: simplfied FE PAW}, $\boldsymbol{\Delta}_{\btH}$ and $\boldsymbol{\Delta}_{\btS}$ are the block diagonal matrices of size $n_{\text{proj}} \times n_{\text{proj}}$ with each block corresponding to the atom-dependent coupling matrices  $\boldsymbol{\Delta}^a_{\btH}$ and $\boldsymbol{\Delta}^a_{\btS}$ respectively as introduced in Eqn.~\eqref{eqn:deltaHanddeltaS}.

 Finally, the discretized governing equations corresponding to Eqn.~\eqref{eqn:PAW GDE} that must be solved in PAW-FE are given by:
\begin{gather}
\bH^{\bk}\feNodalVectorUtilde_{i,\bk} = \varepsilon^h_{i,\bk}\bS^{\bk}\feNodalVectorUtilde_{i,\bk} \nonumber \\
 2\sum_n^N{\fint_{BZ}f(\varepsilon^h_{i,\bk},\mu)d\bk} = N_v,\; f(\varepsilon^h_{i,\bk},\mu) = \frac{1}{1+ \text{exp}\left(\frac{\varepsilon^h_{i,\bk}-\mu}{\sigma} \right)} \nonumber \\
\nTilde^h(\bx) = 2\sum_i^N{\fint_{BZ}f(\varepsilon^h_{i,\bk},\mu)|\fePsips^h_{i,\bk}(\bx)|^2d\bk},\;\widehat{\boldsymbol{\mathsf{U}}}^{a,\bk}_{i} = \bC^{a,\bk\dagger}\feNodalVectorUtilde_{i,\bk}  \nonumber \\
D^{a,h}_{\alpha\beta} = 2\sum_i^N{\fint_{BZ}{f(\varepsilon^h_{i,\bk},\mu)\widehat{{\mathsf{U}}}^{a,\bk^*}_{\alpha i}\widehat{{\mathsf{U}}}^{a,\bk}_{\beta i} d\bk}}\nonumber \\
\bTilde^{a,h}(\bx) = \kappa^a\gTilde^a_0(\bx) + \sum_{lm,\alpha\beta}{\Delta^a_{lm\alpha\beta}D^{a,h}_{\alpha\beta}\gTilde^a_{lm}(\bx)} \nonumber \\
\bTilde^h(\bx) = \sum_a{\bTilde^{a,h}(\bx)},\; \rhoTilde^h(\bx) = \nTilde^h(\bx)+\bTilde^h(\bx),\; \bL \bPhi= \bc
\label{eq: FE PAW governing equations}
\end{gather}
where $\widehat{{\mathsf{U}}}^{a,\bk}_{\beta i}$ refers to the vector components of $\widehat{\boldsymbol{\mathsf{U}}}^{a,\bk}_{i}$ and $D^{a,h}_{\alpha\beta}$ are the matrix components of the discretized atom-centered spherical density matrix $\bD^{a,h}$.

We note that the discretized generalized Hermitian eigenproblem (GHEP) in the above Eqn.~\eqref{eq: FE PAW governing equations} is nonlinear in nature and needs to be solved for $N$ smallest eigenvalue-eigenvector pairs, which in turn can be used to evaluate the pseudo smooth density $\nTilde^{h}(\bx)$ and the atom-centred density matrix $\{ D^{a,h}_{\alpha,\beta}\}$. In the current work, this nonlinear GHEP is self-consistently solved
in conjunction with the discretized Poisson equation subject to the applied boundary conditions to obtain the ground-state electronic fields and the energy. As noted before, replacing the periodic unit-cell domain $\Omega_p$ with a non-periodic simulation domain $\Omega$ and considering $\bk=0$, $\bL_r=0$, we recover the finite-element discretized equations corresponding to non-periodic boundary conditions from Eqn.~\eqref{eq: FE PAW governing equations}. Subsequently, the total all-electron energy in the PAW method, $E^h$ in terms of the finite-element discretized ground-state solution fields $\left (\varepsilon^h_{n,\bk},\nTilde^h(\bx), D^{a,h}_{\alpha\beta}, \varphi^{h,p_{\text{el}}} \right)$ is given by
    \begin{align}
        E^h =& 2\sum_{i=1}^N{\fint_{BZ}{f(\varepsilon^h_{i,\bk},\mu)\varepsilon^h_{i,\bk} d\bk}} - \sum_a{\sum_{\alpha\beta}\Delta h^a_{\alpha\beta}D^{a,h}_{\alpha\beta}} \nonumber \\
        &- \int_{\Omega_p}{\left[ \nTilde^h(\bx)V^h_{\text{eff}}(\bx) + \bV^h_{\text{eff,GGA}}(\bx) \cdot \nabla \nTilde^h(\bx) \right] d\bx} \nonumber \\
        &+ \frac{1}{2}\int_{\Omega_p}{\rhoTilde^h(\bx)\varphi^{h,p_{\text{el}}}(\bx)d\bx} + E_{\text{xc}}\left[\nTilde^h(\bx)\right] \nonumber \\
        &+ \sum_a{\Delta E^a\left[\{D^{a,h}_{\alpha\beta}\}\right]} \label{eqn:FEAllElecEnergy} 
    \end{align}
The various computational aspects involved in efficiently solving the nonlinear generalized eigenvalue problem (Eqn.~\eqref{eq: FE PAW governing equations}) using a self-consistent field (SCF) iteration procedure are discussed in the subsequent section. In the following section, we choose to drop the $\bk$ index in the FE discretized electronic fields, vectors and matrices introduced in Eqn.~\eqref{eqn: simplfied FE PAW} and Eqn.~\eqref{eq: FE PAW governing equations} for brevity, and we refer to them as $\tilde{u}_i^{h}(\bx)$, $\feNodalVectorUtilde_i$, $\bH$, $\bS$ and $\bP$ without the $\bk$ index.  
%\cred somewhere we need to mention that we are solving lowest eigenvalue/eigenvector pairs\cn

\section{Self consistent field iteration}\label{sec:scf}
In this section, we describe the self-consistent field iteration (SCF) procedure adopted in our work to solve the FE discretized nonlinear PAW generalized eigenvalue problem. We begin with the real-space mixing strategy employed in our work suited towards solving the Kohn-Sham fixed point problem inherent in the PAW method. Subsequently, we describe the Chebyshev filtering subspace iteration (ChFSI) approach and focus on aspects related to adapting ChFSI for efficiently solving the underlying generalized eigenvalue problem within the given SCF iteration. Additionally, we elaborate on the key computational strategies that leverage the underlying sparse finite-element structure and the compact support of the PAW projectors in real-space to facilitate fast and scalable calculations of computationally intensive steps in the ChFSI approach.

The finite-element discretized governing equations~\eqref{eq: FE PAW governing equations} in the PAW method involves the solution of a nonlinear generalized eigenvalue problem that needs to be solved self-consistently as alluded to before. We note that the Kohn-Sham fixed point iteration map in the PAW method can be expressed as  $[\nTilde_{\text{out}}(\bx),\{\bD_{\text{out}}^a \}]= F([\nTilde_{\text{in}}(\bx),\{\bD_{\text{in}}^a \}])$. To this end, we start with an input guess for $[\nTilde_{\text{in}}(\bx),\{\bD_{\text{in}}^a \}]$ and the map $F([\nTilde_{\text{in}}(\bx),\{\bD_{\text{in}}^a \}])$ involves evaluating the effective potential and the non-local Hamiltonian coupling matrix to solve for the $N$ smallest eigenpairs of the PAW generalized eigenproblem which are in turn used to compute the output $[\nTilde_{\text{out}}(\bx),\{\bD_{\text{out}}^a \}]$. The above procedure is summarised in Steps 2 to 5 of the Algo.~\eqref{alg: SCF problem} that describes the SCF iteration, and these steps are continued until convergence.

The computationally intensive step involved in Algo.~\eqref{alg: SCF problem} is solving the discretized generalized eigenvalue problem (Step-4) which has an asymptotic cubic scaling with system size, making it crucial to employ efficient numerical strategies to delay the onset of cubic scaling and also reduce the number of SCF iterations. 
We now describe the mixing strategy employed in our work to accelerate the convergence of the SCF procedure.
\begin{figure}
%    \removelatexerror
\begin{algorithm}[H]
    \caption{SCF iteration in \pawfe} \label{alg: SCF problem}
%\KwIn{Initial subspace $\Psips_{\text{in}}$}
\begin{enumerate}
\item Provide initial guess for $\nTilde_0^h(\bx)$ and $\{\bD^{a,h}_{0} \}$ from PAW dataset as the input for the first SCF iteration ($\nTilde_{\text{in}}^h(\bx)=\nTilde_0^h(\bx)$,$\{\bD^{a,h}_{\text{in}}\} = \{\bD^{a,h}_0$\}).
\item Compute the total electrostatic potential $\varphi^{h,p_{\text{el}}}(\bx)$ for the charge density $(\nTilde^h(\bx)+\bTilde^h(\bx))$ by solving the discrete Poisson equation $\bL \bPhi= \bc$
\item Compute the effective potential ($\bV^h_{\text{eff,GGA}}(\bx), V^h_{\text{eff}}(\bx)$) and Hamiltonian coupling matrix($\boldsymbol{\Delta}^a_{\bH}$) to evaluate the Hamiltonian ($\bH$)
\item Solve the linearized PAW generalized eigenvalue problem $\bH\feNodalVectorUtilde_i = \varepsilon_i\bS\feNodalVectorUtilde_i$ to obtain the eigenfunctions $\fePsips^h_{i,\bk}(\bx)$ for $N$ smallest eigenpairs
\item Compute the new output electron density $\nTilde_{\text{out}}^h(\bx)$ and $\{\bD^{a,h}_{\text{out}}\}$ using Eqn.~\eqref{eq: FE PAW governing equations}
\item If $||\rhoTilde^h_{\text{out}}(\bx) - \rhoTilde^h_{\text{in}}(\bx) || <$ tol, exit SCF loop; \\else, compute the new $\nTilde^h_{\text{in}}(\bx),\{\bD^{a,h}_{\text{in}}\}$ using Anderson mixing scheme as described in Section IVA  and go to step 2.

\end{enumerate}
\end{algorithm}
\end{figure}
\subsection{Total charge density mixing}
In contrast to DFT calculations involving norm-conserving pseudopotentials and all-electron methods, where pseudo-valence electron density or the all-electron density is mixed during the SCF iteration, in the PAW method, the total charge density ($\rhoTilde(\bx) = \nTilde(\bx) + \bTilde(\bx)$) needs to be updated self consistently. This is because the compensation charge $\bTilde(\bx)$ depends on $\{\bD^a\}$ which in turn depends on pseudo-smooth wavefunctions (see Eqn.~\eqref{eq: compensation charge}) and hence needs to be mixed at every SCF iteration.

We employ a $n$-stage Anderson~\cite{Anderson1965547} mixing scheme to accelerate the SCF iteration in the current work. Denoting input total charge densities and output total charge densities at the $n^{th}$ SCF iteration as $\rhoTilde^{(n)}_{\text{in}}(\bx)$ and $\rhoTilde^{(n)}_{\text{out}}(\bx)$ respectively, the input total charge density for $(n+1)^{\text{th}}$ SCF iteration $\rhoTilde^{(n+1)}_{\text{in}}(\bx)$ is computed as
\begin{equation}
\rhoTilde^{(n+1)}_{\text{in}} = \gamma_{\text{mix}} \underline{\rhoTilde}_{\text{out}} + (1-\gamma_{\text{mix}}) \underline{\rhoTilde}_{\text{in}} \label{eq:gammamix}
\end{equation}
where $\underline{\rhoTilde}_{\text{in}(\text{out})}$ is expressed as a linear combination of input and output total charge densities from $n$ previous SCF iterations. To this end, we have
\begin{equation}
   \underline{\rhoTilde}_{\text{in}(\text{out})} = \rhoTilde^{n}_{\text{in}(\text{out})} + \sum_{k=1}^{n-1} c_k (\rhoTilde^{(n-k)}_{\text{in}(\text{out})} - \rhoTilde^{(n)}_{\text{in}(\text{out})}) \label{eq:mixcoeff}
\end{equation}
where we determine the linear combination coefficients $c_k$ by minimizing the weighted $L_2$-norm square of the residual $R = ||\rhoTilde_{\text{res}}||^2_W = ||\underline{\rhoTilde}_{\text{out}}(\bx) - \underline{\rhoTilde}_{\text{in}}(\bx)||^2_{W}$ with the weighted norm defined as 
\begin{equation}
    ||\rhoTilde_{\text{res}}||^2_W := \braket{\rhoTilde_{\text{res}}|\mathcal{I}+\kappa_c|\rhoTilde_{\text{res}}}\label{eq:weightnorm}
\end{equation}
 where $\kappa_c$ in $\bx$ basis is the Coulomb kernel $1/|\bx - \bx'|$ and $\mathcal{I}$ is the Identity operator. $\kappa_c$ in this weighted norm definition penalizes the term $\braket{\rhoTilde_{\text{res}}|\rhoTilde_{\text{res}}}$ during the Anderson minimization step at longer wavelength changes of total charge density (i.e. at smaller magnitude of wavevectors, $|\bq| \to 0$ in Fourier representation) in order to suppress charge sloshing in material systems with vanishing bandgaps as  explored in prior works~\cite{GPAW2010,PhysRevB.54.11169,Kim_2020}. Since the objective function ~\eqref{eq:weightnorm} in Anderson mixing is non-local in real-space, we now provide a prescription to recast this equation into a local-form using an auxiliary potential corresponding to residual charge density $\rhoTilde_{\text{res}}(\bx)$ obtained by solving the Poisson problem. To this end, we first note that  
\begin{align}
\braket{\rhoTilde_{\text{res}}|\kappa_c|\rhoTilde_{\text{res}}} &=  \int{\int{\frac{\rhoTilde_{\text{res}}(\bx)\rhoTilde_{\text{res}}(\bx')}{|\bx-\bx'|}d\bx}d\bx'} \nonumber \\
&= \int \rhoTilde_{\text{res}}(\bx) \varphi_{\text{res}}(\bx) d\bx
    \label{eq: Colomb norm}
\end{align}
where $\varphi_{\text{res}}(\bx) = \int{\frac{\rhoTilde_{\text{res}}(\bx')}{|\bx-\bx'|}d\bx'}$ denotes the auxiliary potential and is computed by solving the Poisson problem $-\frac{1}{4\pi}\nabla^2 \varphi_{\text{res}}(\bx) = \rhoTilde_{\text{res}}(\bx)$.  
%\begin{equation}
 % -\frac{1}{4\pi}\nabla^2 \varphi_{\text{res}}(\bx) = \rhoTilde_{\text{res}}(\bx)  
 % \label{eq: Residual Poisson solve}
%\end{equation}
Finally, from this Poisson equation and Eqn.~\eqref{eq: Colomb norm}, we have the following
\begin{align}
&\braket{\rhoTilde_{\text{res}}|\kappa_c|\rhoTilde_{\text{res}}} 
= \int \rhoTilde_{\text{res}}(\bx) \varphi_{\text{res}}(\bx) d\bx \nonumber \\
&=  -\frac{1}{4\pi}\int \nabla^2 \varphi_{\text{res}}(\bx) \varphi_{\text{res}}(\bx) d\bx \nonumber, \\
&= -\frac{1}{4\pi} \int \left(\nabla.(\varphi_{\text{res}} \nabla \varphi_{\text{res}}) - |\nabla \varphi_{\text{res}}(\bx)|^2] \right) d\bx \nonumber, \\
&= \frac{1}{4\pi}\int{|\nabla \varphi_{\text{res}}(\bx)|^2 d\bx},
\end{align}
where we used the divergence theorem for zeroing out the boundary term to get the last equality in the above equation.

Employing the finite-element discretization for the underlying electronic fields, we solve the following minimization problem in the Anderson mixing scheme for the PAW method.
\begin{equation}
    ||\rhoTilde^h_{\text{res}}(\bx)||^2_{W}  :=  \frac{1}{4\pi}\int{|\nabla \varphi^h_{\text{res}}(\bx)|^2d\bx} + \int{|\rhoTilde_{\text{res}}^h(\bx)|^2d\bx}
    \label{eq: Anderson norm}
\end{equation}
where, $\varphi^h_{\text{res}}(\bx)$ is the solution of the FE discretized Poisson equation $\bL\bPhi_{\text{res}}=\bc_{\text{res}}$ with $\bL$ denoting the discretized Laplacian operator as given in Eqn.~\eqref{eq: FELaplacian} and
$\bc_{\text{res}}$ is the forcing vector given by $\int \rhoTilde^{h}_{\text{res}}N_{I}^{h,p_{el}} d\bx$ while the entries of the solution vector $\bPhi_{\text{res}}$ denotes the linear combination coefficients of the expansion of $\varphi^h_{\text{res}}(\bx)$ in FE basis. Finally, the Anderson mixing coefficients $c_k$ in Eqn.~\eqref{eq:mixcoeff} obtained by minimizing the weighted norm defined in Eqn.~\eqref{eq: Anderson norm} are employed to determine the input total charge density ${\rhoTilde_{in}}^{h,{(n+1)}}$ using Eqn.~\eqref{eq:gammamix}. In particular, the linear combination coefficients $c_k$ are used to determine $\nTilde^h_{\text{in}}(\bx)$ and $\{ \bD^{a,h}_{\text{in}}\}$ for the subsequent ${(n+1)}^{th}$ SCF iteration. 

Using $\rhoTilde^{h,(n+1)}_{in}(\bx)$, the Hamiltonian($\bH$) is computed and subsequently we solve the FE discretized generalized eigenvalue problem $\bH\feNodalVectorUtilde_{i}=\varepsilon_{i}\bS\feNodalVectorUtilde_{i}$. The following subsections describes the details of the efficient solution strategies employed in the current work.

\subsection{Subspace iteration procedure for PAW generalized eigenvalue problem}
We employ the Chebyshev filtered subspace iteration approach(ChFSI)~\cite{ChFISalgo2014,SaadChFSIBook,ChFSISAAD2006,LEVITT201598,MOTAMARRI2013308} to solve the FE discretized generalized nonlinear eigenvalue problem in Eqn.~\eqref{eq: FE PAW governing equations} for the smallest $N$ eigenvector and eigenvalue pairs. ChFSI belongs to a class of iterative orthogonal projection methods and has been successfully employed in conducting large-scale real space DFT calculations over the past few years~\cite{dftfe1.0,Sparc2017,LIOU2020107330}. This approach relies on a Chebyshev filtering procedure that constructs a subspace rich in the desired eigenvectors exploiting the fast growth property of Chebyshev polynomials outside the interval $[-1,1]$, followed by a Rayleigh-Ritz step. The ChFSI approach can be viewed as a nonlinear subspace iteration procedure in the context of a self-consistent field (SCF) iteration employed to solve the DFT nonlinear eigenvalue problem. This approach allows the reuse of the filtered subspace $\mathcal{V}$ spanned by the eigenvectors of a given SCF iteration for a subsequent iteration, progressively improving it. To this end, the eigenspace of interest is refined at every SCF iteration, leading to a close approximation of the desired eigenspace of the nonlinear eigenvalue problem as the SCF iterations proceeds towards convergence. Most importantly, the ChFSI method is naturally amenable for the efficient use of high-performance computing architectures, since the construction of the filtered subspace does not involve any coupling among the vectors, allowing for matrix-multivector multiplications to be performed blockwise or in parallel over a few blocks, reducing the peak memory requirement~\cite{Das2019FastSystem,Das2023}. 

Unlike most of the prior works~\cite{ChFSISAAD2006,MOTAMARRI2013308,Sparc2017} that employed ChFSI for standard eigenvalue problems, the problem at hand involves the solution of a large sparse generalized Hermitian eigenvalue problem (GHEP) as discussed in Eqn.~\eqref{eq: FE PAW governing equations}. To this end, we describe how we adapt the ChFSI approach to solve the FE-discretized GHEP arising in the PAW method. \vspace{0.05in}
\paragraph{Construction of Chebyshev filtered subspace:} To solve the GHEP in Eqn.~\eqref{eq: FE PAW governing equations}, we consider the matrix 
$\bS^{-1}\bH$ that has the same eigenpairs as $\bH\bfePsips=\epsilon \bS\bfePsips$. To enrich a given trial subspace with eigenvectors corresponding to $N$ smallest eigenvalues (wanted spectrum) and dampen the unwanted spectrum, we seek to construct a Chebyshev polynomial filter involving  $\bS^{-1}\bH$. This requires us to efficiently evaluate the inverse of $\bS$, a large FE discretized sparse matrix. Recall from Eqn.~\eqref{eqn: simplfied FE PAW} that $\bS$ is of the form $\bS = \bM + \bP\boldsymbol{\Delta}_{\btS}{\bP}^{\dagger}$, a rank-$n_{\text{proj}}$ perturbation of $\bM$.  Using the Woodbury formula~\cite{woodberryformula} to compute the inverse of $\bS$, we have
\begin{align}\label{eqn:Sinv}
  {{\bS}}^{-1} &= {\textbf{M}}^{-1} - {\bM}^{-1}\bP\boldsymbol{\Delta} _{\btS_{\text{\tiny{inv}}}}{\bP}^{\dagger}{\bM}^{-1} \\
  \text{where}\;\; \boldsymbol{\Delta}_{\btS_{\text{\tiny{inv}}}} &= {\left(\boldsymbol{\Delta}_{\btS}^{-1} + {\bP}^{\dagger}{\bM}^{-1}\bP \right)}^{-1} \nonumber
\end{align}
We note that the above Equation~\eqref{eqn:Sinv} requires the computation of inverse of the large sparse FE basis overlap-matrix ($\bM$) of dimension $M \times M$, and the inverse of a smaller matrix $\boldsymbol{\Delta}_{\btS_{\text{\tiny{inv}}}} = \left(\boldsymbol{\Delta}_{\tiny{\bS}}^{-1} + {\bP}^{\dagger}{\bM}^{-1}\bP \right)$ of size $n_{\text{proj}}\cross n_{\text{proj}}$ that scales with number of atoms. In order to evaluate $\bM^{-1}$, Gauss-Lobatto-Legendre (GLL) quadrature rules coincident with the finite-element nodes of spectral finite-elements are employed in this work. These GLL points are reduced-order quadrature rules and are used only to construct $\bM$ in Eqn.~\eqref{eqn:mij}, rendering the matrix $\bM$ diagonal denoted by $\bM_{\text{D}}$. This leads to a trivial evaluation ~\cite{dftfe1.0,MOTAMARRI2013308} of $\bM^{-1}$ by inverting the diagonal entries of $\bM_{\text{D}}$ to be denoted as $\bM_{\text{D}}^{-1}$ subsequently. The other term in ${\bS}^{-1}$ that requires careful treatment is the inverse of  $\left(\boldsymbol{\Delta}_{\btS}^{-1} + {\bP}^{\dagger}{\bM}^{-1}\bP \right)$. Recall from Eqns.~\eqref{eqn:deltaHanddeltaS} and \eqref{eqn: simplfied FE PAW}, 
$\boldsymbol{\Delta}_{\btS}$ is block diagonal with each block corresponding to each atom `$a$' and hence $\boldsymbol{\Delta}_{\btS}^{-1}$ can be evaluated as the inverse of individual blocks $\boldsymbol{\Delta}_{\btS}^{a}$. Furthermore, using the fact that the PAW augmentation spheres are localized in real-space and seldom overlap, the matrix ${\bP}^{\dagger}\bM_{\text{D}}^{-1}\bP$ can be  approximated  as a block diagonal matrix with each block corresponding to an atom in the given material system. Hence, the block diagonal form of the $\boldsymbol{\Delta}_{\btS_{\text{\tiny{inv}}}}$ matrix in real-space allows us to compute its inverse atom-by-atom. Thus, the approximate ${\bS}^{-1}$ can be rewritten as follows:
\begin{align}\label{eqn:Sinvapprox}
\widehat{\bS}^{-1} &= \bM_{\text{D}}^{-1} - \sum_a{{\bM_{\text{D}}^{-1}}\bC^a{\boldsymbol{\Delta}^a _{\bS_{\text{inv}}}}\bC^{a\dagger}} \bM_{\text{D}}^{-1} \\
\text{where}\;\; \boldsymbol{\Delta}^a _{\bS_{\text{inv}}}&= {\left({\boldsymbol{\Delta}_{\bS}^{a^{-1}}} + {\bC^{a}}^{\dagger} \bM_{\text{D}}^{-1}\bC^{a} \right)}^{-1} \nonumber
\end{align}
In this work, we seek to construct a subspace rich in the desired eigenvectors of the PAW generalized eigenvalue problem $\bH\feNodalVectorUtilde=\epsilon \bS\feNodalVectorUtilde$ in Eqn.~\eqref{eq: FE PAW governing equations}  by building a Chebyshev polynomial filter of $\widehat{\bH} =\widehat{\bS}^{-1}\bH$ a close approximation to ${\bS}^{-1}\bH$. To this end, the filtering procedure first relies on mapping the unwanted spectrum of $\widehat{\bH}$ to $[-1,1]$ and the desired spectrum to $(-\infty,-1)$. This is accomplished by computing a shifted and scaled matrix $\widetilde{\bH}$ through the affine transformation given by $\widetilde{\bH} = \frac{2}{b-a}\widehat{\bH} -\frac{b+a}{b-a}\boldsymbol{I}$ where $a$ and $b$ denote the upper bounds of wanted and unwanted spectrum of $\widehat{\bH}$ respectively. The upper bound $b$ of the unwanted spectrum is obtained inexpensively by using a few steps of generalized Lanczos iteration~\cite{GeneralLanczos1982} as discussed in the Algo.~\eqref{alg: Generalized Lanczos} which employs $\widehat{\bS} = \bM_{\text{D}} + \sum_a\bC^a\boldsymbol{\Delta}^a_{\bS}{\bC^a}^{\dagger}$ and its inverse $\widehat{\bS}^{-1}$ as discussed in Eqn.~\eqref{eqn:Sinvapprox}. A good approximation to the upper bound $a$ of the wanted spectrum is obtained as the highest generalized Rayleigh quotient of $\bH$ in the Chebyshev filtered subspace of the previous SCF iteration.

We now construct the filtered subspace by computing the action of a degree-$m$ Chebyshev polynomial filter $T_m(\widetilde{\bH})$ on an input subspace $\Psips_{\text{in}}$ where the subspace $\Psips_{\text{in}}$ is usually the eigenspace from the previous SCF iteration and is a matrix of size $M \times N$. This action is accomplished using the three-term recurrence relation of Chebyshev polynomials  as given below:
\begin{equation*}
   \Psips_\text{f} = T_m(\widetilde{\bH})\Psips_{\text{in}} \;\text{where}\; T_{m+1}(\bX) = 2\bX T_{m}(\bX) - T_{m-1}(\bX)
   \label{eq: ChFSI recurrance}
\end{equation*}
The rapid growth property of Chebyshev polynomials outside $[-1,1]$ amplifies the components in $\Psips_{\text{in}}$ along the direction of the desired eigenvectors (occupied states) while dampening the components along the unwanted eigenvectors (unoccupied states), leading to the filtered subspace $\Psips_{\text{f}}$ to be a close approximation of the eigenspace of interest of the PAW eigenproblem $\bH\bfePsips=\epsilon \bS\bfePsips$ in a given SCF iteration.
 \vspace{0.08in}
\paragraph{Rayleigh-Ritz step:} Upon computing the filtered subspace $\Psips_{\text{f}}$ in a given SCF iteration, we project the Hamiltonian($\bH$) and PAW-overlap($\bS$) onto the filtered subspace $\Psips_{\text{f}}$ to construct projected matrices $\bH^p$ and $\bS^p$ of size $N\cross N$ as shown in step 5 of algo.\eqref{alg: ChFSI}. The Rayleigh-Ritz step now involves solving a subspace eigenvalue problem for $N$ eigenpairs and is of the form
\begin{equation}\label{eqn:subevp}
\bH^p\bX = \bS^p\bX\boldsymbol{\Lambda}
\end{equation}
where $\bX$ and $\boldsymbol{\Lambda}$  denote the eigenvector matrix and the diagonal eigenvalue matrix, respectively, of the smaller subspace eigenvalue problem Eqn.~\eqref{eqn:subevp}. Finally, a subspace rotation step (see Step 7 of Algo.\eqref{alg:ChFSI}) is performed to obtain the $\bS$-orthonormal eigenvectors $\Psips$ in the finite-element subspace which is in turn used to compute the output total charge density $\rhoTilde^{h}(\bx) = \nTilde^{h}(\bx) + \bTilde^h(\bx)$.

A good approximation of the eigenspace of interest in the ChFSI approach employed in this work relies on the choice of the degree $m$ of the Chebyshev polynomial filter. This choice depends on the separation of eigenvalues within the desired region of the eigenspectrum as well as the magnitude of the largest eigenvalue ($\lambda_{\text{max}}$) of the matrix ${\bS}^{-1}\bH$ that determines the ratio of spectral widths of the wanted to the unwanted regions of the eigenspectrum. We note that this $\lambda_{\text{max}}$ is dictated by the fineness of the mesh size of the underlying finite element discretization, i.e., it increases with decreasing mesh size. In the PAW method within the frozen core approximation, the electronic fields $\{\tilde{\psi}_i\}$ (or equivalently $\{\tilde{u}^{\bk}_i\}$ in the periodic case) and $\varphi$ smoothly vary compared to the electronic fields solved in the case of norm-conserving ONCV pseudopotentials for most atomic species. This allows for the use of coarser finite-element meshes in the case of PAW formalism, leading to the choice of Chebyshev polynomial degree between 10 and 30 for accurately approximating the desired eigenspace. The reduced degrees of freedom and the lower Chebyshev polynomial degree required to reach the desired chemical accuracies in the proposed finite-element framework for the PAW approach can provide significant computational advantage compared to current state-of-the-art finite-element code for DFT using ONCV pseudopotentials (\DFTFE). As demonstrated in the subsequent section, these computational gains translate to approximately 10-fold providing a pathway for efficient medium to large-scale calculations. 

The ChFSI approach described here involves computationally intensive operations notably in two key steps: first, constructing a subspace by computing $\bA\bX$, where $\bA$ represents any one of the matrices  $\bS$, $\bH$, $\widehat{\bS}^{-1}$, and second, performing a Rayleigh-Ritz step by computing matrix-matrix products of the form $\bX^\dagger\bY$. We will now elucidate efficient strategies that exploit the compact support nature of the finite-element basis to perform these operations in a computationally efficient manner.
\begin{figure}[h!]
%    \removelatexerror
\begin{algorithm}[H]
    \caption{ChFSI algorithm in \pawfe}\label{alg:ChFSI}
\KwIn{Initial subspace $\Psips_{\text{in}}$}
\KwOut{$\bS$-orthonormalised filtered subspace $\Psips$}
\begin{enumerate}
\item Compute $b$ from  k-step Generalized Lanczos as per Algo.~\eqref{alg: Generalized Lanczos} 
\item Determine the upper bound of wanted spectrum $a$ from the previous SCF iteration.
\item Scale and shift the Hamiltonian: $\widetilde{\bH} = \frac{2}{b-a}\widehat{\bH} -\frac{b+a}{b-a}\boldsymbol{I}$, where $\widehat{\bH} = \widehat{\bS}^{-1}\bH$
\item \texttt{CF}: Compute Chebyshev filtered subspace:    $\Psips_\text{f} = T_m(\widetilde{\bH})\Psips_{\text{in}}$ [$\mathcal{O}(MN)$]
\item \texttt{RR-Projection}: Compute the projected Hamiltonian ($\bH^{p}$) and projected PAW overlap matrices ($\bS^{p}$):
\\$\bH^{p} =\Psips_{\text{f}}^{\dagger}\bH\Psips_{\text{f}} $, $\bS^{p} = \Psips_{\text{f}}^{\dagger}\bS\Psips_{\text{f}}$ [$\mathcal{O}(MN^2)$]
\item \texttt{RR-GHEP} Solve the GHEP:\\ $\bH^{p}\bX = \bS^{p}\bX\bLambda$ [$\mathcal{O}(N^3)$]
\item \texttt{RR-SR} Perform subspace rotation:\\ $\Psips = \Psips_{\text{f}}\bX$. [$\mathcal{O}(MN^2)$]
\end{enumerate}
\label{alg: ChFSI}
\end{algorithm}
\end{figure}

\subsection{Efficient computational methodologies}
In the proposed ChFSI approach in Algo.~\eqref{alg:ChFSI}, the efficient computation of matrix-matrix products of the form $\bA\bX$ and $\bX^\dagger\bY$ is critical for enabling fast, accurate and scalable PAW-FE calculations on parallel computing architectures. We first consider the case of evaluating the matrix multivector products of the form $\bA\bX$ arising in the construction of Chebyshev filtered subspace, where matrix $\bA$ can be one of the discretized Hamiltonian matrices($\bH$), PAW overlap($\bS$) matrix or its inverse. We note here that the matrix $\bA$ is a large sparse matrix of dimension $M \times M$ with $M \approx \order{10^3} - \order{10^8}$ while $\bX$ is a large dense matrix of dimension $M \times N$  with $N \approx \order{10} - \order{10^4}$.  A naive strategy to compute $\bA\bX$ often adopted in the finite-element literature is to construct the FE discretized global sparse matrix $\bA$ using sparse-matrix storage data-structures and then compute the sparse matrix-dense matrix product $\bA\bX$\cite{hughes2012finite,Kirby}. This strategy is highly inefficient on modern parallel computing systems involving multi-threaded and multi-node architectures owing to the expensive data movement costs involved in the underlying sparse-matrix algorithms. To this end, we employ a cell-matrix approach~\cite{dftfe0.6,dftfe1.0,Kronbichler2012AApplication,JPDC} that avoids the construction of global sparse-dense matrix-products and instead employs dense-dense matrix-products at FE cell level to reduce memory requirements, enhance cache locality, and improve the arithmetic intensity of computation. Below, we describe the key mathematical aspects used in this implementation strategy to efficiently evaluate $\bA\bX$  during the ChFSI procedure for solving the PAW generalized eigenvalue problem.

%Furthermore, the operation $\bA\bX$ can be decomposed as $\bA\bX = \bA_{\text{loc}}\bX + \bA_{\text{nloc}}\bX$.
%(where $\bA := \bH, \bS, \widehat{\bS}^{-1}$)

Denoting the simulation domain by $\Omega_p$, we begin by partitioning $\Omega_p$ into subdomains $\Omega^{(t)}\; \forall t = 1,2,...,n_t$, where $n_t$ is the number of subdomains with each subdomain $\Omega^{(t)}$ assigned to an MPI task $t$. Further, let $E_t$ be the number of finite-element (FE) cells and $m_t$ be the number of basis functions (number of finite-element nodes) in each subdomain $\Omega^{(t)}$ such that $\Omega^{(t)} = \cup^{E_t}_{e=1}\Omega^{(e,t)}$, where the notation $(e,t)$ refers to the index of the FE cell '$e$' in an MPI task '$t$'. This allows us to decompose the integrals over the domain $\Omega_p$ in Eqns.~\eqref{eqn:hij}, \eqref{eqn:mij} to a series of integrals evaluated over smaller domains $\Omega^{(e,t)}$ as $\int_{\Omega_p}{f(\bx)d\bx} = \sum_t{\sum_e{\int_{\Omega^{(e,t)}}f(\bx)d\bx}}$. 
% follows:
% \begin{equation}
%     \int_{\Omega_p}{f(\bx)d\bx} = \sum_t{\sum_e{\int_{\Omega^{(e,t)}}f(\bx)d\bx}}
%     \label{eq: subdomain Integration}
% \end{equation}

Subsequently, the integrals over each FE-cell domain $\Omega^{(e,t)}$ are computed using a suitable quadrature rule as:
\begin{align}
    \int_{\Omega^{(e,t)}}{f(\bx)d\bx} =& \cb\int_{\widehat{\Omega}}{f(\widehat{\bx})\det \bJ^{(e,t)}d\widehat{\bx}} \nonumber \\  
    \approx & \cn \sum^{n_q}_q{w_qf(\bx_q)\det \bJ^{(e,t)}\biggr\rvert_{\widehat{\bx}_q}}
    \label{eq: Quadrature Integration}
\end{align}
\cb where $\widehat{\Omega}=[-1,1]^3$ refers to the reference cell, $n_q$ is the total number of quadrature points in the domain $\widehat{\Omega}$ and the $3\times 3$ matrix $\bJ^{(e,t)}$ is the Jacobian of the mapping from the current FE-cell $\Omega^{(e,t)}$ to the reference cell $\widehat{\Omega}$ whose elements are $J^{(e,t)}_{ij} = \nicefrac{\partial x^{(e,t)}_i}{\partial \widehat{x}_j}$ as is usually done in finite-element based approaches \cite{hughes2012finite}. 
\cn 3D Gauss Legendre quadrature rules that are tensor products of 1D quadrature rules~\cite{glrules} are employed in this work, and the order of these rules employed are determined such that errors due to the quadrature rule are of higher order than the discretization error. The flexibility in the choice of quadrature rules for evaluating various integrals involving PAW atomic data in  Eqns.~\eqref{eqn:hij},\eqref{eq: FE nonlocal contribution} allows the use of coarser finite-element grids to represent electronic fields and yet capture the relevant atom-centered data, a notable advantage compared to the finite-difference approach, a popular real-space discretization DFT method, where special schemes need to be devised to retain the use of coarser grid resolution for smoother electronic fields without egg-box effects~\cite{parsecleo}. 
\\We note that the operation $\bA\bX$ can be decomposed as $\bA\bX = \bA_{\text{loc}}\bX + \bA_{\text{nloc}}\bX$ (where $\bA := \bH, \bS, \widehat{\bS}^{-1}$ and $\bA_{\text{loc}}:= \bH_{\text{loc}}, \bM, \bM_{\text{D}}^{-1}$ respectively). To this end, the mathematical expressions for $\bA\bX$ relevant to our implementation strategy within the finite-element framework can be written as 
\begin{widetext} 
\begin{equation}
    \bH\bX = \left[\sum_t^{n_t}{{\bB^{(t)}}^T {\bQ^{(t)}}^T\sum_e^{E_t}{{\bZ^{(e,t)}}^T\left(\bH^{(e,t)}_{\text{loc}} + \sum_a{\bC^{(a,e,t)}\boldsymbol{\Delta}^a _{\bH}\left(\sum_{t^a}{\sum_{e^a}{{\bC^{(a,e^a,t^a)}}^\dagger\delta_{e^ae}\delta_{t^at}}}\right) }\right)\bZ^{(e,t)}\bQ^{(t)}\bB^{(t)} }}\right]\bX \\
    \label{eqn: HX kernel}
\end{equation}
\begin{equation}
     \bS\bX = \left[\sum_t^{n_t}{{\bB^{(t)}}^T {\bQ^{(t)}}^T\sum_e^{E_t}{{\bZ^{(e,t)}}^T\left(\bM^{(e,t)} + \sum_a{\bC^{(a,e,t)}\boldsymbol{\Delta}^a _{\bS}\left(\sum_{t^a}{\sum_{e^a}{{\bC^{(a,e^a,t^a)}}^\dagger\delta_{e^ae}\delta_{t^at}}}\right) }\right)\bZ^{(e,t)}\bQ^{(t)}\bB^{(t)} }}\right]\bX \\
    \label{eqn: SX kernel}   
\end{equation}
\begin{equation}
     \widehat{\bS}^{-1}\bX = \bM^{-1}_{\text{D}}\bX - \left[\sum_t^{n_t}{{\bB^{(t)}}^T {\bQ^{(t)}}^T\sum_e^{E_t}{{\bZ^{(e,t)}}^T  {\bM^{-1}_{\text{D}}}^{(e,t)}\sum_a{\bC^{(a,e,t)}\boldsymbol{\Delta}^a _{\bS_{\text{inv}}}\left(\sum_{t^a}{\sum_{e^a}{{\bC^{(a,e^a,t^a)}}^\dagger\delta_{e^ae}\delta_{t^at}}}\right) }\bZ^{(e,t)}\bQ^{(t)}\bB^{(t)} }}\right]\bM^{-1}_{\text{D}}\bX
    \label{eqn: SinvX kernel}     
\end{equation}      
\end{widetext}
In the Eqns.~\eqref{eqn: HX kernel}-\eqref{eqn: SinvX kernel}, the Boolean sparse matrix $\bB^{(t)}$ represents the \textbf{partitioner matrix} that acts on $\bX$ (or $\bM^{-1}_{\text{D}}\bX$) to provide the subdomain level matrix $\bX^{(t)}$ while preserving the continuity of the nodal field across the subdomain boundaries. The $m_t \cross m_t$ sparse matrix $\bQ^{(t)}$ represents the \textbf{constraint matrix} that constrains the values of the $m_t \times N$ matrix $\bX^{(t)}$ at the specific FE nodes. These constraints either satisfy the necessary boundary conditions (periodic, non-periodic, or semi-periodic) imposed on the discretized electronic wavefunctions or address constraints from non-conforming meshes~\cite{Bangerth2009DataSoftware}. Furthermore, the $(p+1)^3\times m_t$ Boolean matrix $\bZ^{(e,t)}$~denotes the \textbf{sub-domain to FE cell map} associated with the sub-domain $\Omega^{t}$ whose action on $\bQ^{(t)}\bX^{(t)}$ results in the extraction of the finite-element cell-level entries from $\bQ^{(t)}\bX^{(t)}$ while ensuring the continuity condition of the discretized electronic wavefunction across the finite-element cells within the partitioned sub-domain $\Omega^{(e,t)}$. The efficient computation of $\bA\bX$ illustrated in Eqns.~\eqref{eqn: HX kernel}-\eqref{eqn: SinvX kernel} can be summarized as a sequence of the following steps:
\begin{enumerate} [(i)]
    \item Precompute the entries of the cell-level matrices $\bH^{(e,t)}_{\text{loc}}$, $\bS^{(e,t)}_{\text{loc}}$ $\bC^{(a,e,t)}$ and $\boldsymbol{\Delta}^a_{\bA}$, and the atom dependent coupling matrices ($\boldsymbol{\Delta}^a_{\bA}$:= $\boldsymbol{\Delta}^a_{\btH}$, $\boldsymbol{\Delta}^a_{\btS}$, $\boldsymbol{\Delta}^a_{\bS_{\text{inv}}})$ using the quadrature rules as discussed in Eqn.~\eqref{eq: Quadrature Integration}
    \item Extract the finite-element cell-level matrix:\\ $\bX^{(e,t)}= \bZ^{(e,t)}\bQ^{(t)}\bB^{(t)}\bX,\forall (e,t)$
    \item Evaluate the partial non-local operator action at the FE cell level: 
    \\$\Delta^{a}_ {\bY} = \sum_{t^a}{\sum_{e^a}{{\bC^{(a,e^a,t^a)}}^\dagger\delta_{e^ae}\delta_{t^at}}}\bX^{(e,t)}$ where $\delta_{e^a\,e}$ indicates that this computation is done in only those FE cells $e^a$ lying in the augmentation sphere ($\Omega_a$) centered around an atom `$a$'. Furthermore, $\delta_{t^a\,t}$ indicates that this computation is performed only on MPI tasks $t^{a}$ that are associated with $\Omega_a$.
    \item Evaluate the local operator action along with the impending non-local operator action:\\
    $\bY^{(e,t)} = \bA^{(e)}_{\text{loc}}\bX^{(e,t)} + \sum_a{\bC^{(a,e,t)}{\Delta}^a _{\bA}\Delta^{a}_ {\bY}}$
    \item Assembly of the global vector $\bY = \bA \bX$:\\
    $\bY = \sum_t^{n_t}{{\bB^{(t)}}^T {\bQ^{t}}^T\sum_e^{E_t}{{\bZ^{(e,t)}}^T\bY^{(e,t)}}}$
\end{enumerate}
We remark that the finite-element cell-level matrices $\bH^{(e,t)}_{\text{loc}}$ and the coupling matrix $\boldsymbol{\Delta}^a_{\bH}$ are computed at every SCF iteration while $\bS^{(e,t)}_{\text{loc}}$ is computed for a given finite-element mesh. Further $\boldsymbol{\Delta}^a _{\bS}$ is computed only once while $\boldsymbol{\Delta}^{a}_{\btS_{\text{\tiny{inv}}}}$ is evaluated
only once per ground-state calculation.

Following the evaluation of $\bY = \bA\bX$, the key computationally intensive operation in the construction of Chebyshev filtered subspace, the subsequent step is the Rayleigh-Ritz step that requires the evaluation of matrix-matrix products of the form $\bX^\dagger \bY$ (see Algorithm 3). To this end, the matrices $\bX,\bY$ are partitioned row-wise into $\bX_p,\bY_p$ across various MPI tasks. Subsequently, the local $\bX_p^\dagger \bY_p$ is evaluated to obtain the smaller matrix $N\cross N$ in each MPI task and finally, the local contribution is summed across all MPI tasks to obtain $\bX^\dagger\bY = \sum_p{\bX_p^\dagger\bY_p}$. 

In this section, we discussed the various computational methodologies employed in the proposed \pawfe~approach that can enable accurate large-scale DFT calculations on material systems up to tens of thousands of electrons in a computationally efficient manner. The following section discusses the accuracy and performance results of the proposed computational approach while benchmarking with the state-of-the-art plane wave (PW) codes Abinit\cite{Abinit2016} and Quantum Espresso\cite{qe}.

\section{Results and Discussion}\label{sec:results}
We demonstrate here the accuracy, performance and parallel scalability of the proposed \pawfe~method on representative benchmark problems involving periodic, semi-periodic and non-periodic boundary conditions. We first discuss the accuracy of our implementation on non-periodic systems involving molecular systems and small metallic nano-clusters by comparing the ground-state energies with plane-wave-based PAW implementations. We then demonstrate the performance of \pawfe~in terms of both computational node-hours and minimum wall time with respect to plane-wave calculations for increasingly large sizes of metallic Cu nanoclusters containing up to 17,537 electrons (923 atoms). Subsequently, we demonstrate the accuracy of our method compared to plane-wave calculations on semi-periodic and fully periodic systems involving unit cells, large super-cells, and surfaces. We further discuss the performance of our method in terms of computational node-hrs and minimum wall time for increasingly large sizes of Cu$_{3}$Pt periodic supercells containing up to 34,304 electrons (2048 atoms) compared to plane-wave PAW calculations. Finally, we present the computational efficiency and parallel scalability of \pawfe~compared to norm-conserving ONCV pseudopotential calculations using \DFTFE ~\cite{dftfe0.6,dftfe1.0} on large-scale material systems involving $\sim$50,000 electrons.

All DFT calculations reported in this work employ GGA~\cite{GGA1997} exchange-correlation functional of the PBE~\cite{PBE} form. Furthermore, we employ the PAW data-sets from the pseudo-dojo database~\cite{pseudoDojoONCV,pseudoDojoPAW}. In all our simulations, we use Fermi-Dirac smearing with $T=500 K$.  Additionally, for the mixing of the total charge density involving ($\nTilde(\bx)$) and $\{\bD^a\}$, we employ the $n$-stage Anderson mixing~\cite{Anderson1965547} as discussed in the Section~\ref{sec:scf}.

For all accuracy validation studies in this work, we employ refined finite element (FE) meshes with FE interpolating polynomial ($p$) of degree 7. These meshes are constructed such that the differences in the all-electron ground-state energies (Eqn.~\eqref{eq: PAW energy}) between successively sub-divided meshes reach $\order{10^{-6}}$ Ha/atom. In addition, the choices of quadrature integration rules and FE interpolating polynomial for electrostatics $p_{\text{el}}$ are made such that the energy variation with respect to these parameters is an order of magnitude lower than this discretization error. In the case of plane-wave calculations, we use \abinit\footnote{\abinit~version \texttt{9.10.3} is used for the accuracy validation} code and the cut-off energy for wavefunctions $E^{\text{wfc}}_{\text{cut}}$ is chosen so that the ground-state energy is converged up to $\order{10^{-6}}$ Ha/atom while simultaneously ensuring that the energy change with respect to the cut-off energy for the total charge density $E^{\text{rho}}_{\text{cut}}$ is an order of magnitude lower.

A comparative study involving the performance of our \pawfe~implementation with that of plane-wave calculations has been conducted by solving a given problem to the same level of accuracy. To this end, we choose discretization parameters such that the discretization error with respect to a very refined calculation in each code is around $\sim2 \times 10^{-4}$ Ha/atom for our comparisons. We employ \texttt{Quantum espresso}(\qe) code\footnote{\qe~version \texttt{7.2} with ELPA is used for the performance benchmarking study. The authors observed \qe~was computationally efficient than \abinit~on a few benchmark examples.} to carry out our performance studies using plane wave DFT calculations. Furthermore, we use two metrics to compare the performance of \texttt{PAW-FE} and the plane-wave calculations: (i) computational cost ($\eta_{\text{c}}$) in node-hrs, and (ii) minimum wall time ($\tau^{\text{min}}_{\text{c}}$) in secs. The computational cost $\eta_{\text{c}}$ is obtained by multiplying the minimum number of compute nodes required to fit a given problem with the average wall-time per SCF iteration\footnote{Average wall-time per SCF is obtained by dividing the wall time taken for the self-consistent field iteration procedure with the number of SCF iterations taken to reach a default convergence threshold for self-consistency}. In addition, the minimum wall time ($\tau^{\text{min}}_{\text{c}}$) is obtained by computing the average wall time per SCF by increasing the number of compute nodes until a parallel scaling efficiency of around $35\%$ is reached. All simulations involved in accuracy and performance benchmarking studies are performed on high-memory CPU nodes of PARAM Pravega\footnote{PARAM Pravega is one of India's fastest supercomputers stationed at Indian Institute of Science comprising of 156 High Memory Intel Xeon Cascade-Lake based CPU nodes (7,488 Cores) where each node consists of 48 Intel Xeon 6248 processor, 768 GB memory and Mellanox HDR Infiniband interconnect between all the nodes for fast MPI communication.}. 

\subsection{Non-periodic systems}
In this subsection, we consider the case of non-periodic systems for our accuracy and performance benchmarking. In particular, for accuracy validation studies, we consider isolated molecules such as CO, CH\textsubscript{4}, H\textsubscript{2}O and also metallic nano-clusters such as Cu\textsubscript{13}, Al\textsubscript{13}. We benchmark the accuracy of \pawfe~ with \abinit~by comparing the energy vs bond length curves and the ground state energies computed at the minimum energy configuration. Following the accuracy validation study, we assess the performance of our implementation by comparing the computational cost ($\eta_{\text{c}}$) and minimum wall time($\tau^{\text{min}}_{\text{c}}$) of \pawfe~ with Quantum Espresso(\qe), a widely used code employing plane-wave basis. In this performance benchmarking study, we consider three-dimensional icosahedron\cite{Icosahedron} shaped Cu nanoparticles of various sizes ranging from 55 atoms (1045 electrons) to 923 atoms (17537 electrons). Homogeneous boundary conditions have been imposed on the electrostatic potential ($\varphi^h(\bx)$) for these calculations using \pawfe, and suitable vacuum is used till the electronic fields($\varphi^h(\bx),\{\psips^h(\bx)\}$) decays to 0. While in plane-wave codes, periodic boundary conditions (PBCs) are only admissible and hence PBCs are employed using a suitable vacuum to minimise image-image interactions. 

\subsubsection{Accuracy benchmarking}
\begin{figure}[!h]
\centering
    \includegraphics[scale=0.42]{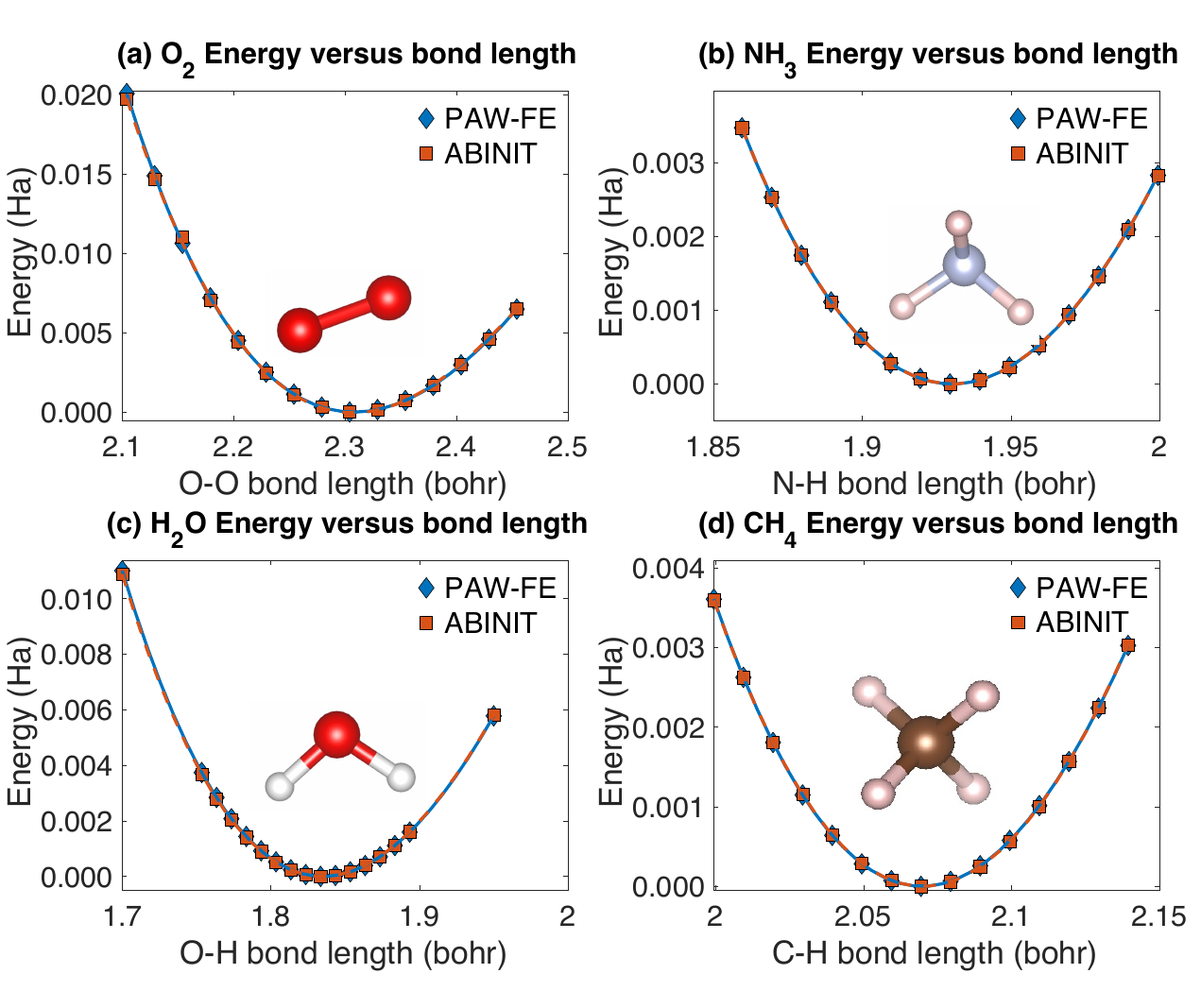}
    \caption{Comparison between \abinit~and \pawfe: Energy vs bond length for (a) Oxygen (b) Ammonia (c) Water (d) Methane. The energy shown is relative to the minimum ground-state energy obtained in each code.}
    \label{fig:EOS NP}
\end{figure}

 In all the accuracy benchmarking studies reported in this subsection, we choose refined finite-element (FE) meshes with FE interpolating polynomial degree $p$ to be $7$ and the mesh size $h$ to be around 1.0~\textrm{Bohr} while in the case of \abinit\footnote{\label{note1} We use the \texttt{usexcnhat}=0 in \abinit~to enforce Bl\"{o}chl formulation of exchange-correlation(XC) contribution with \texttt{pawxcdev}=0 to avoid any approximation in the XC correction term computation.}, we use a plane-wave cut-off energy $E^{\text{wfc}}_{\text{cut}}=60$~\textrm{Ha}. Figure~\eqref{fig:EOS NP} reports the energy as a function of inter-atomic distance obtained from \abinit~and \pawfe~ for the molecular systems O\textsubscript{2},  H\textsubscript{2}O, NH\textsubscript{3} and CH\textsubscript{4}. Since the scale of ground-state energies is different in the two implementations compared here, the energies in the plots of Fig.~\eqref{fig:EOS NP} are measured relative to the minimum ground-state energy obtained in this study using \abinit~and \pawfe~for each of the above molecular systems. We now describe the details of the simulations conducted for our comparative study: \textit{(i) Oxygen:} The energy as a function of O-O bond length is plotted in Fig.~\eqref{fig:EOS NP}(a) and the bond length for the minimum energy configuration is found to be $2.30$~\textrm{Bohr} in \pawfe~ which is in excellent agreement with \abinit, \textit{(ii) Ammonia:} The trigonal pyramidal geometry of ammonia is considered where the N-H bond lengths are varied such that the H-N-H bond angle is maintained to be $106.2^{\circ}$. In the Fig.~\eqref{fig:EOS NP}(b), we plot the energy variation as a function of this N-H bond length and we observe the N-H bond length at the minimum energy configuration to be around $1.93$~\textrm{Bohr} using \pawfe~ which matches very closely to the value obtained with \abinit, \textit{(iii) Water:} The O-H bond lengths in the water molecule are varied while ensuring that the planar H-O-H bond angle of $104.46^{\circ}$ is maintained. Energy of the molecule is plotted as a function of this O-H bond length in Fig.~\eqref{fig:EOS NP}(c) and we observe from our \pawfe~ calculations that the O-H bond length at the minimum energy configuration is $1.83$~\textrm{Bohr} which is in close agreement with \abinit, \textit{(iv) Methane:} The tetrahedral geometry of methane is considered where the C-H bond lengths are varied such that H-C-H angle is maintained at $109.5^{\circ}$. Fig.~\eqref{fig:EOS NP}(d) shows the comparison between \pawfe~and \abinit~calculations. These results indicate that the C-H bond length at the minimum energy configuration is $2.07$~\textrm{Bohr} in both cases.
 \vspace{-0.02in}
 
 To summarize the observations shown in Fig.~\eqref{fig:EOS NP}, we find excellent agreement in all the energy vs bond-length curves between \pawfe~and \abinit~for the representative molecules considered in this work. Furthermore, the bond length at the minimum energy configuration obtained from our implementation shows a very close match with that of plane-wave calculations as summarized in Table~\eqref{tab: bond length}. We now compare the ground-state energies of a few representative systems obtained using \pawfe~with those obtained using \abinit~in Table~\eqref{tab:Accuracy BM table}. To this end, we consider the above four molecular systems at the minimum energy configuration obtained from Fig.~\ref{fig:EOS NP}. Additionally, we consider the 13-atom icosahedron nanoclusters Cu$_{13}$ and Al$_{13}$ with 4.8~\textrm{Bohr} and 5.3~\textrm{Bohr} as the nearest neighbour distances, respectively~\cite{Jain2013}. For a one-on-one comparison of ground-state energies with \abinit~for these non-periodic systems considered in this work, we compute the pseudo valence energy in \pawfe~using the expression in Eqn.~\eqref{eq: PAW valence energy} as opposed to the all-electron energy in Eqn.~\eqref{eqn:FEAllElecEnergy}. We note that Table~\eqref{tab:Accuracy BM table} shows an excellent match between the ground-state energies obtained from \pawfe~ and \abinit, and the differences in the energies are $\order{10^{-5}}$\textrm{Ha}/atom. Having established the agreement of \pawfe~with a plane-wave code in terms of ground-state energies, we now focus on benchmarking the performance of \pawfe~against plane-wave implementations.
 
  \begin{table}
     \centering
     \begin{tabular}{|c|c|c|c|}
         \hline
         Molecule & Bond Type & Bond length  & Bond length\\
         -  & - & (\textrm{Bohr})\pawfe~ & (\textrm{Bohr})\abinit\\        
         \hline
         O\textsubscript{2} & O-O  &\cb2.3072\cn&\cb 2.3073\cn\\
         \hline
          H\textsubscript{2}O& O-H &\cb1.8349\cn&\cb 1.8347\cn\\
          \hline
          NH\textsubscript{3}& N-H &\cb1.9292\cn&\cb1.9295\cn\\
          \hline
          CH\textsubscript{4}& C-H &\cb2.0706\cn &\cb2.0704\cn\\
          \hline  
     \end{tabular}
     \caption{Comparison of bond length at the minimum energy configuration between \abinit~and \pawfe~ for O\textsubscript{2}, H\textsubscript{2}O, NH\textsubscript{3} and CH\textsubscript{4} obtained from Fig.~\eqref{fig:EOS NP}.}
     \label{tab: bond length}
 \end{table}
 
\begin{table}[H]
    \centering
    \begin{tabular}{|c|c|c|c|}
        \hline
        System & Abinit Energy  & PAW-FE Energy  & Error in Energy\\
         -& (\textrm{Ha}/atom)  & (\textrm{Ha}/atom) & (\textrm{Ha}/atom) \\
         \hline
        H\textsubscript{2}O & -5.7771402 & -5.7771352 & $4.96\cross 10^{-6}$ \\
        \hline
         NH\textsubscript{3} & -2.9544162 & -2.9544196 & $3.39 \cross10^{-6}$\\
         \hline
         O\textsubscript{2}& -16.0550787 & -16.0550659 & $1.28\cross 10^{-5}$ \\        
         \hline
         CH\textsubscript{4}& -1.6254853 &-1.6254649  & $2.04\cross 10^{-5}$ \\
         \hline
         Cu\textsubscript{13}&-197.9146105&-197.9146504& $4.0\cross 10^{-5}$ \\
         \hline
         Al\textsubscript{13}& -2.0699280 & -2.0699363 & $8.37\cross10^{-6}$ \\
         \hline
    \end{tabular}
    \caption{Ground state energy comparison between \abinit~and \pawfe~for the molecular systems (H\textsubscript{2}O,NH\textsubscript{3},O\textsubscript{2},CH\textsubscript{2}) and nanoclusters(Cu\textsubscript{13},Al\textsubscript{13}).}
    \label{tab:Accuracy BM table}
\end{table}

\subsubsection{Performance benchmarking}
We benchmark the performance of \pawfe~with Quantum espresso \qe\cite{qe}, a widely used open-source plane-wave (PW) code, by comparing the average CPU time per SCF in terms of computational node-hrs($\eta_{\text{c}}$) and minimum wall time($\tau^{\text{min}}_{\text{c}}$) with increasing sizes of Cu nanoclusters\cite{Icosahedron} solved to a similar level of accuracy between the two approaches. These icosahedron-shaped nanoclusters considered in this study are constructed with a $4.8$~\textrm{Bohr} nearest neighbour distance\cite{Jain2013}, and the nanocluster size is increased by varying the number of shells. To this end, non-periodic DFT calculations using the PAW method are conducted on Cu\textsubscript{3-shell} (2793 electrons), Cu\textsubscript{4-shell} (5871 electrons), Cu\textsubscript{5-shell} (10,659 electrons), Cu\textsubscript{6-shell} (17,537 electrons). The two metrics ($\eta_{\text{c}},\tau^{\text{min}}_{\text{c}}$) used to benchmark the performance results are discussed at the beginning of this section. In \pawfe, the degree of the FE interpolating polynomial $p$ is chosen to be 5, and the mesh size $h=1.2$~\textrm{Bohr} is used with a Chebyshev polynomial degree of 20 in the ChFSI step. In the case of \qe~, the plane-wave cut-off energy $E^{\text{wfc}}_{\text{cut}}=25 \text{\textrm{Ha}}$ is used with $\Gamma$-point for k-point sampling and default settings are used for the eigensolver. These discretization parameters are chosen such that the discretization error in the ground-state energies obtained in both \pawfe~ and \qe~ is $\sim 2 \times 10^{-4}$\textrm{Ha}/atom. Furthermore, we used a vacuum of approximately $20$~\textrm{Boh} and $15$~\textrm{Boh} from the farthest atom of the nanocluster in \pawfe~and \qe~respectively, ensuring that ground-state energies are converged up to $\order{10^{-5}}$\textrm{Ha}/atom with vacuum size.

Table~\eqref{tab:Computational Cost NP} reports the average CPU time per SCF iteration in terms of computational node-hrs ($\eta_{\text{c}}$) and the number of basis functions in \pawfe~and \qe~for various sizes of Cu nanoclusters studied here. From this table, we find that for system sizes greater than $\sim 1000$ electrons, \pawfe~is more efficient than \qe. We observe that the efficiency gains for \pawfe~increases with system size over \qe, achieving gains of around 3.1$\times$, 4.3$\times$ and 5.1$\times$ for Cu$_{\text{4-shell}}$, Cu$_{\text{5-shell}}$ and Cu$_{\text{6-shell}}$ respectively, the benchmark systems considered for CPU-time comparison. This increase in computational gains is attributed to the necessity of using more processors to satisfy the peak memory requirement, where the efficient parallel scalability of \pawfe~provides the necessary advantage. Furthermore, in this case of non-periodic systems, the spatial adaptivity of the finite-element basis allows the FE mesh to be coarse-grained into the vacuum in contrast to the uniform spatial resolution of the plane-wave basis in \qe, which is also a factor for the superior gains observed in \pawfe. Additionally, we estimate the computational complexity of \pawfe~from Table~\eqref{tab:Computational Cost NP} and find it to be of $\order{N_e^{1.8}}$ while for \qe~we observe it to be of $\order{N_e^{2.6}}$. Hence, we expect to see higher gains of \pawfe~over \qe~with further increase in system size.
 
 Finally, the Table~\eqref{tab: Min Wall time NP} compares the average time per SCF in terms of minimum wall time($\tau^{\text{min}}_{\text{c}}$) obtained in both \pawfe~and \qe. We note that the minimum wall time of \qe~is measured at $\sim 30-40\%$ scaling efficiency while a scaling efficiency of $\sim 50-55\%$ is reached for \pawfe\footnote{ \pawfe~could be scaled further on more nodes, however, the maximum available 140 high-memory nodes on PARAM-Pravega limited the further reduction of minimum wall time}. Even for a small system involving Cu\textsubscript{3-shell} comprising of 2793 electrons, we observe a gain of around $3.2\cross$ in the minimum wall time for \pawfe~compared to \qe, and we see almost a staggering speedup of $10\cross$ for Cu\textsubscript{6-shell} comprising of 17,537 electrons, the largest system considered in this study. The increased speedup with system sizes is attributed to the superior scaling of \pawfe~which requires only nearest neighbour communication and the various efficient computational methodologies employed in solving the PAW eigenvalue problem using the Chebyshev filtered subspace iteration approach (ChFSI) in \pawfe~(see Section~\ref{sec:scf}).
% \begin{figure}
%     \centering
%     \includegraphics[scale=0.40]{CuNanoclusterGGA.pdf}
%     \caption{Icosahedron Cu nanoclusters and simulation domain considered for performance benchmark study}
%     \label{fig:Cu cluster}
% \end{figure}
\begin{widetext}

\begin{table}
     \begin{tabular}{|c|c|c|c|c|c|c|c|}
      \hline
       System  & atoms& electrons  & QE  & QE($\eta_{\text{c}}$)  & PAW-FE  & PAW-FE($\eta_{\text{c}}$)  & Speedup \\
        - & - & -&$\#\text{basis fns.}$ & Node-hrs & $\#\text{basis fns.}$  & Node-hrs & - \\
        \hline
         Cu\textsubscript{2-shell}&55 &1045 & 746,429 & 0.0076 & 950,950 & 0.0086 & 1.0 \\
         \hline
         Cu\textsubscript{3-shell}&147 &2793  & 1,289,759 & 0.08 & 1,912,201  & 0.04 & 2.0\\
         \hline
         Cu\textsubscript{4-shell}& 309&5871 & 2,047,417 & 0.39 & 3,399,083 &  0.13& 3.1 \\
         \hline
         Cu\textsubscript{5-shell}& 561&10659 & 3,056,521  & 1.45 & 5,516,293 & 0.34 & 4.3 \\
         \hline
         Cu\textsubscript{6-shell}&923 &17537  & 4,352,211 & 4.74 & 8,439,912 & 0.93 & 5.1 \\
         \hline
    \end{tabular}
    \caption{Computational cost ($\eta_{\text{c}}$) comparison between \pawfe~and \qe~in node-hrs (discretization error $\sim2\cross10^{-4}$~\textrm{Ha}/atom). $\eta_{\text{c}}$ is computed as the product of the minimum number of nodes required to fit a given material system and the average wall time taken per SCF iteration. \textbf{Case Study}: Cu nanoclusters of varying sizes}
    \label{tab:Computational Cost NP}
\end{table}
\end{widetext}

\begin{table}
    \centering
    \begin{tabular}{|c|c|c|c|}
   \hline
      System   & QE ($\tau^{\text{min}}_{\text{c}}$) & PAW-FE ($\tau^{\text{min}}_{\text{c}}$) & Speedup\\   
      -   & sec & sec & -\\
    \hline  
      Cu\textsubscript{3-shell}   & 31.2 & 9.8 & 3.2\\
    \hline     
    Cu\textsubscript{4-shell}     & 96.3 & 15.5 & 6.2\\
    \hline     
    Cu\textsubscript{5-shell}     & 150.7 & 20.9 & 7.2 \\
    \hline     
     Cu\textsubscript{6-shell}    & 490.0 & 48.6 & 10.1\\
    \hline     
    \end{tabular}
    \caption{Comparison between \pawfe~ and \qe~ in terms of minimum wall time per SCF iteration ($\tau^{\text{min}}_\text{c}$) \textbf{Case Study}: Cu nanoclusters of varying sizes}
    \label{tab: Min Wall time NP}
\end{table}  
\subsection{Periodic/Semi-periodic systems}
In this subsection, we consider the case of periodic and semi-periodic systems for benchmarking accuracy and performance. In particular, we benchmark the accuracy of \pawfe~with \abinit~by comparing the energy versus lattice parameters of representative bulk systems involving Li\textsubscript{2}O, Cu, Cu\textsubscript{3}Pt and Li unit cells. Furthermore, we compute and benchmark with \abinit~the formation energy of Li\textsubscript{2}O and the surface energies of the semi-periodic systems -- Zr(001) and La(011) terminated Lithium Lanthanum Zirconium Oxide (LLZO), an electrolyte material. Finally, we compare the band structure of C diamond primitive cell obtained from \pawfe~and \abinit.  Following the accuracy benchmarking study, we assess the performance of \pawfe~with \qe~ on two metrics ($\eta_\text{c}$, $\tau^{\text{min}}_{\text{c}}$) for increasing sizes of Cu\textsubscript{3}Pt supercells.

%mention how was the problem set up in ABINIT and PAW-FE (boundary conditions) in carrying out these comparisons.

\subsubsection{Accuracy benchmarking}
\begin{figure}[!h]
    \includegraphics[scale=0.42]{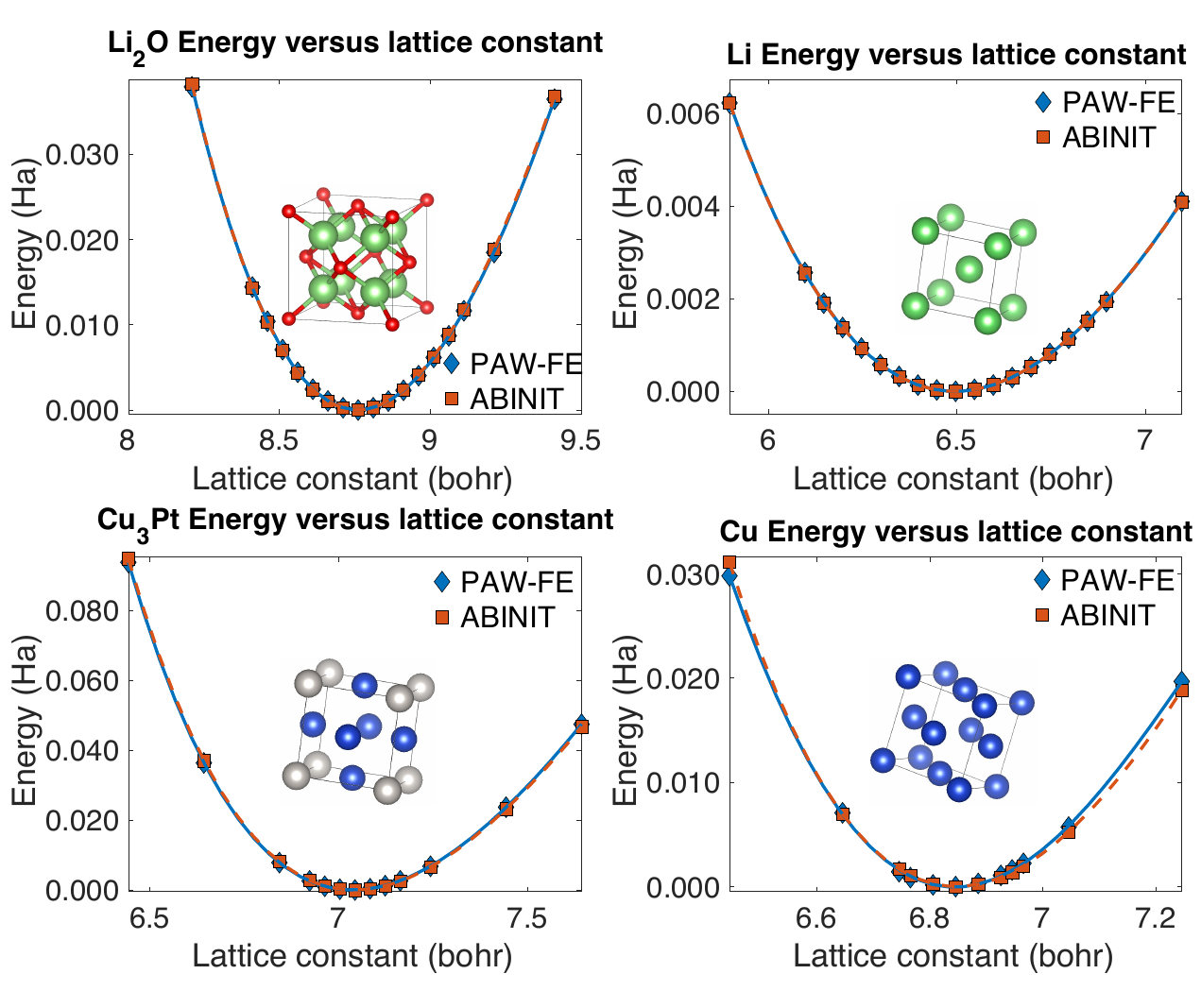}
    \caption{Comparison between \abinit~and \pawfe~:\\ Energy vs lattice parameter for (a) Li\textsubscript{2}O cubic, (b) Li bcc, (c)Cu\textsubscript{3}Pt fcc (d) Cu fcc unit-cells. The energy shown is relative to the minimum ground-state energy obtained in each code.}
    \label{fig:EOS Per}
\end{figure}

In all the accuracy benchmarking studies of periodic/semi-periodic systems reported here, we compute the all-electron PAW ground state energy using Eqn.~\eqref{eqn:FEAllElecEnergy} in \pawfe. Furthermore, we choose refined finite-element (FE) meshes with an FE interpolating polynomial degree of 7 and mesh size $h$ to be around $1.0$~\textrm{Boh} while in the case of \abinit~, we use a plane-wave cut-off energy $E^{\text{wfc}}_{\text{cut}}=60~\text{\textrm{Ha}}$.\\[0.05in]
\paragraph{Energy vs lattice parameter:} Figure~\eqref{fig:EOS Per} shows the variation of ground-state energy as a function of lattice parameter for the cases of Li\textsubscript{2}O(cubic), Cu(fcc), Cu\textsubscript{3}Pt(fcc) and Li(bcc) unitcells. We use orthogonal unit-cells in each case and employ shifted $8\cross8\cross8$ Monkhorst-Pack k-point grid for sampling the Brillouin zone~\cite{mpgrid}. In Fig.~\eqref{fig:EOS Per}, we plot the energies relative to the minimum ground-state energy obtained in this study using \pawfe~and \abinit~for each of the above material systems. This is done to account for different energy scales in the two codes considered here. From Fig.~\eqref{fig:EOS Per}, we observe an excellent correspondence between \pawfe~and \abinit. Furthermore, we also observe from Table~\eqref{tab: lattice constant} that the lattice parameters at the minimum energy configurations show an excellent match for all the representative periodic systems considered here.

 \begin{table}
 
     \centering
     \begin{tabular}{|c|c|c|c|}
         \hline
         System & Crystal type & Lattice parameter  & Lattice parameter  \\
         -  & - & (\textrm{Boh}) \pawfe & (\textrm{Boh}) \abinit\\        
         \hline
         Li\textsubscript{2}O & cubic  & \cb 8.7569 \cn & \cb 8.7571 \cn\\
         \hline
          Li & bcc & \cb 6.4968 \cn & \cb 6.4969 \cn\\
          \hline
          Cu\textsubscript{3}Pt& fcc & \cb7.0415\cn & \cb 7.0416 \cn\\
          \hline
          Cu & fcc & \cb 6.8487\cn& \cb 6.8488 \cn\\
          \hline  
     \end{tabular}
     \caption{Comparison of lattice parameters at the minimum energy configuration between \abinit~and \pawfe~ for Li\textsubscript{2}O(cubic), Li(bcc), Cu\textsubscript{3}Pt(fcc) and Cu(fcc) unit-cells obtained from Fig.~\eqref{fig:EOS Per}.}
     \label{tab: lattice constant}
 \end{table}
\paragraph{Formation energy:} Using the ground-state energies at the minimum energy configuration obtained from Fig.~\ref{fig:EOS Per}, we compute the formation energy $\Delta E^{\text{Li\textsubscript{2}O}}_{\text{f}}$ of Li\textsubscript{2}O as 
Hartree-per-formula unit(\textrm{Ha}/f.u.) using the following expression:
\begin{equation}
\Delta E^{\text{Li\textsubscript{2}O}}_{\text{f}} = \frac{1}{4}E_{Li\textsubscript{2}O} - E_{\text{Li}} - \frac{1}{2}E_{O\textsubscript{2}}
\end{equation}
To this end, $\Delta E^{\text{Li\textsubscript{2}O}}_{\text{f}}$ is found to be -0.223095 \textrm{Ha}/f.u. in \abinit, whereas in the case of \pawfe, we obtain a value of -0.223126 \textrm{Ha}/f.u. This results in a difference of $\sim 3.2\cross10^{-5}$ \textrm{Ha}/f.u., establishing a good agreement in the formation energy computed between the two codes.
\paragraph{Surface energies:} Next, we compute and validate the surface energies of LLZO (\textit{Lithium Lanthanum Zirconium Oxide}), a material system of interest for solid electrolytes, focusing on La(011) and Zr(001) terminations. We obtain the coordinates from a study of LLZO particle morphology\cite{LLZOMorphology}. The slab is placed at the center of the simulation domain with $25$~\textrm{Boh} vacuum on either side. In \pawfe, semi-periodic boundary conditions are employed for the electronic fields, with homogeneous Dirichlet boundary conditions for $\varphi^h(\bx)$ on the boundaries parallel to the surface. While in \abinit, periodic boundary conditions are employed. The surface energy($\gamma$) is computed using the following expression in the units of $\nicefrac{J}{m^2}$ 
\begin{equation}
  \gamma = \frac{1}{2A_{\text{surface}}}\left( E_{\text{surface}} - E_{\text{bulk}} \right)  
\end{equation}
where $A_{\text{surface}}$ is the surface area, $E_{\text{surface}}$ is the ground-state (GS) energy of the material systems with the surface of interest and $E_{\text{bulk}}$ is the GS energy of the corresponding bulk system and the values obtained from \pawfe~and \abinit~are shown in Table~\eqref{tab:Surface energy}. From this table, we observe a close match in surface energies computed from both codes with the difference in surface energies to be $\sim 2\cross10^{-3}~\nicefrac{J}{m^2}$. Additionally, it is observed that the Zr(001) surface has higher surface energy in both codes, consistent with observations from previous studies\cite{LLZOMorphology}.
\begin{table}
    \centering
    \begin{tabular}{|c|c|c|}
    \hline
    \multirow{3}{*}{System} & \abinit& \pawfe~\\
    \cline{2-3}
       & Surface Energy  & Surface Energy \\
      & ($\nicefrac{J}{m^2}$) & ($\nicefrac{J}{m^2}$) \\
    \hline
    La(011) & 0.925 & 0.924 \\
    \hline
    Zr(001) & 1.737 & 1.735 \\
    \hline
    \end{tabular}
    \caption{Comparison of LLZO surface energy computed using \pawfe~with semi-periodic boundary conditions and \abinit~with periodic boundary conditions. La(011) and Zr(001) surfaces are considered in this study with the Zr(001) surface being of higher energy. The difference in surface energy between the two codes is $\sim 2\cross10^{-3}\nicefrac{J}{m^2}$.}
    \label{tab:Surface energy}
\end{table}\\[0.05in]

\begin{figure}[htbp]
        \centering
        \includegraphics[width=1.0\linewidth]{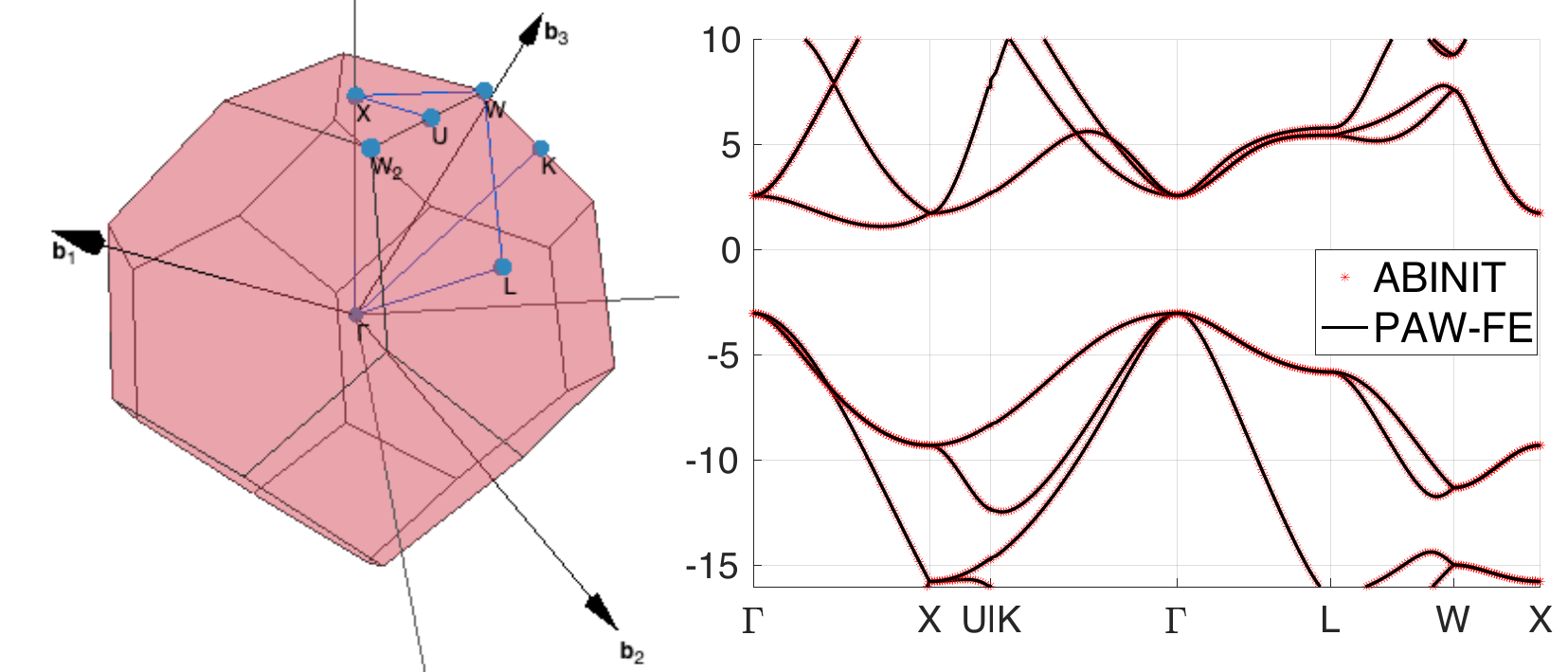}
        \caption{Band structure comparison between \abinit~and \pawfe. The left figure shows the k-path used to generate the band structure while the right compares the band structure between \abinit(red)~and \pawfe~(black).\textbf{Case Study:} Carbon diamond primitive cell }
        \label{fig:Band Structure}
    \end{figure}

\paragraph{Band structure comparisons:} We now compute the band-structure of the Carbon primitive cell(diamond crystal structure) with lattice constant $3.373$~\textrm{Bohr}. The Brioullin Zone(BZ) and the k-point path are shown in Fig.~\eqref{fig:Band Structure} and were obtained using seeK-path\cite{HINUMA2017140}. A shifted $8\cross8\cross8$ Monkhorst-Pack k-point grid was used for the ground state calculation for sampling the Brillouin zone, following which a non-self-consistent field (NSCF) calculation was performed using the k-points specified along the high symmetry points of Fig.~\eqref{fig:Band Structure} to extract the band structure. Figure~\eqref{fig:Band Structure} shows the overlay of the band structure computed using \abinit~and \pawfe.  We observe an excellent match in the eigenvalues and, hence, the band structure between the two codes.  

The comprehensive benchmarking studies discussed here establish the accuracy of our \pawfe~computational methodology and implementation for periodic/semi-periodic systems. We now discuss the performance aspects of \pawfe~for periodic systems with increasing system sizes.
% \begin{table}
%     \centering
%     \begin{tabular}{|c|c|c|}
%     \hline
%          System&Abinit  & PAW-FE \\
%          -&(Ha)  & (Ha) \\
%          \hline
%          O\textsubscript{2}&-32.105767  &-150.329951 \\
%          \hline
%          Li&-15.043692 & -15.043733\\
%          \hline
%          Li\textsubscript{2}O& -125.278682& -361.726961\\
%          \hline
%          Formation Energy& -0.223095 & -0.223032\\
%          \hline
%     \end{tabular}
%     \caption{Formation energy computation of Li\textsubscript{2}O}
%     \label{tab:Formation energy}
% \end{table}

\subsubsection{Performance benchmarking}
Here, we benchmark the performance of \pawfe~with \qe~ by comparing the computational efficiency and minimum wall time on increasingly large supercells of Cu\textsubscript{3}Pt crystal solved to similar levels of accuracy between the two codes \pawfe~ and \qe. The supercells are constructed by stacking of Cu\textsubscript{3}Pt unit cells along the three lattice dimensions and to this end, we consider supercells of sizes $4\times 4 \times 4$ (4288 electrons), $5 \times 5 \times 5$ (8375 electrons), $6 \times 6 \times 6$ (14472 electrons), $7 \times 7 \times 7$ (22981 electrons), $8 \times 8 \times 8$ (34304 electrons). We perform a $\Gamma-$point calculation with periodic boundary conditions imposed on all electronic fields in both \pawfe~and \qe. Further, the discretization parameters are chosen such that the discretization error in the ground-state energies obtained in both cases is $~2\cross10^{-4}$\textrm{Ha}/atom. In particular, in \pawfe, the degree of the FE interpolating polynomial $p$ is chosen to be 6, and a mesh size of $h=1.8$~\textrm{Bohr} is employed with Chebyshev polynomial degree of 20 in the ChFSI step. In the case of \qe~, the plane-wave cutoff energy $E^{\text{wfc}}_\text{cut} = 25$~\textrm{Ha} is used with the default settings for the eigensolver.

Table~\eqref{tab:Computational Cost Per} and \eqref{tab: Min Wall time Per} presents the computational cost\footnote{The computing system used for conducting the performance benchmarking studies are high memory nodes (~768 GB per node) which favours \qe~ as the minimum number of nodes required to fit peak memory requirement is lesser.} and the minimum wall time comparison between \pawfe~and \qe.  Recall that the computational cost ($\eta_{\text{c}}$) is computed as the product of the minimum number of nodes required to fit the problem and the average wall time per SCF iteration. From  Table~\eqref{tab:Computational Cost Per}, we observe that the number of basis functions required by \pawfe~is a factor of $6\cross$ higher than that of \qe~and we begin to see computational gains in \pawfe~for systems more than 8000 electrons. The efficiency gains we achieve with increasing system sizes are around 1.7$\times$, 2.3$\times$ and 3.2$\times$ for $6 \times 6 \times 6$, $7 \times 7 \times 7$, $8 \times 8 \times 8$ Cu\textsubscript{3}Pt supercells respectively, the benchmark systems considered for CPU-time comparison in this work. We note that the computational gains obtained in the periodic case are lesser than the case of non-periodic systems because of the degrees of freedom advantage that plane-wave (PW) codes provide for bulk systems ($\sim$ 500 PW basis functions per atom) compared to non-periodic systems ($\sim$ 5000 PW basis functions per atom) to reach the desired accuracy.  However, on estimating the computational complexity of \pawfe~from Table~\eqref{tab:Computational Cost Per} we find that \pawfe~scales as $\order{N_e^{2.2}}$ whereas \qe~scales as $\order{N_e^{2.8}}$ indicating that we will obtain an increase in computational gains for \pawfe~compared to \qe~as system size increases.

Finally, Table~\eqref{tab: Min Wall time Per} compares the average time per SCF in terms of the minimum wall time ($\tau^{\text{min}}_{\text{c}}$) between \pawfe~and \qe. To this end, we observe a gain of around $2.3\times$ for Cu\textsubscript{3}Pt $5 \times 5 \times 5$ supercell with 8375 electrons and the computational gains increase to almost $7.7\times$ for a system size comprising  34,304 electrons. Even for fully periodic calculations, these increased gains with the system size underscore the importance of fast and scalable computational methods developed in this work.
 
\begin{widetext}

\begin{table}[H]
    \centering
    \begin{tabular}{|c|c|c|c|c|c|c|c|}
      \hline
       Cu\textsubscript{3}Pt Supercell   & atoms &electrons  & QE  & QE($\eta_{\text{c}}$)  & PAW-FE  & PAW-FE($\eta_{\text{c}}$)  & Ratio \\
        - & -&- & $\#\text{basis}$ & Node-hrs & $\#\text{basis}$   & Node-hrs & - \\
        % \hline
        %  Cu\textsubscript{55}&  &  &  &  &  & \\
         \hline
         $4\cross4\cross4$& 256 &4288  & 128,923 & 0.02 & 753,571  & 0.03 & 0.5\\
         \hline
         $5\cross5\cross5$& 500 &8375  & 251,727 & 0.113 & 1,520,875  & 0.091 & 1.23\\         
         \hline
         $6\cross6\cross6$& 864&14472 & 434,257 & 0.51 & 2,685,619 & 0.3 & 1.7 \\
         \hline
         $7\cross7\cross7$& 1372&22981 &  690,981 &1.84 & 4,330,747 &0.79  & 2.3 \\
         \hline
         $8\cross8\cross8$&2048 &34304& 1,032,445 & 6.16 &  5,929,741& 1.92 &3.2  \\
         \hline
    \end{tabular}
    \caption{Computational cost ($\eta_{\text{c}}$) comparison between \pawfe~and \qe~in node-hrs ($\sim2\cross10^{-4}$~\textrm{Ha}/atom discretization error). $\eta_{\text{c}}$ is computed as the product of minimum nmber of nodes required to fit a given material system and the average wall time taken per SCF iteration. \textbf{Case Study}: Cu\textsubscript{3}Pt supercells of varying size}
    \label{tab:Computational Cost Per}
\end{table}
\end{widetext}

\begin{table}[H]
    \centering
    \begin{tabular}{|c|c|c|c|}
   \hline
      Cu\textsubscript{3}Pt Supercell    & QE ($\tau^{\text{min}}_{\text{c}}$) & PAW-FE ($\tau^{\text{min}}_{\text{c}}$) & Speedup\\   
      -   & sec & sec & -\\
    \hline  
      $5\cross5\cross5$   & 42.1 & 18 & 2.3\\
    \hline     
    $6\cross6\cross6$     & 96.7 & 26.8 & 3.61\\
    \hline     
    $7\cross7\cross7$    & 282.9 & 45.7 & 6.2 \\
    \hline     
     $8\cross8\cross8$    & 836.1 & 109 & 7.67\\
    \hline     
    \end{tabular}
    \caption{Comparison between \pawfe~and \qe~in terms of minimum wall time per SCF iteration($\tau_{c}^{min}$). \textbf{Case Study}: Cu\textsubscript{3}Pt supercells of varying sizes}
    \label{tab: Min Wall time Per}
\end{table} 
 
\subsection{\cb Scalability benchmarking with norm-conserving pseudopotential DFT-FE calculations\cn}
In this subsection, we compare the computational cost and parallel scaling efficiency of the proposed computational methodology \pawfe~relative to the norm-conserving pseudopotential (ONCV) calculations~\cite{oncv2013} using the \cb\DFTFE~code\cite{dftfe0.6,dftfe1.0}\cn solved to a similar level of accuracy. We remark here that the transferability of ONCV pseudopotential DFT calculations is on par with PAW approaches~\cite{lejaeghere} for most material systems in contrast to the norm-conserving Troullier-Martins (TM) pseudopotentials~\cite{tm91} and have become one of the popular methods of choices in most DFT implementations including \DFTFE. However, as alluded to before, these ONCV pseudopotentials can become very hard, requiring a large number of degrees of freedom to reach the desired accuracy for most material systems involving elements with $d$ or $f$ electrons. Towards this study, we consider $9\cross9\cross9$ Cu\textsubscript{3}Pt supercell and an icosahedral-shaped Pt\textsubscript{8-shell} nanocluster~\cite{Icosahedron} with the nearest neighbour distance being 5.27~\textrm{Bohr}, as representative systems for periodic and non-periodic calculations. We obtain the PAW data set and ONCV pseudopotential data from pseudodojo~\cite{pseudoDojoONCV,pseudoDojoPAW} repository for our comparative study. The finite-element mesh parameters in both cases are determined such that the discretization error is around $~2\cross10^{-4}$~\textrm{Ha}/atom. Table~\eqref{tab:Comparison Systems} shows the details of number of electrons, finite-element (FE) mesh parameters, number of FE basis functions between the ONCV based \DFTFE~and \pawfe~calculations and the Chebyshev polynomial degree employed during the ChFSI step. The simulations for this study are conducted on the OLCF supercomputer Frontier \footnote{Frontier is the world's first exa-scale computer. Each node of Frontier has  64-core AMD Epyc 7713 processor with 512GB.}

\begin{widetext}

\begin{figure}[H]
    \centering
    \begin{subfigure}{0.48\textwidth}
        \centering
        \includegraphics[width=1.0\linewidth]{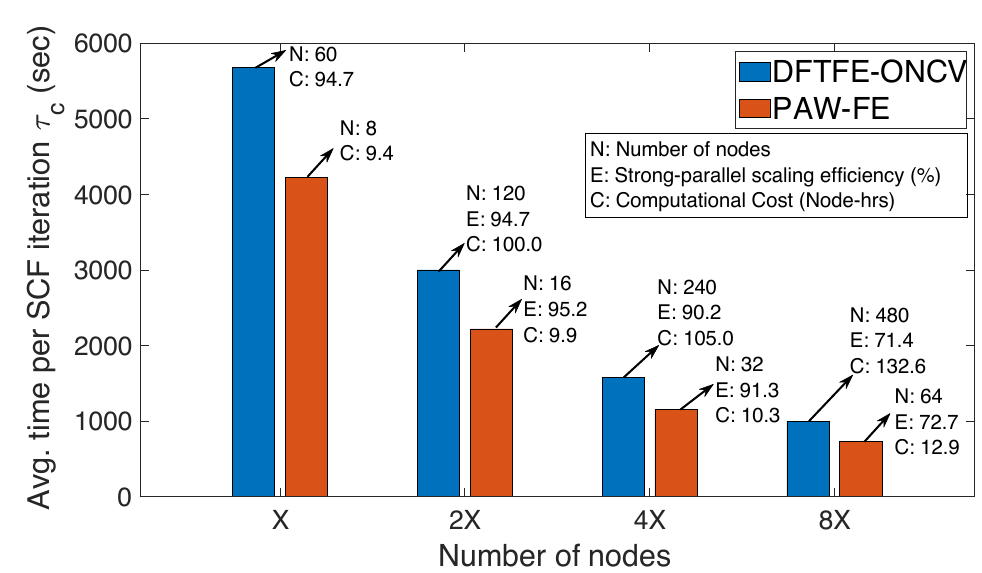}
        \caption{\centering \textbf{Case study}: Cu\textsubscript{3}Pt~$9\cross9\cross9$-supercell.(2,916 atoms) \\The DoFs/task at X-nodes is $\sim20$k}
    \end{subfigure}%
    \begin{subfigure}{0.48\textwidth}
        \centering
        \includegraphics[width=1.0\linewidth]{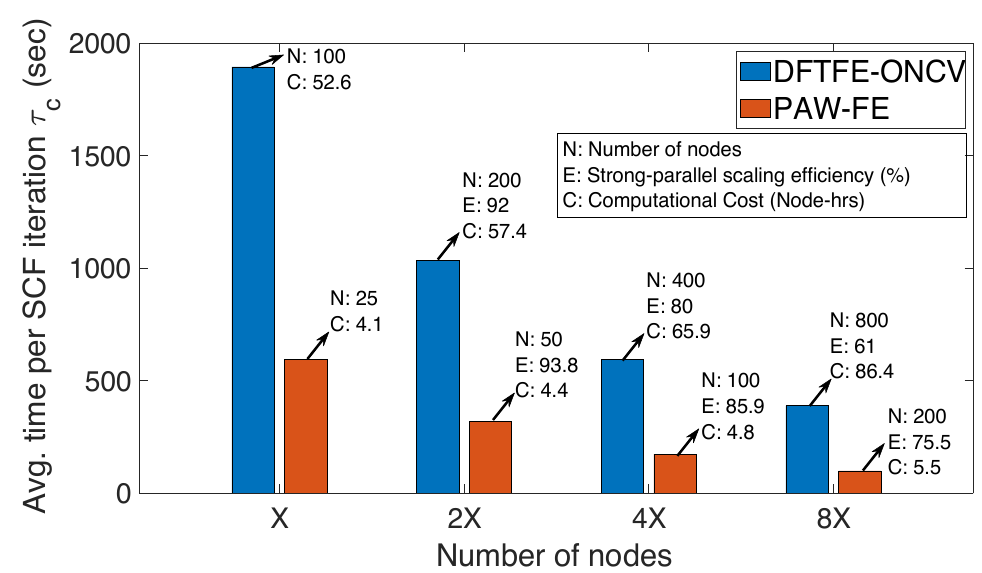}
        \caption{\centering \textbf{Case study}: Pt\textsubscript{8-shell} nanocluster.(2,057 atoms) \\The DoFs/task at X-nodes is $\sim15$k}
    \end{subfigure}
    \caption{Comparison of ONCV pseudopotential \DFTFE~against \pawfe~on Frontier CPUs. The figures compare the average wall time per SCF, scaling efficiency, computational cost between \DFTFE~and \pawfe~for $9\cross9\cross9$~Cu\textsubscript{3}Pt supercell and Pt\textsubscript{8-shell} nanocluster}
    \label{fig:ComparisonONCVPAW}
\end{figure}    
\end{widetext}

To assess the strong scaling behaviour between \DFTFE~using ONCV pseudopotentials and  \pawfe~calculations, we plot the average wall time per SCF iteration $\tau_\text{c}$ against the number of compute nodes as shown in Fig.~\eqref{fig:ComparisonONCVPAW}.
We observe that \pawfe~is almost an order of magnitude ($\sim$ 10$\times$ to 12$\times$) faster than  \DFTFE~calculations using the ONCV pseudopotentials in terms of computational node-hrs for the large-scale representative systems considered here. Furthermore, we observe similar or better parallel scaling efficiency while requiring 4$\times$ to 7$\times$ fewer compute nodes to fit the given problem in the case of \pawfe. The efficiency gains obtained in \pawfe~can be attributed to (i) coarser finite-element mesh size leading to much lesser degrees of freedom compared to the ONCV pseudopotential-based \DFTFE~calculations to reach the desired chemical accuracy, (ii) reduced Chebyshev polynomial degree during the ChFSI step due to the coarser mesh size in \pawfe~calculations, and  (iii)  a lesser valence electrons within the frozen-core approximation of the PAW method in the case of Pt leading to better parallel scaling efficiency of ChFSI due to reduced communication costs in the Rayleigh-Ritz(RR) step. We further note that the computational gains of \pawfe~compared to \DFTFE~using ONCV pseudopotential calculations may vary depending on the pseudopotential repository used, and the current comparisons use the data sets obtained from the pseudo-dojo\cite{Pseudodojo} repository, as mentioned before.
\vspace{0.01in}
\begin{table}[H]
    \centering
    \begin{tabular}{|c|c|c|c|}
    \hline
       System  & -  & ONCV & PAW\\
       \hline
       \multirow{2}{*} {Cu\textsubscript{3}Pt} & No of electrons($N_e$)  & 54,685 & 48,933\\
        \cline{2-4}
         & $p,p_{\text{el}},h$~(\textrm{Boh}) & $6,6,0.92$ & $6,7,1.85$ \\\cline{2-4}
         & Cheby. poly. degree $(m)$ & 50  & 20 \\\cline{2-4}         
       2916 atoms  & No. of basis/atom & 23,463 &2,954 \\
         \hline
     \multirow{2}{*}{Pt\textsubscript{8-shell}}    & No. of electrons($N_e$)  & 37,026 & 20,570\\\cline{2-4}
         & $p,p_{\text{el}},h$~(\textrm{Boh}) & $6,6,0.85$ &$5,7,1.2$ 
         \\\cline{2-4}
          & Cheby. poly. degree $(m)$ & 53 & 20 \\\cline{2-4}         
     2057 atoms    & No. of basis/atom & 41,031 &10,306 \\
         \hline
    \end{tabular}
    \caption{Comparison of ONCV based \DFTFE~ and \pawfe~on periodic(Cu\textsubscript{3}Pt) and non-periodic(Pt\textsubscript{\text{8-shell}}) systems}
    \label{tab:Comparison Systems}
\end{table}

\section{Summary}\label{sec:concl}
In this work, we present a computationally efficient real-space methodology for large-scale density functional theory (DFT) calculations using finite-element (FE) discretization of the projector-augmented wave (PAW) method. This is achieved for the first time through the synergy of the PAW method's ability to transform the underlying all-electron DFT problem into a form that involves smoother electronic fields and the potential of systematically convergent higher-order finite-element basis to achieve significant computational gains. The methodology exploits the compactly supported nature of FE basis functions and further takes advantage of the adaptive resolution offered by the FE basis to handle generic boundary conditions efficiently. Furthermore, the proposed work demonstrates a substantial computational advantage over the current state-of-the-art plane-wave-based PAW implementations for medium to large-scale material systems, thereby pushing the boundaries of length-scales that can be realised from \emph{ab initio} calculations than possible today. 

The proposed approach is based on the amalgamation of the following key ideas. First, we deduced the finite-element (FE) discretized governing equations corresponding to Bl\"{o}chl's PAW method by using a locally reformulated electrostatics functional involving a pseudo smooth electron density and compensation charge. Furthermore, higher-order quadrature rules were utilized to accurately evaluate the underlying integrals involving atom-centered functions in these governing equations, thereby allowing the use of coarser FE grids for the various electronic fields. Second, we employed a self-consistent field (SCF) iteration approach based on Anderson mixing of pseudo smooth electron density and atom-dependent spherical density matrix. This was accomplished by minimizing the weighted L\textsubscript{2}-norm of the total charge density residual, which was locally reformulated using the gradients of the potential corresponding to this residual. Next, we developed an efficient strategy to compute the inverse of the FE-discretized PAW overlap matrix, which is essential for solving the underlying generalized eigenvalue problem. By leveraging this strategy, we designed a Chebyshev polynomial-based filter within each SCF iteration to compute the desired eigenspace. The FE-cell-level structure of both the discretized Hamiltonian and the inverse PAW overlap matrix were exploited to devise computationally efficient strategies that employ cache-friendly dense matrix-matrix products to evaluate the action of the Chebyshev filter on the trial subspace during the course of the Chebyshev-filtered subspace iteration procedure.

The accuracy of the proposed method was validated against a plane-wave implementation of PAW across various non-periodic, semi-periodic, and fully periodic representative benchmark systems. Compared to plane-wave calculations, our proposed methodology(\pawfe) showed excellent agreement in ground-state energies, energy variation with bond length and lattice parameters, formation energies, surface energies, and band structures. Furthermore, compared to the state-of-the-art plane-wave implementation, the efficient computational strategies employed in \pawfe~enabled a $5\cross$ reduction in computational cost in non-periodic systems($\sim18,000$ electrons) and a $3\cross$ reduction in periodic systems ($\sim34,000$ electrons) with more significant gains possible as the system size increased. Additionally, in terms of the minimum wall time we observed substantial speedups for both periodic and non-periodic systems ($\sim7\times-10\times$) using \pawfe. In addition, the proposed computational strategies exhibit excellent parallel scaling efficiency ($\sim70\%$) even at 1,200 degrees of freedom per MPI task and achieve a substantial ten-fold reduction in computational cost over the current state-of-the-art \DFTFE~code using norm-conserving pseudopotentials for system sizes close to 50,000 electrons.

The proposed work, \pawfe, offers a systematically convergent basis capable of handling generic boundary conditions while substantially reducing the number of basis functions required to achieve the desired chemical accuracy compared with the current state-of-the-art finite element-based DFT methods using norm-conserving pseudopotentials. Furthermore, the computational methodologies developed are well-suited for modern GPU architectures, which can significantly accelerate \pawfe~calculations and reduce computational time—a focus of a future investigation. This work constitutes a significant step towards enabling \emph{ab initio} calculations of complex systems at larger length scales, thereby aiding in tackling a diverse set of key scientific and technological problems. Additionally, it paves the way for longer time-scale molecular dynamics simulations with \emph{ab initio} accuracy by providing a more comprehensive energy landscape for robust statistical and machine learning-based computational models.

\section{Acknowledgements}
The authors would like to thank Prof. Vikram Gavini for helpful discussions and valuable suggestions. The authors gratefully acknowledge the seed grant from the Indian Institute of Science and SERB Startup Research Grant from the Department of Science and Technology India (Grant Number: SRG/2020/002194). The research used the resources of PARAM Pravega at the Indian Institute of Science, supported by the National Supercomputing Mission (NSM) R\&D for exa-scale grant (DST/NSM/R\&D Exascale/2021/14.02). This research also used resources from the Oak Ridge Leadership Computing Facility at the Oak Ridge National Laboratory, which is supported by the Office of Science of the U.S. Department of Energy under Contract No. DE-AC05-00OR22725. K.R. acknowledges the Prime Minister Research Fellowship(PMRF) from the Ministry of Education India for financial support. S.D acknowledges the support from the Toyota Research Institute. P.M. acknowledges Google India Research Award 2023 and Indo-Korean Science and Technology center (IKST) Bengaluru for financial support during the course of this work.  

\begin{widetext}

\appendix
\cb
\section{Glossary}
This section summarizes various notations and dimensions of various matrices, vectors used in the current work while describing the real-space formulation and finite-element (FE) discretization of the PAW method in the main manuscript.\newline \newline
% \phantom{~}\noindent
\small{
\centering{
\begin{tabular}{c p{0.59\textwidth}r}
\hline
Symbol&Description& Eqn. no. \\
\hline
$N$&Number of DFT eigenstates to be solved\\
$N_v$&Number of valence electrons \\
$n^{a}_{\text{pj}}$ & Number of projectors for atom '$a$'. This also specifies the number of all-electron and pseudo-smooth partial wavefunctions ($\phiae^a(\bx),\phips^a(\bx)$). \\
$n_{\text{proj}}$ & Total number of projectors in the system $n_{\text{proj}} = \sum_a{n^{a}_{\text{pj}}}$  \\
$E\left[\{ \psiae_i \}, \{ \bR^a\} \right]$&All-electron DFT energy &\eqref{eqn:allElecKS}\\
$\rho(\bx)$ & Total charge density comprising all-electron density ($n(\bx)$) and nuclear charge density ($b(\bx)$) & \\
$J\left[\rho(\bx),\{\bR^a \}\right]$&All-electron electrostatic energy that includes electron-electron, electron-nuclear and nuclear-nuclear repulsion&\eqref{eqn:allElecKS}\\
$T_s\left[\{ \psiae_i \} \right]$&Non interacting kinetic energy in the all-electron DFT problem&\eqref{eqn:Ts}\\
$E_{\text{xc}}[n(\bx)]$&Exchange-correlation energy in the all-electron DFT problem&\eqref{eqn:exc}\\
$E_\text{el}\left[\rho\right ]$&Electrostatic energy associated with $\rho(\bx)$ in the all-electron DFT problem&\eqref{eqn:Eel}\\
$\phiae^a_\alpha(\bx)$&All-electron partial wavefunction corresponding to atom '$a$'&\eqref{eqn:wfcdecomp}\\
$\phips^a_\alpha(\bx)$&Pseudo smooth(PS) partial wavefunction of atom '$a$'&\eqref{eqn:wfcdecomp}\\
$\psiae_i(\bx)$&All-electron DFT wavefunction&\eqref{eqn:wfcdecomp}\\
$\psips_i(\bx)$&Pseudo smooth(PS) PAW wavefunction&\eqref{eqn:wfcdecomp}\\
$\pTilde^a_\alpha(\bx)$&Atom-centered projector of atom '$a$'&\eqref{eqn:wfcdecomp}\\
$\nTilde(\bx)$&Pseudo smooth electron density&\eqref{eq: PS electron density}\\
$n^a(\bx)$&Atom-centered all-electron density corresponding to atom '$a$' &\eqref{eq: atom all electron density}\\
$\nTilde^a(\bx)$&Atom-centered pseudo smooth electron density corresponding to atom '$a$'&\eqref{eq: atom PS density}\\
$\bD^a$, $D^a_{\alpha\beta}$&Spherical density matrix of dimension $n^{a}_{\text{pj}}\times n^{a}_{\text{pj}}$&\eqref{eq: Dij entries}\\
$\boldsymbol{\Delta T}^a$, $\Delta T^a_{\alpha\beta}$&Atom-dependent kinetic energy correction matrix of dimension $n^{a}_{\text{pj}}\times n^{a}_{\text{pj}}$&\eqref{eqn:atomCenteredCorrectionKEandXC},\eqref{Tijalphabeta}\\
$\Delta E_{\text{xc}}^a$&Atom-dependent exchange-correlation energy correction term&\eqref{eqn:atomCenteredCorrectionKEandXC},\eqref{eqn: Delta Exc}\\
\hline
\end{tabular}}
\newpage
\centering{
\begin{tabular}{c p{0.59\textwidth}r}
\hline
Symbol&Description& Eqn. no. \\
\hline
$\bTilde^a(\bx)$&Atom-centered compensation charge&\eqref{eq: compensation charge} \\
$\bTilde(\bx)$&Total compensation charge, computed as $\sum_a{\bTilde^a(\bx)}$&\eqref{eq: compensation charge} \\
$\boldsymbol{\Delta_{lm}}^a$, $\Delta ^a_{lm\alpha\beta}$&Multipole expansion coefficients to compute $\bTilde^a(\bx)$. For each $\{l,m\}$ index the matrix is of dimension $n^{a}_{\text{pj}} \times n^{a}_{\text{pj}}$&\eqref{eqn: Kappa_a and Deltalm}\\
$\rhoTilde(\bx)$ & Total charge density corresponding to $\nTilde(\bx)+\bTilde(\bx)$ & \eqref{eqn:electrosmooth} \\

% $E^a_{\text{el}}\left[\rho^a(\bx)\right]$&ijkl&\eqref{eq: AtomCentered Electrostatics energy}\\
% $E^a_{\text{el}}\left[\rhoTilde^a(\bx)\right]$&ijkl&\eqref{eq: AtomCentered Electrostatics energy}\\
$E_\text{el}\left[\rhoTilde(\bx)\right ]$&Electrostatics energy associated with total charge density $\rhoTilde(\bx)$&\eqref{eqn:electrosmooth},\eqref{eqn:localelectro}\\
$\Delta E^a_\text{el}$&Atom-dependent electrostatics energy correction term&\eqref{eq: AtomCentered Electrostatics energy},\eqref{eqn:atomcenterDeltaEel}\\
$\widetilde{E}[\{ \psips_i\}, \{ \bR^a\}]$&PAW energy functional&\eqref{eq: PAW energy}\\
$\Delta E^a[\left\{D^a_{\alpha\beta}\right\}]$&Atom-dependent total energy correction&\eqref{eq: PAW energy}\\
$\mathcal{H}$&PAW Hamiltonian operator &\eqref{eqn:pawghep},\eqref{eqn:pawHamiltonian}\\
$\mathcal{S}$&PAW overlap operator&\eqref{eqn:pawghep},\eqref{eqn:pawOverlap}\\
 $\Delta h^a_{\alpha\beta}$&Atom-dependent PAW Hamiltonian coupling matrix of dimension $n^{a}_{\text{pj}}\times n^{a}_{\text{pj}}$ arising in the non-local part of $\mathcal{H}$&\eqref{eqn:pawHamiltonian},\eqref{eq: Hamiltonian coupling}\\

 $\Delta s^a_{\alpha\beta}$&Atom-dependent PAW overlap coupling matrix of dimension $n^{a}_{\text{pj}}\times n^{a}_{\text{pj}}$ arising in the non-local part of $\mathcal{S}$&\eqref{eqn:pawHamiltonian},\eqref{eq: overlap coupling}\\

$\hat{U}^a_{\alpha i}$&Atom-dependent vector of dimension $n^{a}_{\text{pj}}\times N$ with each entry corresponding to inner product of $\pTilde_{\alpha}^a(\bx)$ and $\psips(\bx)$&\eqref{eq: Governing PAW PDE}\\
$\fePsips^h_{i,\bk}(\bx)$&FE discretized Bloch wavefunction&\eqref{eqn:FEbasisexp}\\
$M$, $M_{\text{el}}$ & Number of degrees of freedom (grid points) in the FE mesh corresponding to electronic wavefunctions and electrostatic potential respectively\\
$p$, $p_{\text{el}}$&Polynomial degree of the FE basis function corresponding to the discretized electronic wavefunction and electrostatic potential respectively &\\
$\varphi^{h,p_{\text{el}}}(\bx)$&FE discretized electrostatic potential&\eqref{eqn:FEbasisexp}\\
$N^{h,p}_I(\bx)$, $N_I^{h,p_{\text{el}}}(\bx)$&FE basis function corresponding to FE node $I$ (Lagrange polynomials of degree $p$ and $p_{\text{el}}$ respectively) &\\
$\nabla N^{h,p}_I(\bx)$, $\nabla N_I^{h,p_{\text{el}}}$&Gradient of FE basis  function corresponding to node $I$&\eqref{eqn:hij},\eqref{eqn:mij}\\
$\bH^{\bk}$&FE discretized PAW Hamiltonian matrix of dimension $M\times M$&\eqref{eqn: simplfied FE PAW}\\
$\boldsymbol{\Delta}_{\btH}$&Block diagonal coupling matrix arising in the non-local part of $\bH^{\bk}$. Each block corresponds to atom '$a$' of dimension $n^{a}_{\text{pj}}\times n^{a}_{\text{pj}}$&\eqref{eqn:deltaHanddeltaS}\\
$\bM$&FE basis overlap matrix of dimension $M\times M$&\eqref{eqn:mij}\\
$\bS^{\bk}$&FE discretized PAW overlap matrix of dimension $M\times M$&\eqref{eqn: simplfied FE PAW}\\
$\boldsymbol{\Delta}_{\btS}$&Block diagonal coupling matrix arising in the non-local part of $\bS^{\bk}$. Each block corresponds to atom '$a$' of dimension $n^{a}_{\text{pj}}\times n^{a}_{\text{pj}}$&\eqref{eqn:deltaHanddeltaS}\\
$\bC^{a,\bk}$&Atom-dependent matrix of dimension $M\times n^{a}_{\text{pj}}$ whose entries are inner products of atom projectors and FE-basis functions&\eqref{eq: FE nonlocal contribution}\\
$\feNodalVectorUtilde_{i,\bk}$, $\fePsips^J_{i,\bk}$&Nodal vector of dimension $M\times1$ corresponding to the FE discretized $i$-th electronic wavefunction&\eqref{eq: FE PAW governing equations},\eqref{eqn:FEbasisexp}\\
% $\widehat{\boldsymbol{\mathsf{U}}}^{a,\bk}_{i}$,$\widehat{{\mathsf{U}}}^{a,\bk}_{\beta i}$&ijkl&\eqref{eq: FE PAW governing equations}\\
$\bPhi$, $\varphi^J$&Nodal vector of dimension $M_{\text{el}}\times1$ corresponding to the FE discretized electrostatic potential &\eqref{eqn:FEbasisexp}\\
$\bL$&FE discretized Laplace operator of dimension $M_{\text{el}}\times M_{\text{el}}$&\eqref{eqn: simplfied FE PAW}\\
$\bc$&Right-hand-side vector of dimension $M_{\text{el}}\times1$ in the FE discretized Poisson equation&\eqref{eqn: simplfied FE PAW}\\
${{\bS}}^{-1}$&FE discretized PAW inverse overlap matrix of dimension $M\times M$&\eqref{eqn:Sinv}\\
$\boldsymbol{\Delta}_{\btS_{\text{\tiny{inv}}}}$&Block diagonal coupling matrix arising in the inverse PAW  overlap matrix. Each block corresponds to atom '$a$' of dimension $n^{a}_{\text{pj}}\times n^{a}_{\text{pj}}$&\eqref{eqn:Sinv}\\
$\widehat{\bS}^{-1}$& Approximate ${{\bS}}^{-1}$ of dimension $M\times M$&\eqref{eqn:Sinvapprox}\\
$\bM_{\text{D}}$&Diagonal approximation of $\bM$ &\eqref{eqn:Sinvapprox}\\
$\boldsymbol{\Delta}^a _{\bS_{\text{inv}}}$&Atom-dependent coupling matrix of dimension $n^{a}_{\text{pj}}\times n^{a}_{\text{pj}}$ used to the evaluate $\widehat{\bS}^{-1}$ \eqref{eqn:Sinvapprox}\\
$n_q$ & The total number of quadrature points in the reference cell $\widehat{\Omega}$.&\eqref{eq: Quadrature Integration}\\
$\bJ^{(e,t)}$&Jacobian matrix of dimension $3\times3$ corresponding to the mapping from FE-cell to reference cell&\eqref{eq: Quadrature Integration}\\
$\det \bJ^{(e,t)}\biggr\rvert_{\widehat{\bx}_q}$&Determinant of the Jacobian matrix $\bJ^{(e,t)}$ evaluated at quadrature point $\widehat{\bx}_q$ &\eqref{eq: Quadrature Integration}\\
$E_t$ & Total number of FE-cells in MPI task '$t$'.&\eqref{eqn: HX kernel},\eqref{eqn: SX kernel}, \eqref{eqn: SinvX kernel}\\
$n_t$ & Total number of MPI tasks.&\eqref{eqn: HX kernel},\eqref{eqn: SX kernel}, \eqref{eqn: SinvX kernel} \\
$m_t$ & Number of degrees of freedom in MPI task '$t$'& \\
$\bB^{(t)}$&Partitioner matrix of dimension $m_t\times M$&\eqref{eqn: HX kernel},\eqref{eqn: SX kernel},\eqref{eqn: SinvX kernel}\\
$\bQ^{(t)}$&Constraint matrix of dimension $m_t\times m_t$&\eqref{eqn: HX kernel},\eqref{eqn: SX kernel},\eqref{eqn: SinvX kernel}\\
$\bZ^{(e,t)}$&MPI task level sub-domain ($\Omega^{t}$) to FE cell ($\Omega^{(e,t)}$) map \\&of dimension $(p+1)^3\times m_t$&\eqref{eqn: HX kernel},\eqref{eqn: SX kernel},\eqref{eqn: SinvX kernel}\\
\hline
\end{tabular}}}
\cn
\section{Energy evaluation}\label{sec: appendix delta Ea}
The reformulated all-electron energy functional Eqn.\eqref{eqn:FEAllElecEnergy} in the PAW method is expressed in terms of $\widetilde{E}[\{ \psips_i\}, \{ \bR^a\}]$ and $\Delta E^a[\{D^a_{\alpha\beta}\}]$.  To this end, the atom-centered correction term is expressed as
\begin{equation}
\Delta E^a[\{D^a_{\alpha\beta}\}] = \Delta E^a_{\text{xc}} + \Delta E^a_{\text{el}} + \sum_{\alpha\beta}\Delta T^a_{\alpha\beta}D^a_{\alpha\beta} + T^a_{\text{core}}     
\end{equation}
  In this section, we discuss the evaluation of various terms involved in computing $\Delta E^a[\{D^a_{\alpha\beta}\}]$ for completeness which is along the lines of prior works\cite{rostgaard2009projectoraugmentedwavemethod,GPAW2010}. 
\paragraph{Kinetic Energy correction:} 
\subparagraph{}
\begin{align}\label{Tijalphabeta}
  \Delta T^a_{\alpha\beta} = & \left(\frac{1}{2}\sum_{l_{\alpha} l_{\beta}}{\sum_{m_\alpha m_\beta}}{\left(\int{\frac{d\phi_{\alpha}(r)}{dr}\frac{d\phi_{\beta}(r)}{dr}r^2dr}\delta_{l_{\alpha} l_{\beta}}\delta_{m_{\alpha} m_{\beta}} + \int_{\Omega_a}{{\phi_{\alpha}(r)\phi_{\beta}(r)}dr}(l_\alpha+1)(l_{\alpha})\delta_{l_\alpha l_\beta}\delta_{m_\alpha m_\beta} \right)} \right. \nonumber \\
  &-\left. \frac{1}{2}\sum_{l_\alpha l_\beta}{\sum_{m_\alpha m_\beta}}{\left(\int_{\Omega_a}{\frac{d\phips_{\alpha}(r)}{dr}\frac{d\phips_{\beta}(r)}{dr}r^2dr}\delta_{l_\alpha l_\beta}\delta_{m_\alpha m_\beta} + \int_{\Omega_a}{{\phips_{\alpha}(r)\phips_\beta(r)}dr}(l_\alpha+1)(l_\alpha)\delta_{l_\alpha l_\beta}\delta_{m_\alpha m_\beta} \right)}\right)  
\end{align}
\paragraph{Electrostatics energy correction}
\subparagraph{}
\begin{equation}\label{eqn:atomcenterDeltaEel}
    \Delta E^a_\text{el} = 
E^a_{\text{el}}\left[\rho^a(\bx)\right] - E^a_{\text{el}}\left[\rhoTilde^a(\bx)\right]  - E^a_{\text{self}}\left[\bR^a\right] = \Delta \mathcal{C}^a + \sum_{\alpha\beta}\Delta \mathcal{C}^a_{\alpha\beta}D^a_{\alpha\beta} + \sum_{\alpha\beta\alpha'\beta'}{D^{a*}_{\alpha\beta}\Delta \mathcal{C}^{a}_{\alpha\beta\alpha'\beta'}D^a_{\alpha'\beta'}}
\end{equation} 
\begin{align}\label{el1}
\text{where,}\;\;\Delta \mathcal{C}^a = & \frac{1}{2}\left( \int_{\Omega_a}{\int_{\Omega_a}{\frac{n_c^a(\bx)n_c^a(\by)}{|\bx-\by|}d\bx}d\by} - \int_{\Omega_a}{\int_{\Omega_a}{\frac{\tilde{n}_c^a(\bx)\tilde{n}_c^a(\by)}{|\bx-\by|}d\bx}d\by} - (\kappa^a)^2\int_{\Omega_a}{\int_{\Omega_a}{\frac{\gTilde_{00}^a(\bx)\gTilde_{00}^a(\by)}{|\bx-\by|}d\bx}d\by} \right)  \nonumber \\
  & - \kappa^a \int_{\Omega_a}{\int_{\Omega_a}{\frac{\tilde{n}_c^a(\bx)\gTilde_{00}^a(\by)}{|\bx-\by|}d\bx}d\by} -\sqrt{4\pi}\mathcal{Z}^a\int_{\Omega_a}{\frac{n^a_c(r)}{r}dr}
\end{align}
\begin{align}\label{el2}
 \;\;\;\;\;\Delta \mathcal{C}^a_{\alpha\beta} =& \int_{\Omega_a}{\int_{\Omega_a}{\frac{\phi^a_i(\bx)\phi^a_j(\bx)n^a_c(\by)}{|\bx-\by|}d\bx}d\by} - \int_{\Omega_a}{\int_{\Omega_a}{\frac{\phips^a_i(\bx)\phips^a_j(\bx)\tilde{n}^a_c(\by)}{|\bx-\by|}d\bx}d\by}  -\mathcal{Z}^a\int_{\Omega_a}{\frac{\phi^a_i(\bx)\phi^a_j(\bx)}{|\bx|}d\bx} \nonumber \\
 & -\kappa^a \int_{\Omega_a}{\int_{\Omega_a}{\frac{\phips^a_i(\bx)\phips^a_j(\bx)g_{00}^a(\by)}{|\bx-\by|}d\bx}d\by} - \Delta^a_{00\alpha\beta}\left(\kappa^a \int_{\Omega_a}{\int_{\Omega_a}{\frac{\gTilde_{00}^a(\bx)\gTilde_{00}^a(\by)}{|\bx-\by|}d\bx}d\by} + \int_{\Omega_a}{\int_{\Omega_a}{\frac{\tilde{n}^a_c(\bx)\gTilde_{00}^a(\by)}{|\bx-\by|}d\bx}d\by} \right)
\end{align}
\begin{align}\label{el3}
  \text{and}\;\; \Delta \mathcal{C}^a_{\alpha\beta\alpha'\beta'} =& \frac{1}{2}\left[ \int_{\Omega_a}{\int_{\Omega_a}{\frac{\phi^a_\alpha(\bx)\phi^a_\beta(\bx)\phi^a_{\alpha'}(\by)\phi^a_{\beta'}(\by)}{|\bx-\by|}d\bx}d\by} - \int_{\Omega_a}{\int_{\Omega_a}{\frac{\phips^a_{\alpha}(\bx)\phips^a_{\beta}(\bx)\phips^a_{\alpha'}(\by)\phips^a_{\beta'}(\by)}{|\bx-\by|}d\bx}d\by}   \right]  \nonumber \\
 & -\sum_L{\left[ \frac{1}{2}\Delta^a_{lm\alpha'\beta'}\int_{\Omega_a}{\int_{\Omega_a}{\frac{\phips^a_\alpha(\bx)\phips^a_\beta(\bx)\gTilde_{lm}^a(\by)}{|\bx-\by|}d\bx}d\by} + \frac{1}{2}\Delta^a_{lm\alpha\beta}\int_{\Omega_a}{\int_{\Omega_a}{\frac{\phips^a_{\alpha'}(\bx)\phips^a_{\beta'}(\bx)\gTilde_{lm}^a(\by)}{|\bx-\by|}d\bx}d\by}  \right]}  \nonumber \\
 &-\sum_{lm}{ \frac{1}{2}\Delta^a_{lm\alpha\beta}\int_{\Omega_a}{\int_{\Omega_a}{\frac{\gTilde_{lm}^a(\bx)\gTilde_{lm}^a(\by)}{|\bx-\by|}d\bx}d\by}\Delta^a_{lm\alpha'\beta'}} 
\end{align}
where, $r = |\bx-\bR^a|$ and $\kappa^a,\Delta^a_{lm\alpha\beta}$ are defined in Eqn.\eqref{eqn: Kappa_a and Deltalm}. We note that $\frac{1}{|\bx-\by|}$ in the above equations can be expanded in terms of real spherical harmonics as
\begin{equation}
\frac{1}{|\bx-\by|} =
\sum_\ell \sum_{m=-\ell}^\ell
\frac{4\pi}{2\ell+1}
\frac{\min(r,r')^\ell}{\max(r,r')^{\ell+1}}
S_{\ell,m}(\widehat{\boldsymbol{\theta}})S_{\ell,m}(\widehat{\boldsymbol{\theta'}})  
\label{eq: Expansion potential}
\end{equation}
where $r = |\bx-\bR^a|$ and $\widehat{\boldsymbol{\theta}} \in \mathbb{R}^2$ represents the azhimuthal and polar coordinates of $\bx$ in spherical coordinates, with $r' = |\by-\bR^a|$ and $\widehat{\boldsymbol{\theta'}}$ represents the azhimuthal and polar coordinates of $\by$ in spherical coordinates. Using the above Eqn.\eqref{eq: Expansion potential}, the integrals involved in Eqns.~\eqref{el1},~\eqref{el2},~\eqref{el3} can be
evaluated in spherical coordinate system as discussed in detail in the Appendix of \cite{abinitPAW2008}.
\vspace{0.2in}
\paragraph{Exchange Correlation Energy correction:}
\subparagraph{}
The evaluation of the atom-centered exchange-correlation term is computed on a radial grid using the Lebedev quadrature rule\cite{LEBEDEV197610} for the angular integration and Simpson's rule for the radial integration.
\begin{equation}
     \Delta E^a_{\text{xc}} = \int_{\Omega_a}{\left(\epsilon_{\text{xc}}[n^a,\nabla n^a]-\epsilon_{\text{xc}}[\nTilde^a,\nabla \nTilde^a]\right)d\bx}
\end{equation}
where
\begin{equation}
    n^a(\bx) = n^a_c(r)S_{00}(\widehat{\boldsymbol{\theta}}) + \sum_{\alpha\beta}{\phi^a_\alpha(r)\phi^a_\beta(r)S_{l_\alpha m_\alpha}(\widehat{\boldsymbol{\theta}})S_{l_\beta m_\beta}(\widehat{\boldsymbol{\theta}})D^a_{\alpha\beta}}
\end{equation}
\begin{equation}
    \nabla n^a(\bx) = \begin{bmatrix}
        \frac{\partial n^a_c(r)}{\partial r}S_{00}(\widehat{\boldsymbol{\theta}}) + \sum_{\alpha\beta}{\left(\frac{\partial \phi^a_\alpha(r)}{\partial r}\phi^a_\beta(r) + \frac{\partial \phi^a_{\beta}(r)}{\partial r}\phi^a_\alpha(r) \right)S_{l_\alpha m_\alpha}(\widehat{\boldsymbol{\theta}})S_{l_\beta m_\beta}(\widehat{\boldsymbol{\theta}})D^a_{\alpha\beta}}\\
        \sum_{\alpha\beta}{\frac{1}{r}\left(\phi^a_{\alpha}(r)\frac{\partial S_{l_\alpha m_\alpha}(\widehat{\boldsymbol{\theta}})}{\partial \vartheta}\phi^a_{\beta}(r)S_{l_\beta m_\beta}(\widehat{\boldsymbol{\theta}}) + \phi^a_{\beta}(r)\frac{\partial S_{l_\beta m_\beta}(\widehat{\boldsymbol{\theta}})}{\partial \vartheta}\phi^a_{\alpha}(r)S_{l_\alpha m_\beta}(\widehat{\boldsymbol{\theta}})\right)D^a_{\alpha\beta}} \\
 \sum_{\alpha\beta}{\frac{1}{r \sin(\vartheta)}\left(\phi^a_{\alpha}(r)\frac{\partial S_{l_\alpha m_\alpha}(\widehat{\boldsymbol{\theta}})}{\partial \varsigma}\phi^a_{\beta}(r)S_{l_\beta m_\beta}(\widehat{\boldsymbol{\theta}}) + \phi^a_{\beta}(r)\frac{\partial S_{l_\beta m_\beta}(\widehat{\boldsymbol{\theta}})}{\partial \varsigma}\phi^a_{\alpha}(r)S_{l_\alpha m_\beta}(\widehat{\boldsymbol{\theta}})\right)D^a_{\alpha\beta}}       
    \end{bmatrix}
    \label{eq: atom centered gradiant density}
\end{equation}
where $\widehat{\boldsymbol{\theta}} = [\vartheta,\varsigma] \in \mathbb{R}^2$ represents the polar and azhimuthal coordinates in spherical coordinate system. Further Eqn.\eqref{eq: atom centered gradiant density} can be simplified  using the below expressions
\begin{equation}
    \frac{\partial S_{lm}(\widehat{\boldsymbol{\theta}})}{\partial \vartheta} = - \frac{\sqrt{2l+1}}{\sqrt{2l+3}}\sqrt{(l+1)^2 - m^2}\frac{S_{l+1,m}(\widehat{\boldsymbol{\theta}})}{\sin(\vartheta)}
\end{equation}
\begin{equation}
 \frac{\partial S_{lm}(\widehat{\boldsymbol{\theta}})}{\partial \varsigma} = -|m|S_{l\bar{m}}(\widehat{\boldsymbol{\theta}}), \; \text{where,}\;\;\bar{m} = -m   
\end{equation}
The expression for $\nTilde^a(\bx)$ can be derived following the above steps similarly. Finally, the exchange-correlation energy correction term is evaluated as
\begin{equation}
    \Delta E^a_{\text{xc}} = \sum_q{w_q\int{\left(\epsilon_{\text{xc}}[n^a(r,\widehat{\boldsymbol{\theta}}_q),\nabla n^a(r,\widehat{\boldsymbol{\theta}}_q)]-\epsilon_{\text{xc}}[\nTilde^a(r,\widehat{\boldsymbol{\theta}}_q),\nabla \nTilde^a(r,\widehat{\boldsymbol{\theta}}_q)]\right) r^2dr}}
    \label{eqn: Delta Exc}
\end{equation}
Further, the atom-centered exchange correlation potential correction ($\Delta V^{a, \alpha\beta}_{\text{xc}}$) is expressed as
\begin{equation}
    \Delta V^{a,\alpha\beta}_{\text{xc}} = \sum_q{w_q\left(\int{ \frac{\delta E^a_{\text{xc}}[n^a,\nabla n^a]}{\delta D^a_{\alpha\beta}}\phi^a_{\alpha}(r)\phi^a_\beta(r) r^2dr} - \int{\frac{\delta E^a_{\text{xc}}[\nTilde^a,\nabla \nTilde^a]}{\delta D^a_{\alpha\beta}}\phips^a_{\alpha}(r)\phips^a_\beta(r) r^2dr}\right)S_{l_\alpha m_\alpha}(\widehat{\boldsymbol{\theta}}_q)S_{l_\beta m_\beta}(\widehat{\boldsymbol{\theta}}_q)}
    \label{eqn: Delta Vxc}
\end{equation}
where, $q$ refers to the Lebedev quadrature points used to sample $\widehat{\boldsymbol{\theta}}$ and $w_q$ are the corresponding quadrature weights while the above radial integrals are evaluated until augmentation radius ($r^a_c$) using the adaptive Simpsons integration rule. We note that the Eqns. \eqref{eqn: Delta Exc}, \eqref{eqn: Delta Vxc}  are efficiently evaluated by recasting the integrands into a dense matrix-matrix multiplication in our \pawfe~implementation.
\vspace{0.2in}
\paragraph{PAW Valence Energy for non-periodic systems:}
Recall from Eqn.\eqref{eqn:FEAllElecEnergy}, the FE discretized all-electron PAW energy for a non-periodic system corresponding to a simulation domain $\Omega$ can be computed using the following expression
\begin{align}
        E^h =& 2\sum_{i=1}^N{{f(\varepsilon^h_{i},\mu)\varepsilon^h_{i}}} - \sum_a{\sum_{\alpha\beta}\Delta h^a_{\alpha\beta}D^{a,h}_{\alpha\beta}} 
        - \int_{\Omega}{\left[ \nTilde^h(\bx)V^h_{\text{eff}}(\bx) + \bV^h_{\text{eff,GGA}}(\bx) \cdot \nabla \nTilde^h(\bx) \right] d\bx} \nonumber \\
        &+ \frac{1}{2}\int_{\Omega}{\rhoTilde^h(\bx)\varphi^{h,p_{\text{el}}}(\bx)d\bx} + E_{\text{xc}}\left[\nTilde^h(\bx)\right]+ \sum_a{\Delta E^a\left[\{\bD^{a,h}\}\right]}\;\; \text{where}\;\; \rhoTilde^h(\bx) = \nTilde^h(\bx) + \bTilde^h(\bx)
    \end{align}
    We note that the above expression for $E^h$ includes the energy contribution of the frozen core, and in order to exclude the frozen core contribution in $E^h$ and compare the ground-state energies one-on-one with \abinit~\cite{Abinit2016} in our benchmarking studies, we introduce the PAW valence energy $E^h_v$ for non-periodic systems as follows
    \begin{align}
      &E^h_{\text{v}} = E^h - \sum_a{E_{\text{xc}}\left[n^a_c(\bx) \right]} -\sum_a{C^a_{\text{corr}}},\;\;\text{where,}\; E_{\text{xc}}\left[n^a_c(\bx) \right] = \int_{\Omega_a}{\epsilon_{\text{xc}}[n^a_c,\nabla n^a_c]d\bx} \nonumber \\
        &\text{and}\; \mathcal{C}^a_{\text{corr}} = \int_{\Omega}{\int_{\Omega}{\frac{\rhoTilde^a_c(\bx)\rhoTilde^a_c(\by)}{|\bx-\by|}d\bx}d\by} - \left(E^a_{\text{el}}[n^a_c(\bx)+b^a(\bx)] - E^a_{\text{el}}[\rhoTilde^a_c(\bx)]\right),\;  \rhoTilde^a_c(\bx) = \nTilde^a_c(\bx) + \kappa^a\gTilde_0^a(\bx)
        \label{eq: PAW valence energy}
    \end{align}
\section{Variational formulation of the PAW energy functional and finite-element discretization}\label{sec: Appendix GEP}
In this section, we present the variational formulation of the PAW energy functional and subsequently derive the finite-element discretized equations to be solved in the PAW method. For brevity, the expressions derived below assume the case of a non-periodic isolated system or a periodic system involving a large supercell with $\Gamma-$point sampling. To this end, we make the assumption that the wavefunctions are real; however, the derivation is trivially extendable for unit-cell calculations with multiple $\bk-$point sampling of the Brillouin zone involving complex wavefunctions. We begin with the all-electron Kohn-Sham energy functional in Eqn.\eqref{eqn:FEAllElecEnergy} to be minimized subject to $\mathcal{S}$-orthonormality constraints. Consequently, we introduce the Lagrange multiplier matrix ($\mathbf{\Lambda}$) for enforcing these constraints and the functional to be minimized to determine the ground-state energy for given positions of nuclei $\{\bR^{a}\}$ in the PAW method can be written as follows:
\begin{equation}
F[\{\psips_i \},\varphi,\mathbf{\Lambda}] = E[\{\psips_i \},\varphi] - \sum_{ij}{\lambda_{ij}(\int{\psips_i(\bx)\mathcal{S}\psips_j(\bx)d\bx}-\delta_{ij})} \label{eqn:KSfunctionalLag}
\end{equation}
Now, taking the variational derivative of the above functional with respect to $\psips_n$, we have
\begin{align}
   &4\,f_n\int{\frac{1}{2}\nabla \psips_n(\bx)\cdot \nabla\delta\psips_n(\bx)d\bx} + 4\,f_n\int{(\varphi(\bx)+\bar{V})\psips_n\delta\psips_n d\bx} + \sum_a{\sum_{\alpha\beta}\Delta^a_{lm\alpha\beta}\int{\frac{\delta D^a_{\alpha\beta}}{\delta \psips_n(\by)}\delta \psips_n(\by)d\by}{\int\gTilde^a_{lm}(\bx){\varphi(\bx)d\bx}}} \nonumber \\ 
  & + \int{\frac{\delta E_{\text{xc}}}{\delta \nTilde(\bx)}\left(\int{  \frac{\delta \nTilde(\bx)}{\delta \psips_n(\by)}\delta \psips_n(\by)d\by}\right)d\bx} + \int{\frac{\delta E_{\text{xc}}}{\delta \nabla \nTilde(\bx)}\cdot\left[\int{\left(\frac{\delta\nabla \nTilde(\bx)}{\delta \psips_n(\by)}\delta \psips_n(\by) + \frac{\delta \nabla \nTilde(\bx)}{\delta \nabla \psips_n(\by)} \delta \nabla \psips_n(\by) \right)d\by}\right] d\bx} \nonumber \\
 & +\sum_a{\sum_{\alpha\beta}{\frac{\delta \Delta E^a}{\delta D^a_{\alpha\beta} }\int{\frac{\delta D^a_{\alpha\beta}}{\delta \psips_n(\by)}\delta \psips_n(\by)d\by}}} = 2\sum_j{\frac{(\lambda_{nj}+ \lambda_{jn})}{2}\int{\delta \psips_n(\bx)\mathcal{S}\psips_j(\bx)d\bx}}
\end{align}
where,
\begin{equation*}
    \frac{\delta \Delta E^a[\{D^a_{\alpha\beta}\}]}{\delta D^a_{\alpha\beta} } = \left(\Delta T^a_{\alpha\beta} + \Delta E^a_{\text{el},\alpha\beta} + 2\sum_{\alpha'\beta'}{\Delta E^a_{\text{el},\alpha\beta \alpha'\beta'}D^a_{\alpha'\beta'}} + \Delta V^{a, \alpha\beta}_{\text{xc}} - \int_{\Omega_a}{\bar{v^a}(r)S_{00}(\widehat{\boldsymbol{\theta}})\phips^a_\alpha(\bx)\phips^a_{\beta}(\bx)d\bx} \right) =  \Delta h^a_{\alpha\beta}
\end{equation*}
with
\begin{equation*}
 \int{\frac{\delta D^a_{\alpha\beta}}{\delta \psips_n(\by)}\delta \psips_n(\by)d\by} = 4\,f_n\int{\pTilde_{\alpha}^a(\bx) \delta \psips_n(\bx)d\bx}\int{\pTilde_{\beta}^a(\by) \psips_n(\by)d\by}
\end{equation*}
Finally, we have
\begin{align}
  &f_n\int{\frac{1}{2}\left(\nabla \delta\psips_n(\bx)\cdot \nabla\psips_n(\bx) +\left(\frac{\delta E_{\text{xc}}}{\delta \nTilde(\bx)} + \varphi(\bx)\right)\delta\psips_n(\bx)\psips_n(\bx)\right)d\bx} + f_n\int{\frac{\delta E_{\text{xc}}}{\delta \nabla \nTilde(\bx) }\cdot\left( \psips_n(\bx)\delta \psips_n(\bx) + \delta \psips_n(\bx)\psips_n(\bx) \right) d\bx} \nonumber\\
  &+ f_n\sum_a{\sum_{\alpha\beta}{\int{\pTilde^a_\alpha(\by)\delta\psips_n(\bx)d\bx}{{\left(\Delta h^a_{\alpha\beta}+\sum_{lm}\Delta^a_{lm\alpha\beta}{\int{\gTilde_{lm}^a(\bx)\varphi(\bx)d\bx}}\right)}}\int{\pTilde^a_\beta(\by)\psips_n(\by)d\by}= \frac{1}{2}\sum_j{\lambda^{S}_{nj}\int{\delta \psips_n(\bx)\mathcal{S}\psips_j(\bx)d\bx}}}}
  \label{eq: weak form GEVP}
\end{align}
where the entries $\lambda^{S}_{nj} = \frac{\lambda_{nj}+\lambda_{jn}}{2}$ are the entries of a symmetric matrix $\Lambda^{S} = \frac{1}{2}(\Lambda + \Lambda^T)$.
Similarly, taking the variational derivative of Eqn.~\eqref{eqn:KSfunctionalLag} with respect to $\varphi(\bx)$, we get
\begin{equation}
    \frac{1}{4\pi}\int{\nabla\varphi(\bx)\cdot\nabla\delta \varphi(\bx)d\bx} = \int{\rhoTilde(\bx)\delta \varphi(\bx)d\bx}\label{eq: weak form poisson}
\end{equation} 
where the simulation domain for Eqns.\eqref{eq: weak form GEVP} and \eqref{eq: weak form poisson} is represented by $\Omega$. The variatonal quantities are such that $\psips_n(\bx) \in \mathbb{X}$ and $\varphi(\bx)\in \gamma$ with $\mathbb{X}=\gamma = H^1_0(\Omega)$ in the case of non-periodic problems and $\mathbb{X}=\gamma=H^1_{\text{per}}(\Omega)$ in the case of periodic problems.
 Now, we let $\mathbb{V}^h \in \mathbb{X}$, $\mathbb{V}^h_{\text{el}}\in \gamma$ denote the finite-dimensional subspaces of dimensions $M$ and $M_{el}$ respectively, and in the discrete setting, we seek the approximate solution $\psips_n^h(\bx)\in\ \mathbb{V}^h$, $\varphi^h(\bx)\in \mathbb{V}^h_{\text{el}}$  such that 
\begin{align}
  &f_n\int_{\Omega}{\frac{1}{2}\left(\nabla \delta\psips_n(\bx)\cdot \nabla\psips^h_n(\bx) +\left(\frac{\delta E_{\text{xc}}}{\delta \nTilde(\bx)} + \varphi^h(\bx)\right)\delta\psips_n(\bx)\psips^h_n(\bx)\right)d\bx} + f_n\int{\frac{\delta E_{\text{xc}}}{\delta \nabla \nTilde(\bx) }\cdot\left( \psips^h_n(\bx)\delta \psips_n(\bx) + \delta \psips_n(\bx)\psips^h_n(\bx) \right) d\bx} \nonumber\\
  &+ f_n\sum_a{\sum_{\alpha\beta}{\int{\pTilde^a_\alpha(\by)\delta\psips_n(\bx)d\bx}{{\left(\Delta h^a_{\alpha\beta}+\sum_{lm}\Delta^a_{lm\alpha\beta}{\int{\gTilde^a_{lm}(\bx)\varphi^{h}(\bx)d\bx}}\right)}}\int{\pTilde^a_\beta(\by)\psips^h_n(\by)d\by}}}= \frac{1}{2}\sum_j{\lambda^{S}_{nj}\int{\delta \psips_n(\bx)\mathcal{S}\psips^h_j(\bx)d\bx}} \nonumber \\
  &\forall\; \delta \psips_n(\bx) \in \mathbb{V}^h 
\label{eq: finite space weak form}
\end{align}
and 
\begin{equation}
    \frac{1}{4\pi}\int{\nabla\varphi^h(\bx)\cdot\nabla\delta \varphi(\bx)d\bx} = \int{\rhoTilde(\bx)\delta \varphi(\bx)d\bx}\;\;\; \forall\; \delta \varphi \in  \mathbb{V}^h_{\text{el}}  \label{eq: weak form poisson2} 
\end{equation}
Further, let the finite dimensional subspace $\mathbb{V}^h$ be spanned by the finite-element basis $\{ N^{h,p}_I(\bx) \}\;\; \text{for} \;\; 1 \leq I \leq  M$ and the subspace $\mathbb{V}^h_{\text{el}}$ be spanned by the FE basis $\{N^{h,p_{\text{el}}}_I(\bx)\},\; \text{for}\; 1 \leq I \leq   M_{\text{el}}$. This allows us to express the approximate solution $\psips^h_n(\bx)$ uniquely as $\psips^h_n(\bx) = \sum_J{N^{h,p}_J(\bx)u_n^J}$ and, subsequently $\varphi^h(\bx)$ as $\varphi^h(\bx) = \sum_J{N_J^{h,p_{\text{el}}}(\bx)\varphi^J}$.
 Moreover, since Eqn.\eqref{eq: finite space weak form} is valid for all $\;\delta \psips_n(\bx) \in \mathbb{V}^h$, choosing $\delta \psips_n(\bx) = N^{h,p}_I(\bx)$ successively for every $I$, we now obtain the FE discretized equations to be solved in the form of the following matrix equations 
\begin{align}
    &f_n\sum_{J}{\int{\frac{1}{2}\nabla N^{h,p}_I(\bx)\cdot \nabla N^{h,p}_J(\bx)u_n^J d\bx}} + f_n\sum_J{\int{V^h_{\text{eff}}(\bx)N_I^{h,p}(\bx)N_J^{h,p}(\bx)u_n^J d\bx}} \nonumber \\ + & f_n\sum_J{\int{\boldsymbol{V}^h_{\text{GGA}}\cdot\left(N^{h,p}_I(\bx)\nabla N_J^{h,p}(\bx) + \nabla N_I^{h,p}(\bx)N^{h,p}_J(\bx)\right)u_n^J d\bx}} \nonumber \\
    &+ f_n\sum_J{\sum_{a,\alpha\beta}\int{N^{h,p}_I(\bx)\pTilde^a_\beta(\bx)d\bx \left({{\Delta h^a_{\alpha\beta}+\sum_{lm}{\Delta^a_{lm\alpha\beta}{\int{\gTilde^a_{lm}(\bx)\varphi^h(\bx)d\bx}}}}}\right)\int{\pTilde_\alpha^a(\by)N_J^{h,p}(\by)u_n^Jd\by}}} \nonumber\\
    &= \frac{1}{2}\sum_{j}{\lambda^{S}_{nj}\sum_J{\left(\int{N^{h,p}_I(\bx)N^{h,p}_J(\bx)u_j^Jd\bx} + {\sum_{a,\alpha\beta}\int{N^{h,p}_I(\bx)\pTilde^a_\beta(\bx)d\bx \sqrt{4\pi}\Delta ^a_{00\alpha\beta}\int{\pTilde_\alpha^a(\by)N_J^{h,p}(\by)u_j^Jd\by}}} \right)} }
\end{align}
Rewriting the above expression in the matrix-form we have $\bH\boldsymbol{\mathsf{U}} = \bS\boldsymbol{\mathsf{U}}\boldsymbol{\Lambda}^S $ 
% \begin{equation}
%     \bH\boldsymbol{\mathsf{U}} = \bS\boldsymbol{\mathsf{U}}\boldsymbol{\Lambda}^S \label{eqn:discgenevp}
% \end{equation}
where, the matrix $\boldsymbol{\mathsf{U}} = [\boldsymbol{\mathsf{u}}_1... \boldsymbol{\mathsf{u}}_n ...\boldsymbol{\mathsf{u}}_N]$ with the entries of the vector $\boldsymbol{\mathsf{u}}_n$ given by $u^{J}_n$ for $1 \leq J \leq  M$ which are the linear combination coefficients of $\psips^h_n(\bx)$ when expressed in FE basis. Now, we note that the matrix $\boldsymbol{\Lambda}^{S}$ is a symmetric matrix which is orthogonally diagonalizable to be of the form $\boldsymbol{\Lambda}^S = \bQ\bD\bQ^T$
 and let $2\epsilon^h_nf_n$ be the entries of the diagonal matrix $\bD$. Furthermore, introducing the unitary transformation of the discretized wavefunction matrix $\boldsymbol{\mathsf{U}}$ using the orthogonal matrix $\bQ$ diagonalizing $\boldsymbol{\Lambda}^{S}$, we have $\bfePsipsFull = \boldsymbol{\mathsf{U}}\bQ$ which can then be used to transform the matrix equation $\bH\boldsymbol{\mathsf{U}} = \bS\boldsymbol{\mathsf{U}}\boldsymbol{\Lambda}^S$ to obtain the following generalized eigenvalue problem in the PAW method
\begin{equation}
    \bH\feNodalVectorUtilde_n = \epsilon^h_n\bS\feNodalVectorUtilde_n
\end{equation}
where the entries of the FE discretized Hamiltonian ($\bH$) and the PAW overlap matrix are given by Eqns.\eqref{eqn:mij} to \eqref{eqn:deltaHanddeltaS} in Section \eqref{sec:fem}. Additionally, since Eqn.\eqref{eq: weak form poisson2} is satisfied $\forall\; \delta \varphi \in  \mathbb{V}^h_{\text{el}}$, we choose $\delta \varphi(\bx) = N_I^{h,p_{\text{el}}}(\bx)$ successively for all $I$ to finally obtain the following FE discretized Poisson's equation
\begin{equation}
  \frac{1}{4\pi}\sum_J{\int{\nabla N_I^{h,p_{\text{el}}}(\bx)\cdot N_J^{h,p_{\text{el}}}(\bx)\varphi^Jd\bx}} = \int{\rhoTilde^h(\bx)N_I^{h,p_{\text{el}}}(\bx)d\bx}
  \end{equation}

\section{Chebyshev filtered subspace iteration}
As discussed in Section \eqref{sec:scf}, we use a self-consistent field iteration (SCF) procedure to solve the nonlinear PAW generalized eigenvalue problem. This involves the solution of a linearized generalized eigenvalue problem at each SCF iteration for which we employ the Chebyshev filtered subspace iteration(ChFSI) approach. As outlined in Algo.\eqref{alg: ChFSI}, ChFSI requires upper and lower bounds of the eigenspectrum corresponding to the FE discretized eigenvalue problem $\bH\feNodalVectorUtilde = \epsilon^h_n\bS\feNodalVectorUtilde$ for which we use the generalized Lanczos procedure. The details of the generalized Lanczos algorithm are discussed below in Algo.\eqref{alg: Generalized Lanczos}.
\begin{figure}[H]
%    \removelatexerror
\begin{algorithm}[H]
    \caption{$k$-step generalized Lanczos iteration} \label{alg: Generalized Lanczos}
%\KwIn{Initial subspace $\Psips_{\text{in}}$}
\KwOut{Estimate of $b$ the upper bound of unoccupied spectrum of $\bH\bfePsips=\epsilon^h \widehat{\bS} \bfePsips$} 
\begin{enumerate}
\item Choose random vector $\bv_0$, such that $||\bv_0||_{\widehat{\bS}}=1$, where $||\bv||_{\widehat{\bS}} = (\bv^*\widehat{\bS}\bv)^{\frac{1}{2}}$
\item Compute entries of $k\cross k$ tridiagonal matrix $\bT$ whose entries are $T_{i,i} = \alpha_i$ and $T_{i+1,i}=T_{i,i+1} = \beta_i$ as follows:
\item $\alpha_1 = (\bv_0^*\bH\bv_0)$
\item $\bu_0 =\widehat{\bS}^{-1}\bH\bv_0 - \alpha_1 \bv_0$
\item for $i=1\; \text{to}\;k\;$ do 
\begin{itemize}
    \item $\beta_i = ||\bu_{i-1}||_{\widehat{\bS}}$
    \item $\bv_i = \frac{1}{\beta_i}\bu_{i-1}$
    \item $\bu_i =\widehat{\bH}\bv_i - \beta_i\bv_{i-1}$
    \item $\alpha_i = (\bv_i^*\bH\bv_i)$
    \item $\bu_i = \bu_i - \alpha_i\bv_i$
\end{itemize}
\item compute $\lambda_T^{\text{max}}$ the max eigenvalue of $\bT$
\item $\beta_{k+1} = ||\bu_{i}||_{\widehat{\bS}}$
\item $b=\lambda_T^{\text{max}}+\beta_{k+1}$
\end{enumerate}
\end{algorithm}
\end{figure}
     
\end{widetext}

\end{document}